\documentclass[a4paper,runningheads]{llncs}

\usepackage[english]{babel}
\usepackage[T1]{fontenc}
\usepackage[utf8x]{inputenc}

\usepackage{dejavu}
\usepackage{montserrat}
\usepackage{carlito}
\usepackage{lmodern}
\usepackage{pifont}

\usepackage{amsmath,amsthm,amssymb}
\usepackage{calrsfs}
\DeclareMathAlphabet{\pazocal}{OMS}{zplm}{m}{n}
\usepackage{subcaption,graphicx,wrapfig}
\usepackage{xfrac,xspace}
\usepackage[dvipsnames]{xcolor}
\usepackage{marvosym}  % \Yinyang
\usepackage[inline]{enumitem}
\usepackage{booktabs,multirow,array} %,colortbl}
\usepackage[normalem]{ulem}
\usepackage[symbol*,hang,marginal,bottom]{footmisc}  % [symbol*,perpage,marginal]
\usepackage{listings,algorithm,algorithmicx,algpseudocode}
\usepackage{tikz,tikzsymbols,pgfplots}
\usepackage{relsize}
\usepackage{breakcites}
\usepackage{verbatim}
\usepackage{lipsum}
\usepackage[breaklinks=true]{hyperref}
\usepackage[nameinlink]{cleveref}

\newboolean{shownotes}
\setboolean{shownotes}{false}

\newboolean{longversion}
\setboolean{longversion}{true}

\usetikzlibrary{arrows,arrows.meta,backgrounds,calc,shapes,pgfplots.groupplots,positioning,automata,plotmarks,shapes.gates.logic.US,shapes.gates.logic.IEC,patterns}
\tikzstyle{every picture}+=[remember picture, overlay, anchor=center]
\tikzstyle{every node}=[initial text=]
\tikzstyle{every picture}=[>=stealth]
\tikzstyle{BE}=[ellipse, draw, fill=normalfill, minimum width=7mm, inner sep=1]
\graphicspath{{images/},{results/},{models/figures/}}  % figs dirs

\Crefname{equation}{eq.}{eqs.}
\crefname{equation}{equation}{equations}
\Crefname{figure}{Fig.}{Figs.}
\crefname{figure}{figure}{figures}
\Crefname{tabular}{Table}{Tables}
\crefname{tabular}{table}{tables}
\Crefname{definition}{Def.}{Defs.}
\crefname{definition}{definition}{definitions}
\Crefname{proposition}{Prop.}{Props.}
\crefname{proposition}{proposition}{propositions}
\Crefname{section}{Sec.}{Secs.}
\crefname{section}{section}{sections}
\Crefname{algorithm}{Algorithm}{Algorithms}
\crefname{algorithm}{algorithm}{algorithms}
\crefname{labelenumi}{}{}
\crefname{labelenumii}{}{}

\crefname{listing}{code}{codes}  % https://tex.stackexchange.com/a/86177

\newcommand{\whitestar}{\ding{73}}

\newcommand{\cmarkbf}{\ding{52}}
\newcommand{\xmark}{\ding{55}}
\newcommand{\xmarkbf}{\ding{56}}

\DefineFNsymbols{lamport}{
  {$\hspace{2pt}$\whitestar$\hspace{2pt}$}
  {$\hspace{1pt}\dagger\hspace{.2pt}$}
  {$\hspace{1pt}\ddagger\hspace{.2pt}$}
  {$\hspace{1pt}\ast\hspace{.2pt}$}
  {\hspace{1pt}\S\hspace{.2pt}}
  {\hspace{1pt}\P\hspace{.2pt}}
  {$\hspace{1pt}\|\hspace{.2pt}$}
  {$\hspace{1pt}\dagger\dagger\hspace{.2pt}$}
  {$\hspace{1pt}\ddagger\ddagger\hspace{.2pt}$}
  {$\hspace{1pt}\ast\ast\hspace{.2pt}$}
}
\setfnsymbol{lamport}

\newcommand{\dejavu}[1]{{\fontfamily{DejaVuSans-TLF}\selectfont{#1}}}
\newcommand{\carlitos}[1]{{\carlitoTOsF#1}}

\ifthenelse{\boolean{shownotes}}{
  \renewcommand{\note}[1][]{%
    ~\par\vspace{.5ex}\noindent\parbox{\linewidth}{
      \color{blue}\sffamily\textbf{NOTE:} \smaller\color{Blue}#1}
    ~\\[-1ex]}
  \newcommand{\sketch}[1][]{\textsl{\color{Brown}\textbf{SKETCH:} #1}\xspace}
  \newcommand{\todo}[1][]{\textbf{\sffamily\textcolor{Red}{TODO:} #1}%
    \marginpar{\textsf{\color{red}\bfseries TODO}}}
  \newcommand{\tocite}[1][REF!]{\carlitos{\color{Red}\bfseries[cite #1]}\xspace}
  \newcommand{\sidenote}[3][WildStrawberry]{\marginpar{\smaller\textit{%
	\color{#1}#2}}\textcolor{#1}{{#3}}}
  \newcommand{\onlysidenote}[2][WildStrawberry]{\marginpar{\smaller\textit{%
	\color{#1}#2}}}
  \newcommand{\onlysidenoter}[2][WildStrawberry]{\reversemarginpar\marginpar{%
	\smaller\textit{\color{#1}#2}}}
  \newcommand{\msnote}[1]{{\it MS: #1}}
  \newcommand{\commentMS}[1]{{\it\color{Red} MS: #1}}
}{
  \renewcommand{\note}[1][]{}
  \newcommand{\sketch}[1][]{}
  \newcommand{\todo}[1][]{}
  \newcommand{\tocite}[1][]{}
  \newcommand{\sidenote}[3][]{#3}
  \newcommand{\onlysidenote}[2][]{}
  \newcommand{\onlysidenoter}[2][]{}
  \newcommand{\msnote}[1]{}
  \newcommand{\commentMS}[1]{}
}

\newcommand{\N}{\ensuremath{\mathbb{N}}\xspace}

\newcommand{\R}{\ensuremath{\mathbb{R}}\xspace}

\newcommand{\AVA}{\ensuremath{\mathrm{AVA}}\xspace}
\newcommand{\UNAVA}{\ensuremath{\mathrm{UNAVA}}\xspace}
\newcommand{\REL}[1][T]{\ensuremath{\mathrm{REL}%
  \if\relax\detokenize{#1}\relax\else_{#1}\fi}\xspace}
\newcommand{\UNREL}[1][T]{\ensuremath{\mathrm{UNREL}%
  \if\relax\detokenize{#1}\relax\else_{#1}\fi}\xspace}
\newcommand{\Tree}{\ensuremath{\triangle}\xspace}
\newcommand{\xbf}{\ensuremath{\boldsymbol{x}}\xspace}
\newcommand{\zbf}{\ensuremath{\boldsymbol{z}}\xspace}
\newcommand{\idx}[1][]{\ensuremath{\mathit{idx}%
  \if\relax\detokenize{#1}\relax\else(#1)\fi}\xspace}
\newcommand{\nodev}{\ensuremath{v}\xspace}
\newcommand{\nodew}{\ensuremath{w}\xspace}
\newcommand{\nodes}[1][\Tree]{\ensuremath{\mathit{nodes}%
  \if\relax\detokenize{#1}\relax\else(#1)\fi}\xspace}
\newcommand{\type}[1][]{\ensuremath{\mathit{type}%
  \if\relax\detokenize{#1}\relax\else(#1)\fi}\xspace}
\newcommand{\child}[1][]{\ensuremath{\mathit{chil}%
  \if\relax\detokenize{#1}\relax\else(#1)\fi}\xspace}
\newcommand{\clocks}{\pazocal{C}\xspace}
\newcommand{\actions}{\pazocal{A}\xspace}
\newcommand{\inactions}{\actions^{\mathsf i}\xspace}
\newcommand{\outactions}{\actions^{\mathsf o}\xspace}
\newcommand{\uractions}{\actions^{\mathsf u}\xspace}
\newcommand{\states}{\pazocal{S}\xspace}
\newcommand{\trans}[1][]{\xrightarrow{{#1}}}

\newcommand{\safe}[1][]{\ifthenelse{\equal{#1}{}}{\mathrm{safe}}{\mathrm{safe\hspace{1pt}}(#1)}}

\newcommand{\leaf}[1]{\ensuremath{\mathsf{#1}}\xspace}
\newcommand{\gate}[1]{\ensuremath{\mathsf{\MakeUppercase{#1}}}\xspace}
\newcommand{\ANDgate}{\gate{and}}
\newcommand{\ORgate}{\gate{or}}
\newcommand{\VOTgate}{\gate{vot}}
\newcommand{\PANDgate}{\gate{pand}}
\newcommand{\SPAREgate}{\gate{spare}}
\newcommand{\FDEPgate}{\gate{fdep}}
\newcommand{\BE}{\gate{be}}
\newcommand{\SBE}{\gate{sbe}}

\newcommand{\RBOX}{\gate{rbox}}
\newcommand{\code}[1]{\ensuremath{\mathtt{\smaller#1}}\xspace}
\newcommand{\acronym}[1]{\ensuremath{\text{\normalfont\scshape{\larger #1}}}\xspace}
\newcommand{\ci}{{\normalfont{\acronym{ci}}}\xspace}
\newcommand{\pdf}{{\normalfont{\acronym{pdf}}}\xspace}
\newcommand{\iosa}{{\normalfont{\acronym{iosa}}}\xspace}
\newcommand{\fta}{\acronym{fta}}
\newcommand{\ft}{\acronym{ft}}
\newcommand{\dft}{\acronym{dft}}
\newcommand{\rft}{\acronym{rft}}
\newcommand{\smc}{\acronym{smc}}
\newcommand{\res}{\acronym{res}}
\newcommand{\ISPLIT}{\acronym{isplit}}
\newcommand{\restart}{\acronym{restart}}
\newcommand{\fe}{\acronym{fe}}
\ifthenelse{\boolean{longversion}}
  {%
  \newcommand{\fen}[1]{\mbox{\acronym{fe}-{#1}}\xspace}
  \newcommand{\rstn}[1]{\mbox{\acronym{rst}-{#1}}\xspace}
  \newcommand{\rstes}{\mbox{\acronym{rst-es}}\xspace}
  }{%
  \newcommand{\fen}[1]{\acronym{fe}{\textsubscript{#1}}\xspace}
  \newcommand{\rstn}[1]{\acronym{rst}{\textsubscript{#1}}\xspace}
  \newcommand{\rstes}{\acronym{rst}{\textsubscript{\sc es}}\xspace}
  }
\newcommand{\avoid}{\ensuremath{\text{\smaller\color{red!75!black}\,\xmarkbf}}}
\newcommand{\goal}{\ensuremath{\text{\smaller\color{green!55!black}\,\cmarkbf}}}
\newcommand{\trunc}{\ensuremath{\text{\smaller\color{Yellow!70!black}\,\xmark}}}

\newcommand{\IFUN}[1][]{\ensuremath{\pazocal{I}%
  \if\relax\detokenize{#1}\relax\else\left(#1\right)\fi}\xspace}

\newcommand{\expnum}[3]{\ensuremath{#1{\mathsmaller\times\hspace{-.7pt}}%
  10^{\text{\hspace{.5pt}\raisebox{.1pt}{#2}\hspace{-.2pt}}#3}}\xspace}
\newcommand{\rarep}[2]{\expnum{#1}{-}{#2}}
\newcommand{\fig}{\normalfont{\textsmaller{\montserratalt{FIG}}}\xspace}
\newcommand{\cs}[1][]{\ensuremath{\mathsmaller{\pazocal{C\hspace{-1.5pt}S}}%
  \if\relax\detokenize{#1}\relax\else_\mathfrak{#1}\fi}\xspace}

\newcommand{\algo}{\textit{algo}\xspace}

\makeatletter
\renewcommand{\paragraph}{\@startsection{paragraph}{5}{0em}%
  {.7ex plus .2ex minus .1ex}%
  {-.5em}%
  {\bfseries}}
\makeatother
\ifthenelse{\boolean{shownotes}}{
  \usepackage[pagewise]{lineno}  % [pagewise,switch,modulo]
  \linenumbers
  
}{}

\begin{document}

\iftrue  % LNCS first page
	\title{Rare event simulation for\\non-Markovian repairable fault trees%
	\thanks{%
	This work was partially funded by NWO, NS, and ProRail project 15474 (\emph{SEQUOIA}), ERC grant 695614 (\emph{POWVER}), EU project 102112 (\emph{SUCCESS}), ANPCyT PICT-2017-3894 (\emph{RAFTSys}), and SeCyT project 33620180100354CB (\emph{ARES}).
	}}
	\titlerunning{%
		\texorpdfstring{\res}{RES} for non-Markovian repairable
		\texorpdfstring{\ft{s}}{FTs}}
	\author{%
		Carlos E.~Budde\inst{1}\and
		Marco Biagi\inst{2}\and
		Ra\'ul E.~Monti\inst{1}\and
		Pedro R.~D'Argenio\inst{3,4,5}\and
		Mari\"elle Stoelinga\inst{1,6}
	}
	%
	%\authorrunning{Budde et al.}
	%\authorrunning{\hspace{12mm}Budde, Biagi, Monti, D'Argenio, Stoelinga}  % ugly :/
	%
	\institute{
		Formal Methods and Tools, University of Twente, Enschede, the Netherlands
			\email{\{c.e.budde,r.e.monti,m.i.a.stoelinga\}@utwente.nl}
		\and
		Department of Information Engineering, University of Florence, Florence, Italy
			\email{marco.biagi@unifi.it}
		\and
		FAMAF, Universidad Nacional de C\'ordoba, C\'ordoba, Argentina
			\email{dargenio@famaf.unc.edu.ar}
		\and
		CONICET, C\'ordoba, Argentina
		\and
		Department of Computer Science, Saarland University, Saarbr\"ucken, Germany
		\and
		Department of Software Science, Radboud University, Nijmegen, the Netherlands
	}
\else  % IEEE first page
	\title{Rare event simulation for \\ non-Markovian repairable fault trees}
	\author{
	\IEEEauthorblockN{
		Carlos E.\ Budde\IEEEauthorrefmark{1}
		Marco Biagi\IEEEauthorrefmark{2}
		Ra\'ul E.\ Monti\IEEEauthorrefmark{3}\IEEEauthorrefmark{4}
		Pedro R.\ D'Argenio\IEEEauthorrefmark{3}\IEEEauthorrefmark{4}\IEEEauthorrefmark{5}
		Mari\"elle Stoelinga\IEEEauthorrefmark{1}\IEEEauthorrefmark{6}
	}
	\IEEEauthorblockA{\IEEEauthorrefmark{1}{University of Twente,
		Formal Methods and Tools, Enschede, the Netherlands}}
	\IEEEauthorblockA{\IEEEauthorrefmark{2}{University of Florence,
		Department of Information Engineering, Florence, Italy}}
	\IEEEauthorblockA{\IEEEauthorrefmark{3}Universidad Nacional de C\'ordoba,
		FAMAF, C\'ordoba, Argentina}
	\IEEEauthorblockA{\IEEEauthorrefmark{4}CONICET,
		C\'ordoba, Argentina}
	\IEEEauthorblockA{\IEEEauthorrefmark{5}Saarland University,
		Department of Computer Science, Saarbr\"ucken, Germany}
	\IEEEauthorblockA{\IEEEauthorrefmark{6}{Radboud University Nijmegen,
		Department of Software Science, Nijmegen, the Netherlands}}
	{\texttt{\small
		\{c.e.budde,m.i.a.stoelinga\}@utwente.nl
		marco.biagi@unifi.it
		\{dargenio,rmonti\}@famaf.unc.edu.ar}}
	}
\fi

\maketitle
\setcounter{footnote}{1}
% page numbers:
\thispagestyle{plain}
\pagestyle{plain}
% Listings customisation
\renewcommand{\thelstlisting}{\arabic{lstlisting}}  % single # in cter

\begin{abstract}
% Do not put math, special symbols or citations in the abstract

Dynamic fault trees (\dft) are widely adopted in industry to assess the dependability of safety-critical equipment.
Since many systems are too large to be studied numerically, \dft{s} dependability is often analysed using Monte Carlo simulation.
A bottleneck here is that many simulation samples are required in the case of rare events, e.g.\ in highly reliable systems where components fail seldomly.
Rare event simulation (\res) provides techniques to reduce the number of samples in the case of rare events.
We present a \res technique based on importance splitting, to study failures in highly reliable \dft{s}.
Whereas \res usually requires meta-information from an expert, our method is fully automatic:
By cleverly exploiting the fault tree structure we extract the so-called importance function.
We handle \dft{s} with Markovian and non-Markovian failure and repair distributions---for which no numerical methods exist---and show the efficiency of our approach on several case studies.

%%  % Previous version:
%%	Dynamic fault trees (\dft) are widely adopted in industry to assess the dependability of safety-critical systems.
%%	Usually too complex to be performed numerically, \dft analysis is also approached via Monte Carlo simulation.
%%	Studying faults in highly reliable systems and other rare events poses an extra challenge, since infeasible simulation times are needed when e.g.\ components seldom fail.
%%	In these cases, rare event simulation (\res) is a key methodology that provides means to reduce the simulation time.
%%	
%%	We present a \res technique for \dft analysis, that determines simulation splitting points based on the hierarchical tree structure.
%%	Key contributions include % (but may not be limited to)
%%	\begin{enumerate*}[label=(\emph{\roman*})]
%%	\item	an automatic and compositional derivation of the so-called
%%			importance function required to implement (importance splitting) \res;
%%	\item	a framework for \res oriented to quantitative analysis of dependability
%%			metrics for \dft{s};
%%	\item	support for repairable \dft{s};
%%	\item	support for arbitrary fail and repair \pdf{s} of components;
%%	\item	an implementation of our \res framework in a software tool chain.
%%	\end{enumerate*}
%%	We show the efficiency of our approach on several case studies.
%%
%%	\keywords{Rare event simulation, Dynamic fault trees, System reliability analysis}
\end{abstract}

%%%%%%%%%%%%%%%%%%%%%%%%%%%%%%%%%%%%%%%%%%%%%%%%%%%%%%%%%%%%%%%%%%%%%%%%%%%%%%
%% !TEX root =  ../main.tex
\section{Introduction}
\label{sec:intro}

Reliability engineering is an important field that provides methods and tools to assess and mitigate the risks related to complex systems.
Fault tree analysis (\fta) is a prominent technique here.
Its application encompasses a large number of industrial domains that range from automotive and aerospace system engineering, to energy and telecommunication systems and protocols.

\paragraph{Fault trees.}
A fault tree (\ft) describes how component failures occur and propagate through the system, eventually leading to system failures. 
Technically, an \ft is a directed acyclic graph whose leaves model component failures, and whose other nodes (called gates) model failure propagation.
%Technically, an \ft is a tree (or rather, a directed acyclic graph, since subtrees can be shared) whose leaves model component failures, and whose gates model failure propagation, see Figure~\ref{fig:tiny_RWC} for an example. 
Using fault trees one can compute dependability metrics to quantify how a system fares w.r.t.\ certain performance indicators. 
Two common metrics are system \emph{reliability}---the probability that there are no system failures during a given mission time---and system \emph{availability}---the average percentage of time that a system is operational. 

\emph{Static fault trees} (aka standard \ft{s}) contain a few basic gates, like \ANDgate and \ORgate gates.
This makes them easy to design and analyse, but also limits their expressivity. 
\emph{Dynamic fault trees} (\dft{s}~\cite{DBB90,VSDFMR02}) are a common and widely applied extension of standard \ft{s}, catering for more complex dependability patterns, 
like spare management  and causal dependencies.
To model these patterns, \dft{s} come with additional gates, for instance \SPAREgate, \PANDgate, and \FDEPgate.

Such gates make \dft{s} more difficult to analyse.
In static \ft{s} it only matters whether or not a component has failed, so they can be analysed with Boolean methods, such as binary decision diagrams \cite{JGKRS15}.
Dynamic fault trees, on the other hand, crucially depend on the failure order, so Boolean methods are insufficient.
Moreover and on top of these two classes, \emph{repairable fault trees} (\rft \cite{bobbio2004parametric}) permit components to be repaired after they have failed.
This is crucial to model fault-tolerant systems more realistically.
Yet repairs make analyses even harder: it does not suffice to know which components failed, or in which order, but also if they are simultaneously failed.
The general rule is that the more complex the formalism, the more realistic the model, and the harder the analyses.

\Cref{fig:tiny_RWC} is an \rft with a top \ANDgate gate, a \SPAREgate (\leaf{Rcab}), and three leaves.

\paragraph{Fault tree analysis.}
The reliability/availability of a fault tree can be computed via numerical methods, such as probabilistic model checking.
This involves exhaustive explorations of state-based models such as interactive Markov chains \cite{ruijters2015fault}.
Since the number of states (i.e.\ system configurations) is exponential in the number of tree elements, analysing large trees remains a challenge today \cite{JGKRS15,ABCHS18}.
Moreover, numerical methods are usually restricted to exponential failure rates and combinations thereof, like Erlang and acyclic phase type distributions \cite{ruijters2015fault}.

Alternatively, fault trees can be analysed using (standard) Monte Carlo simulation (\smc \cite{GKS+14,ruijters2015fault,RGNS16}, aka statistical model checking).
Here, a large number of simulated system runs (\emph{samples}) is produced.
Reliability and availability are then statistically estimated from the resulting sample set.
Such sampling does not involve storing the full state space.
Therefore, \smc is much more memory efficient than numerical techniques.
Furhermore, \smc is not restricted to exponential probability distributions.
However, a known bottleneck of \smc are rare events: when the event of interest has a low probability (which is typically the case in highly reliable systems), millions of samples may be required to observe it. Producing these samples can take a unacceptably long simulation time.

\paragraph{Rare event simulation.}
To alleviate this problem, the field of rare event simulation (\res) provides techniques that reduce the number of samples \cite{RT09b}.
The two leading techniques are importance sampling and importance splitting.

\ifthenelse{\NOT\boolean{longversion}}{%
%%%%%%%%%%%%%%%%%%%%%%%%%%%%%%%%%%%%%%%%%%%%%%%%%%%%%%%%%%%%%%%%%%%%%%%%
\emph{Importance sampling} tweaks the probabilities in a model, then computes the metric of interest for the changed system, and finally adjusts the analysis results to the original model \cite{Hei95,NSN01}.
Unfortunately it has specific requirements on the stochastic model: in particular, it is generally limited to Markov models.

\emph{Importance splitting}, deployed in this paper, does not have this limitation.
Importance splitting relies on rare events that arise as a sequence of less rare intermediate events \cite{KH51,Bay70}.
We exploit this fact by generating more (partial) samples on paths where such intermediate events are observed.
As a simple example, consider a biased coin whose probability of heads is $p=\sfrac{1}{80}$.
Suppose we flip it eight times in a row, and say we are interested in observing at least three heads.
If heads comes up at the first flip ($H$) then we are on a promising path.
%we generate more samples from this state $H$--- a state is as sequence in $\{H,T\}^{\leq 5}$ that records the outcomes seen thus far. 
%So, rather than continuing the current simulation run, we generate multiple runs from state $H$, and adjust the simulation outcome accordingly: say we start 10 simulation runs from $H$, 
%then each occurrence of the event of interest (i.e. seeing at least three $H$'s) is counted as $1/10$, rather than as 1.
We can then clone (\emph{split}) the current path $H$, generating e.g.\ 7 copies of it, each clone evolving independently from the second flip onwards.
Say one clone observes three heads---the copied $H$ plus two more.
Then, this observation of the rare event (three heads) is counted as $\sfrac{1}{7}$ rather than as 1 observation, to account for the splitting where the clone was spawned.
Now, if a clone observes a new head ($HH$), this is even more promising than $H$, so the splitting can be repeated.
If we make 5 copies of the $HH$ clone, then observing three heads in any of these copies counts as $\frac{1}{35} = \frac{1}{7}\cdot\frac{1}{5}$.
Alternatively, observing tails as second flip ($HT$) is less promising than heads.
One could then decide not to split~such~path.
%%%%%%%%%%%%%%%%%%%%%%%%%%%%%%%%%%%%%%%%%%%%%%%%%%%%%%%%%%%%%%%%%%%%%%%%
}{ %else  Importance Sampling / importance splitting
%%%%%%%%%%%%%%%%%%%%%%%%%%%%%%%%%%%%%%%%%%%%%%%%%%%%%%%%%%%%%%%%%%%%%%%%
\emph{Importance sampling} tweaks the probabilities in a model, then computes the metric of interest for the changed system, and finally adjusts the analysis results to the original model \cite{Hei95,NSN01}.
As a simple example consider a coin whose probability of heads is $p=\sfrac{1}{80}$.
We can increase that probability to $p'=\sfrac{1}{8}$, but count each occurrence of heads as $\sfrac{1}{10}$ rather than as 1.
This is typically denoted \emph{change of measure}. 
Thus, if we draw $n=1000$ samples with the increased probability $p'$, and we see 67 heads coming up, we estimate the probability on heads as $0.067= \frac{67}{1000}\cdot\frac{1}{10}$.
In the limit $n\rightarrow\infty$, the expected number of heads that come up is the same for the original and the tweaked model (after the adjustment).
However, sample outcomes have a lower variance in the tweaked model, so statistical analyses converge faster: few samples yield accurate estimations.

\emph{Importance splitting}, deployed in this paper, relies on rare events that arise as a sequential combination of less rare intermediate events \cite{KH51,Bay70}.
We  exploit this fact by generating more (partial) samples on paths where such intermediate events are observed.
In the coin example, suppose we flip it eight times in a row, and say we are interested in observing at least three heads.
If heads comes up at the first flip ($H$) then we are on a promising path.
%we generate more samples from this state $H$--- a state is as sequence in $\{H,T\}^{\leq 5}$ that records the outcomes seen thus far. 
%So, rather than continuing the current simulation run, we generate multiple runs from state $H$, and adjust the simulation outcome accordingly: say we start 10 simulation runs from $H$, 
%then each occurrence of the event of interest (i.e. seeing at least three $H$'s) is counted as $1/10$, rather than as 1.
We can then clone (\emph{split}) the current path $H$, generating e.g.\ 7 copies of it, 
each copy evolving independently from the second flip onwards.
Say one of them observes three heads---the cloned $H$ plus two more.
Then each observation of the rare event (three heads) is counted as $\sfrac{1}{7}$ rather than as 1, to account for the splitting that spawned the clone.
Now, if a clone observes a new head ($HH$), this is even more promising than $H$, so the splitting mechanism can be repeated.
If we make 5 copies of the $HH$ clone, then observing the event of interest in any of these copies counts as $\frac{1}{35} = \frac{1}{7}\cdot\frac{1}{5}$.
Alternatively, observing tails as second flip ($HT$) is less promising than heads.
One could then decide not to split such path.
%%%%%%%%%%%%%%%%%%%%%%%%%%%%%%%%%%%%%%%%%%%%%%%%%%%%%%%%%%%%%%%%%%%%%%%%
} %end if  Importance Sampling / importance splitting

This example highlights a key ingredient of importance splitting: the \emph{importance function}, that indicates for each state how promising it is w.r.t.\ the event of interest.
This function, together with other parameters such as thresholds~\cite{Gar00}, are used to choose e.g.\ the number of clones spawned when visiting a state.
An importance function for our example could be the number of heads seen thus far.
Another one could be such number, multiplied by the number of coin flips yet to come.
%The idea is that observing the event of interest should be more likely from states with higher importance: the degree to which the selected function reflects this property favours the efficiency of the importance splitting implementation.
The goal is to give \emph{higher importance} to states from which observing the \emph{rare event is more likely}.
%The goal is to give \uline{higher importance} to states from which observing the \uline{rare event is more likely}.
The efficiency of an importance splitting implementation increases as the importance function better reflects such property.

Rare event simulation has been successfully applied in several domains \cite{Rid05,VA07,XLL07,BM09,BBK09,VA18}.
However, a key bottleneck is that it critically relies on expert knowledge.
%for importance sampling,  a domain expert has to decide how to increase the probability of $p=1/80$ to $p'=1/8$. 
%Lower values of $p'$ require more samples, while higher values of $p'$ yield a biased estimator. 
In particular for importance splitting, finding a good importance function is a well-known highly non-trivial task \cite{RT09b,JLST15}.
%A key result in this paper is that, by exploiting the structure of the fault tree, we provide a method to automatically derive an importance function. 

\paragraph{Our contribution: rare event simulation for fault trees.}
This paper presents an importance splitting method to analyse \rft{s}.
In particular, we automatically derive an importance function by exploiting the description of a system as a fault tree.
This is crucial, since the importance function is normally given manually in an ad hoc fashion by a domain or \res expert.
We use a variety of \res algorithms based in our importance function, to estimate system unreliability and unavailability.
%% Our approach is as automatic as analysis by standard Monte Carlo simulation. \commentMS{I do not understand this last sentence}
Our approach can converge to precise estimations in increasingly reliable systems.
%, as shown by our experiments.
This method has four advantages over earlier analysis methods for \rft{s}---which we overview in the related work \cref{sec:relwork}---namely:
\begin{enumerate*}[label=(\arabic*)]
\item	we are able to estimate both the system reliability and availability;
\item	we can handle arbitrary failure and repair distributions;
\item	we can handle rare events; and
\item	we can do it in a fully automatic fashion.
\end{enumerate*}

Technically, we build local importance functions for the (automata-semantics of the) nodes of the tree.
We then aggregate these local functions into an importance function for the full tree.
Aggregation uses structural induction in the layered description of the tree.
%We then derive a composition strategy for these local importance functions, using structural induction in the gate-layered description of the \ft.
Using our importance function, we implement importance splitting methods to run \res analyses.
We implemented our theory in a full-stack tool chain.
With it, we computed confidence intervals for the unreliability and unavailability of several case studies.
Our case studies are \rft{s} whose failure and repair times are governed by arbitrary continuous probability density functions (\pdf{s}).
%%	Our case studies are \rft{s} whose failure and repair times follow exponential, Erlang, uniform, Rayleigh, Weibull, normal, and log-normal distributions.
Each case study was analysed for a fixed runtime budget and in increasingly resilient configurations.
In all cases our approach could estimate the narrowest intervals for the most resilient configurations.

%%	Remarkably, we work with \emph{repairable} fault trees, where components can be repaired after failing.
%%	Albeit crucial in fault-tolerant systems, original \dft{s} did not include repairs \cite{DBB90}; nowadays, recent extensions can handle these \cite{DBLP:conf/dsn/RaiteriIFV04,ruijters2015fault}.
%%	In particular we deploy the framework in \cite{Mon18}, where repairs are modelled through repair boxes, and failure and repair times are governed by arbitrary probability density functions (\pdf{s}).

\ifthenelse{\boolean{longversion}}{%
\paragraph{Organization of the paper.} 
We first introduce the formal concepts used for our mathematical definitions in \Cref{sec:background:fta,sec:background:smc}.
Then, we detail our theory to implement \res for repairable \dft{s} with arbitrary \pdf{s} in \Cref{sec:theory}.
For that, \Cref{sec:theory:ifun} introduces our (compositional) importance function, and \Cref{sec:theory:isplit} explains how to embed it into an automated framework for Importance Splitting \res.
Next, \Cref{sec:impl} describes how we implement this theory in our tool chain.
In \Cref{sec:expe} we show an extensive experimental evaluation that corroborates our expectations.
We finally overview related work in \Cref{sec:relwork}, and conclude our contributions in \Cref{sec:conclu}.
}{%
\paragraph{Paper outline.} 
%Essential mathematical background is provided in \Cref{sec:background:fta,sec:background:smc}.
Background on fault trees and \res is provided in \Cref{sec:background:fta,sec:background:smc}.
We detail our theory to implement \res for \rft{s} in \Cref{sec:theory}.
Using a tool chain, we performed an extensive experimental evaluation that we present in \Cref{sec:expe}.
We overview related work in \Cref{sec:relwork} and conclude our contributions in \Cref{sec:conclu}.
}

\ifthenelse{\boolean{longversion}}{%
\begin{table}[ht]
	\vspace{-4ex}
	\centering
	%% !TEX root =  ../main.tex
%
\begingroup
%
%\smaller
\def\head#1{\textbf{#1}}
\begin{tabular}{@{~~}l@{\hspace{2em}}l@{\qquad}l}
	\toprule
	\head{Scope} \vspace{.5ex}
		& \hspace{-2em}\head{Abbreviation}
		& \head{Meaning}\\
	\toprule
	\multirow{8}{*}{General}%
	& \pdf   & Probability density function\\
	& \ci    & Confidence interval\\
	& \fta   & Fault Tree Analysis\\
	& \ft    & Fault Tree\\
	& \dft   & Dynamic Fault Tree\\
	& \rft   & Repairable (Dynamic) Fault Tree\\
	& \smc   & Standard Monte Carlo simulation\\
	& \res   & Rare Event Simulation\\
	& \iosa  & \parbox[t]{16em}{Input/Output Stochastic Automata\\[-.3ex]
	                            with Urgency \cite{DM18}}\\
	\midrule
	\multirow{6}{*}{\parbox{5em}{Tree gates\\($m$ inputs)}}%
	& \ANDgate       & Conjunction: $m$-ary AND\\
	& \ORgate        & Disjunction: $m$-ary OR\\
	& $\VOTgate_k$   & Voting: $k$ out of $m$\\
	& \PANDgate      & Priority AND\\
	& \SPAREgate     & Spare: 1 primary \BE, $m$-$1$ spare \BE{s}\\
	& \FDEPgate      & \parbox[t]{15em}{Functional dependency:\\[-.3ex]
	                   1 trigger, $m$-$1$ dependent \BE{s}}\\
	\midrule
	\multirow{3}{*}{\parbox{5em}{Other\\tree nodes}}%
	& \BE    & Basic element\\
	& \SBE   & Spare basic element\\
	& \RBOX  & Repair box\\
	\midrule
	\multirow{7}{*}{\parbox{5em}{Case\\studies}}%
	& VOT    & Voting gates (synthetic)\\
	& DSPARE & Double-spare gates (synthetic)\\
	& RWC    & Railway cabinets \cite{GSS15,RRBS17}\\
	& HVC    & High voltage cab.\ (RWC subsys.)\\
	& RC     & Relay cab.\ (RWC subsys.)\\
	& FTPP   & Fault tolerant parallel processor \cite{DBB90}\\
	& HECS   & \parbox[t]{15em}{Hypothetical example computer\\[-.3ex]
	                            system \cite{VSDFMR02}}\\
	\bottomrule
\end{tabular}
\endgroup

	\vspace{2ex}
	\caption{Glossary of acronyms and abbreviations}
	\label{tab:glossary}
\end{table}
}{}

%
%%%%%%%%%%%%%%%%%%%%%%%%%%%%%%%%%%%%%%%%%%%%%%%%%%%%%%%%%%%%%%%%%%%%%%%%%%%%%%
%% !TEX root =  ../main.tex
\section{Fault tree analysis}
\label{sec:background:fta}

A fault tree `\Tree' is a directed acyclic graph that models how component failures propagate and eventually cause the full system to fail.
We consider repairable fault trees (RFTs), where failures and repairs are governed by
%\emph{repair boxes} manage the repair of failed basic elements.
arbitrary probability distributions. 

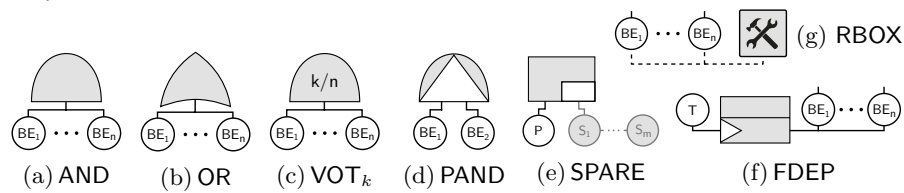
\begin{figure}[h]
	\vspace{-6ex}
	\centering
	%\includegraphics[width=.95\linewidth]{gates}
	%\caption{Nodes in \rft{s}: (spare) basic elements, gates, and repair boxes}
	\def\pic[#1,#2,#3,#4]{\begin{subfigure}{#1}\centering%
		\includegraphics[width=#1]{#2}\vspace{-.5ex}\caption{\!#3}\label{#4}%
		\end{subfigure}}
	\def\picl[#1,#2,#3,#4,#5]{\begin{subfigure}{#1}\centering%
		\includegraphics[width=#2]{#3}\vspace{-.5ex}\caption{\!#4}\label{#5}%
		\end{subfigure}}
	\def\piccap[#1,#2,#3]{\begin{subfigure}{#1}\centering
		\caption{#2}\label{#3}\end{subfigure}}
	\def\picncap[#1,#2]{\includegraphics[width=#1]{#2}}
	\begin{tikzpicture}
		%%%%%%%%%%%%%%%%%%%%%%%
		% Grid for orientation:
		%%	\def\xmin{0}\def\xmax{12}\def\ymin{0}\def\ymax{3}
		%%	\draw [step=0.5, Gray, thick] (\xmin,\ymin) grid (\xmax,\ymax);
		%%	\foreach \i in {\xmin,...,\xmax} {
		%%		\node [Gray,below] at (\i,\ymin) {\sf\i}; }
		%%	\foreach \i in {\ymin,...,\ymax} {
		%%		\node [Gray,left] at (\xmin,\i) {\sf\i}; }
		%%%%%%%%%%%%%%%%%%%%%%%
		\node at (0.0,0.18) {\pic[15mm,gate_and,\ANDgate,fig:gates:and]};
		\node at (1.7,0.18) {\pic[15mm,gate_or,\ORgate,fig:gates:or]};
		\node at (3.4,0.18) {\pic[15mm,gate_vot,$\VOTgate_k$,fig:gates:vot]};
		\node at (5.1,0.18) {\picl[15mm,11.3mm,gate_pand,\PANDgate,fig:gates:pand]};
		\node at (6.9,0.18) {\pic[18.5mm,gate_spare,\SPAREgate,fig:gates:spare]};
		\node at (9.5,0.0) {\pic[31mm,gate_fdep,\FDEPgate,fig:gates:fdep]};
		\node at (8.35,1.3) {\picncap[24mm,gate_rbox]};
		\node at (10.3,1.2) {\piccap[15mm,\RBOX,fig:gates:rbox]};
	\end{tikzpicture}
		\vspace{-2ex}
		\caption{Fault tree gates and the repair box}
	\label{fig:gates}
	\vspace{-4ex}
\end{figure}

\paragraph{Basic elements.} 
The leaves of the tree, called basic events or \emph{basic elements} (\BE{s}), model the failure of components.
A \BE $b$ is equipped with a failure distribution $F_b$ that governs the probability for $b$ to fail before time $t$, and a repair distribution $R_b$ governing its repair time. 
Some \BE{s} are used as spare components: these (\SBE{s}) replace a primary component when it fails.
\SBE{s} are equipped also with a dormancy distribution $D_b$, since spares fail less often when \emph{dormant}, i.e.\ not in use.
Only if an \SBE becomes active, its failure distribution is given by $F_b$.

\paragraph{Gates.}
Non-leave nodes are called \emph{intermediate events} and are labelled with \emph{gates}, describing how combinations of lower failures propagate to upper levels.
\Cref{fig:gates} shows their syntax.
Their meaning is as follows: the $\ANDgate$, $\ORgate$, and $\VOTgate_k$ gates fail if respectively all, one, or $k$ of their $m$ children fail (with \mbox{$1\leqslant k\leqslant m$}).
The latter is called the \emph{voting} or $k$ out of $m$ gate.
Note that $\VOTgate_1$ is equivalent to an \ORgate gate, and $\VOTgate_m$ is equivalent to an \ANDgate.
The \emph{priority-and gate} (\PANDgate) is an \ANDgate gate that only fails if its children fail from left to right (or simultaneously).
\PANDgate{s} express failures that can only happen %after another ones,
in a particular order,
e.g.\ a short circuit in a pump can only occur after a leakage.
\SPAREgate gates have one \emph{primary} child and one or more \emph{spare} children: spares replace the primary when it fails.
The \FDEPgate gate has an input \emph{trigger} and several \emph{dependent events}: all dependent events become unavailable when the trigger fails.
\FDEPgate{s} can model for instance networks elements that become unavailable if their connecting bus~fails.

\paragraph{Repair boxes.}
An \RBOX determines which basic element is repaired next according to a given policy.
Thus all its inputs are \BE{s} or \SBE{s}.
\ifthenelse{\boolean{longversion}}{Repair events of basic elements propagate along the tree analogously to fail events.}{}
Unlike gates, an \RBOX has no output since it does not propagate failures.

\ifthenelse{\NOT\boolean{longversion}}{%
\begin{wrapfigure}[6]{r}[0pt]{.19\linewidth}
	\vspace{-8ex}
	\centering
	\includegraphics[width=.95\linewidth]{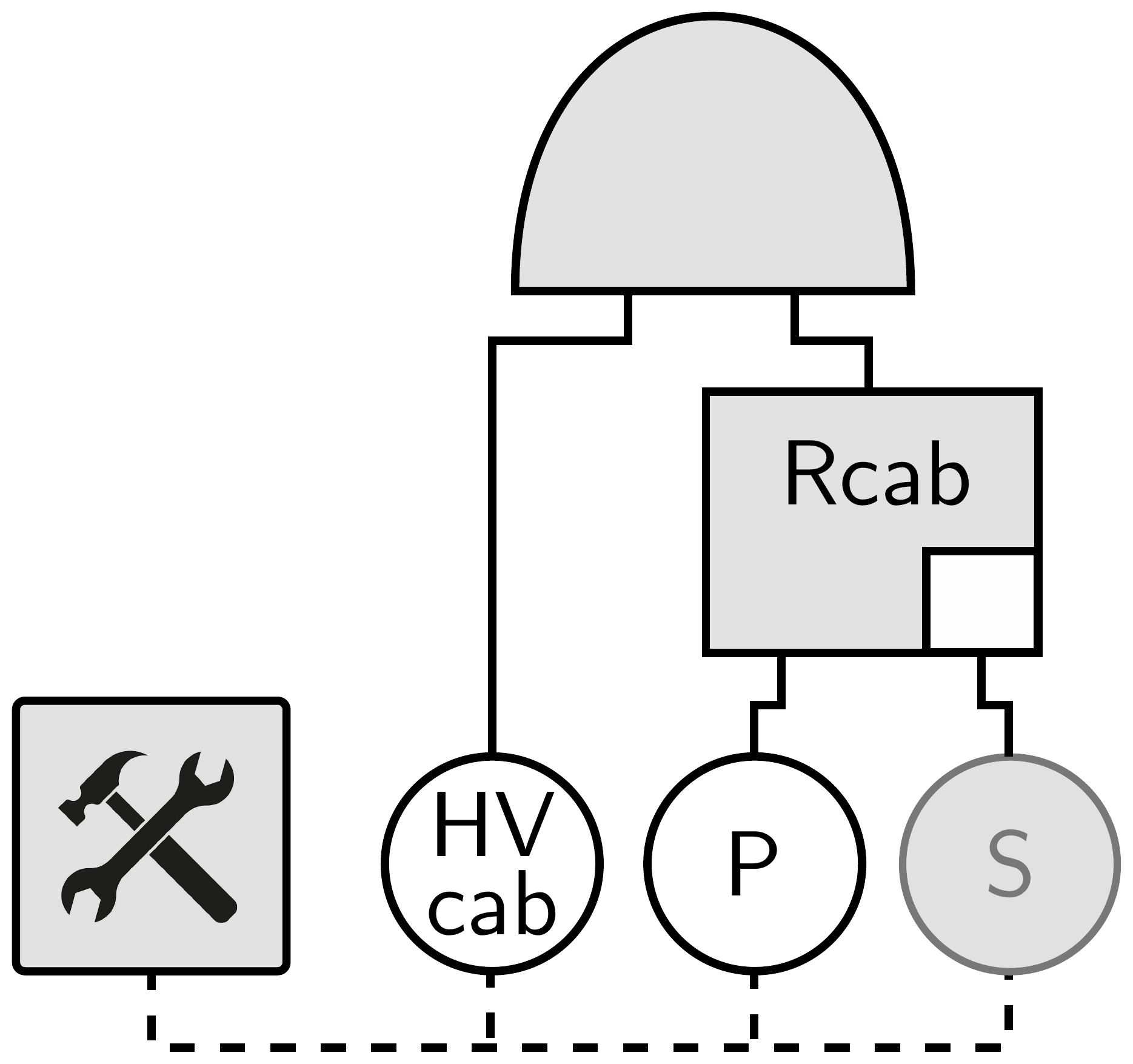}
	\vspace{-1ex}
	\caption{\smaller\!Tiny \rft\!\!}
	\label{fig:tiny_RWC}
	%\vspace{-7.5ex}
\end{wrapfigure}
}{}

\paragraph{Top level event.}
A full-system failure occurs if the \emph{top event} (i.e.\ the root node) of the tree fails. 
%Repair boxes cannot be top events: the root node of the tree is either an intermediate or basic event

\paragraph{Example.}
\ifthenelse{\boolean{longversion}}{%
\begin{wrapfigure}[6]{r}[0pt]{.19\linewidth}
	\vspace{-8ex}
	\centering
	\includegraphics[width=.95\linewidth]{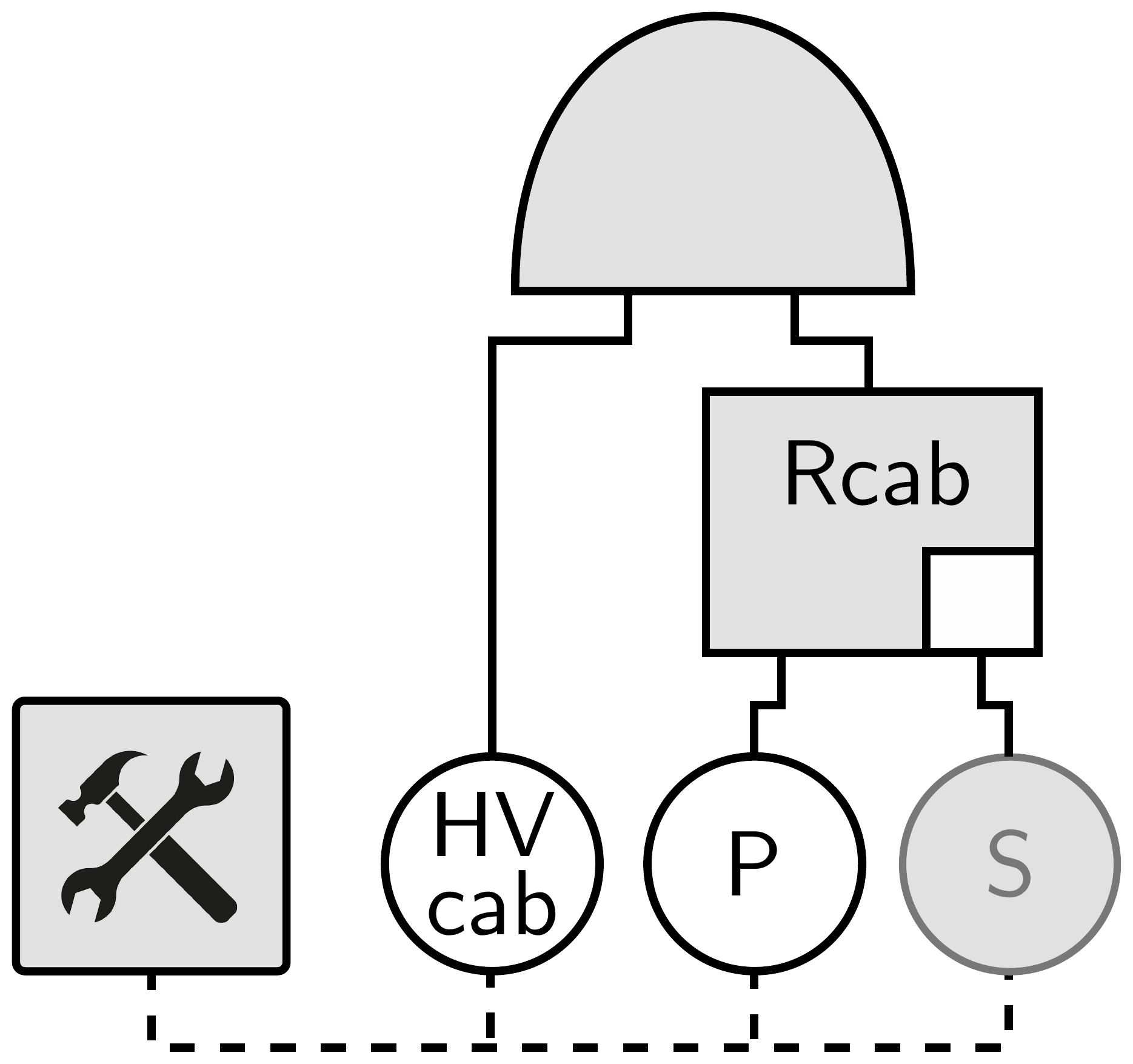}
	\vspace{-1ex}
	\caption{\smaller\!Tiny \rft\!\!}
	\label{fig:tiny_RWC}
	%\vspace{-7.5ex}
\end{wrapfigure}
}{}
The tree in \Cref{fig:tiny_RWC} models a railway-signal system, which fails if its \underline{h}igh \underline{v}oltage and \underline{r}elay \underline{cab}inets fail \cite{GSS15,RRBS17}.
Thus, the top event is an \ANDgate gate with children \leaf{HVcab} (a~\BE) and \leaf{Rcab}.
The latter is a \SPAREgate gate with primary \leaf{P} and spare \leaf{S}.
All \BE{s} are managed by one \RBOX with repair priority $\leaf{HVcab}>\leaf{P}>\leaf{S}$.

\paragraph{Notation.}
The nodes of a tree $\Tree$ are given by \mbox{$\nodes=\{0,1,\ldots,n-1\}$}.
We let $v,w$ range over $\nodes$.
A function
$\type^\Tree\colon\nodes\to\{\BE,\SBE,\ANDgate,\ORgate,
\allowbreak
\VOTgate_k,\PANDgate,\SPAREgate,\FDEPgate,\allowbreak\RBOX\}$
yields the type of each node in the tree.
%, graphically represented as in \Cref{fig:gates}.
A function $\child^\Tree\colon\nodes\to\nodes^*$ returns the ordered list of children of a node.
If clear from context, we omit the superscript $\Tree$ from function names.

\paragraph{Semantics.}
\ifthenelse{\boolean{longversion}}{%
The semantics of static fault trees, i.e.\ trees that only feature the static gates $\ANDgate$, $\ORgate$ and $\VOTgate_k$, can be given as a Boolean function.
For the gates $\PANDgate,\SPAREgate,\FDEPgate$ the order of the failures matter, so a Boolean function does not suffice.
Therefore, the semantics for (repairable) dynamic fault trees is given in terms of stochastic transition models, such as Markov automata, Petri nets, \iosa, etc.}%
{}
Following \cite{Mon18} we give semantics to \rft as Input/Output Stochastic Automata (\iosa), so that we can handle arbitrary probability distributions.
Each state in the \iosa represents a system configuration, indicating which components are operational and which have failed.
Transitions among states describe how the configuration changes when failures or repairs occur.

More precisely, a \emph{state} in the \iosa is a tuple $\xbf=(\xbf_0,\ldots,\xbf_{n-1})\in\states\subseteq\N^n$, where $\states$ is the \emph{state space} and $\xbf_\nodev$ denotes the state of node \nodev in $\Tree$.
The possible values for $\xbf_\nodev$ depend on the type of \nodev.
%% The \emph{output} of a node \nodev is a function of the state $\zbf_\nodev\colon\N^n\to\{0,1\}$, that indicates whether it is operational ($\zbf_\nodev=\zbf_\nodev(\xbf)=0$) or it has failed ($\zbf_\nodev=1$).
%
The \emph{output} $\zbf_\nodev\in\{0,1\}$ of node \nodev indicates whether it is operational ($\zbf_\nodev{=}0$) or failed ($\zbf_\nodev{=}1$)
and is calculated as follows:
\begin{itemize} %[label=\raisebox{1pt}{\textbullet}]
\item	\BE{s}~(white circles in \Cref{fig:gates})
		have a binary state: $\xbf_\nodev=0$ if \BE \nodev is operational
		and $\xbf_\nodev=1$ if it is failed. The output of a \BE is its state:
		$\zbf_\nodev=\xbf_\nodev$.
\item	\SBE{s}~(gray circles in \Cref{fig:gates:spare})
		have two additional states: $\xbf_\nodev=2,3$ if a \emph{dormant}
		\SBE \nodev is resp.\ operational, failed.
		Here $\zbf_\nodev=\xbf_\nodev\bmod2$.
\item	\ANDgate{s}~%(\Cref{fig:gates:and})
		have a binary state. Since the \ANDgate gate \nodev fails
		iff all children fail:
		$\xbf_\nodev = \min_{w\in\child[\nodev]} \zbf_w$.
		An \ANDgate gate outputs its internal state:
		$\zbf_\nodev=\xbf_\nodev$.
\item	\ORgate gates~%(\Cref{fig:gates:or})
		are analogous to \ANDgate gates, but fail iff any child fail, i.e.\ 
		$\zbf_\nodev = \xbf_\nodev = \max_{w\in\child[\nodev]} \zbf_w$
		for \ORgate gate $\nodev$.
\item	\VOTgate gates~%(\Cref{fig:gates:vot})
		also have a binary state: a $\VOTgate_k$ gate fails
		iff $1\leqslant k\leqslant m$ children fail, thus
		$\zbf_\nodev=\xbf_v = 1$ if $k\leqslant\sum_{w\in\child[\nodev]}\zbf_w$,
		and $\zbf_\nodev=\xbf_v = 0$ otherwise.
		%% $\zbf_\nodev=\xbf_v = \min^{m-k+1}_{w\in\child[\nodev]} \zbf_w$,
		%% where $\min^s$ are the smallest $s$ elements.
\item	\PANDgate gates~%(\Cref{fig:gates:pand})
		admit multiple states to represent the failure order of the children. 
		For \PANDgate \nodev with two children we let $\xbf_\nodev$ equal:
		$\boldsymbol0$ if both children are operational;
		$\boldsymbol1$ if the left child failed, but the right one has not;
		$\boldsymbol2$ if the right child failed, but the left one has not;
		$\boldsymbol3$ if both children have failed, the right one first;
		$\boldsymbol4$ if both children have failed, otherwise.	
		The output of $\PANDgate$ gate \nodev is $\zbf_v=1$ if
		$\xbf_v=4$ and $\zbf_v=0$ otherwise.
		\PANDgate gates with more children are handled by exploiting 
		$\PANDgate(w_1,w_2,w_3) = \PANDgate(\PANDgate(w_1,w_2),w_3)$.
\item	\SPAREgate gate~%(\Cref{fig:gates:spare}):
		\nodev leftmost input is its primary \BE. All other (spare) inputs
		are \SBE{s}. \SBE{s} can be shared among \SPAREgate gates.
		When the primary of \nodev fails, it is replaced with an
		\emph{available} \SBE. An \SBE is unavailable if it is failed,
		or if it is replacing the primary \BE of another \SPAREgate.
		The output of $\nodev$ is $\zbf_\nodev=1$
		if its primary is failed and no spare is available.
		Else $\zbf_\nodev=0$.
\item	An \FDEPgate gate~%(\Cref{fig:gates:fdep}):
		has no output.
		All inputs are \BE{s} and the leftmost is the trigger.
		We consider non-destructive \FDEPgate{s} \cite{BCH+08}: if the trigger
		fails, the output of all other \BE is set to $1$,
		without affecting the internal state.
		Since this can be modelled by a suitable combination of \ORgate
		gates~\cite{Mon18}, we omit the details.
%%	\item	\FDEPgate gate~%(\Cref{fig:gates:fdep}):
%%			\begingroup
%%			\def\tr{\ensuremath{\mathit{tr}}\xspace}
%%			\nodev has a \emph{dummy output} $\zbf_\nodev=0$ connected to no node.
%%			All inputs are \BE{s} and the leftmost is the trigger (\tr).
%%			We model non-destructive \FDEPgate{s} \cite{BCH+08}: if \tr
%%			fails, the output of every dependent event \nodew is $\zbf_w=1$,
%%			although $\xbf_w$ is not affected.
%%			When $\zbf_\tr=0$, the output of the dependent events is unaffected.
%%			Thus, $\FDEPgate(\leaf{T},\BE_1,\ldots,\BE_m)$ is a syntax
%%			sugar of $\{\ORgate(\leaf{T},\BE_i)\}_{i=1}^m$.
%%			\endgroup
\end{itemize}

For example, the \rft from \Cref{fig:tiny_RWC} starts with all operational elements, so the initial state is $\xbf^0=(0,0,2,0,0)$.
If then \leaf{P} fails, $\xbf_\leaf{P}$ and $\zbf_\leaf{P}$ are set to 1 (failed) and \leaf{S} becomes $\xbf_\leaf{S}=0$ (active and operational spare), so the state changes to $\xbf^1=(0,1,0,0,0)$.
The traces of the \iosa are given by $\xbf^0\xbf^1\cdots\xbf^n\in \states^\ast$, where a change from $\xbf^j$ to $\xbf^{j+1}$ corresponds to transitions triggered in the \iosa.

\paragraph{Nondeterminism.}
Dynamic fault trees may exhibit nondeterministic behaviour as a consequence of underspecified failure behaviour \cite{CBS07,JGKS16}. 
This can happen e.g.\ when two \SPAREgate{s} have a single shared \SBE: if all elements are failed, and the \SBE is repaired first, the failure behaviour depends on which \SPAREgate gets the \SBE. 
Monte Carlo simulation, however, requires fully stochastic models and cannot cope with nondeterminism. 
To overcome this problem we deploy the theory from \cite{DM18,Mon18}. 
If a fault tree adheres to some mild syntactic conditions, then its \iosa semantics is \emph{weakly deterministic},
meaning that all resolutions of the nondeterministic choices lead to the same probability value.
In particular, we require that 
(1) each \BE is connected to at most one \SPAREgate gate, and
(2) \BE{s} and \SBE{s} connected to \SPAREgate{s} are not connected to \FDEPgate{s}.
In addition to this, some semantic decisions have been fixed, e.g.\ the semantics of \PANDgate is fully specified, and policies should be provided for \RBOX and spare assignments.

\paragraph{Dependability metrics.}
An important use of fault trees is to compute relevant dependability metrics.
\ifthenelse{\boolean{longversion}}{%
Let $\{X_t\}_{t\geqslant0}$ be the stochastic process induced by \Tree \cite{CSD00}, and let $X_{t,\nodev}$ be the random variable that represents the (distribution of the) state of the top event of \Tree at time $t$.
We focus on two popular metrics:
}{%
Let $X_t$ denote the random variable that represents the state of the top event at time $t$ \cite{CSD00}.
Two popular metrics~are:
}
\ifthenelse{\boolean{longversion}}{%
\begin{itemize}[label=\textbullet,itemsep=1ex]
}{%
\begin{itemize}
}
\item	\emph{system reliability}:
		\ifthenelse{\boolean{longversion}}{%
		is the continuity of correct service, i.e.\ }%
		{}
		the probability of observing no top event failure
		before some mission time $T>0$, viz.\
		\ifthenelse{\boolean{longversion}}{%
			\mbox{$\REL=\mathit{Prob}\left(\forall_{t\in[0,T]}\,.\,X_{t,\nodev}=0\right);$}
		}{%
			\mbox{$\REL=\mathit{Prob}\left(\forall_{t\in[0,T]}\,.\,X_t=0\right);$}
		}
\item	\emph{system availability}:
		the proportion of time that the system remains operational
		in the long-run, viz.\
		\ifthenelse{\boolean{longversion}}{%
			\mbox{$\AVA=\lim_{t\to\infty} \mathit{Prob}\left(X_{t,\nodev}=0\right)$}.
		}{%
			\mbox{$\AVA=\lim_{t\to\infty} \mathit{Prob}\left(X_t=0\right)$}.
		}
\end{itemize}
%
%%	% Un-itemised version:
%%	Two popular metrics are \emph{system reliability} and \emph{system availability}.
%%	System reliability is the probability of observing no top event failure before some mission time $T>0$, viz.\ $\REL=\mathit{Prob}\left(\forall_{t\in[0,T]}\,.\,X_t=0\right)$.
%%	System availability is the proportion of time that the system remains operational in the long-run, viz.\ $\AVA=\lim_{t\to\infty} \mathit{Prob}\left(X_t=0\right)$.
System \emph{unreliability} and \emph{unavailability} are the reverse of these metrics. That is: \mbox{$\UNREL=1-\REL$} and \mbox{$\UNAVA=1-\AVA$}.

\section{Stochastic simulation for Fault Trees}
\label{sec:background:smc}

\ifthenelse{\boolean{longversion}}{%
\paragraph{{I}nput-{O}utput {S}tochastic {A}utomata (IOSA).}
\iosa \cite{DM18,DArgenioLM16:formats} are an extension of GSMP \cite{Sch77} amenable to compositional modelling.
An \iosa is a state-transition system where the residence time in a state is governed by a \pdf.
\iosa{s} feature two ingredients that are crucial for our analysis:
(1) residence times can be governed by arbitrary probability distributions described by real-valued clocks, and
(2) discrete transitions are labelled by actions, and allow automata to communicate with each other.

To record the passage of time and control the occurrence of events, \iosa use real-valued variables called \emph{clocks}.
Clocks are set to a positive random value according to the (state-dependent) associated \pdf.
As time evolves, all clocks count down from their respective values at the same rate.
When the value of a clock reaches zero it may trigger some \emph{action}.
Thus, to model \BE $e$ in a fault tree, we associate a clock to $F_e$ and another to $R_e$.
As a matter of fact, each node in an \ft is modelled as an \iosa automaton.
The propagation of fail/repair events in the tree is done by (discrete, instantaneous) action synchronisation among automata.
Formally:
\begin{definition}[\iosa~\cite{DM18}]
	\label{def:iosau}
	An \emph{Input/Output Stochastic Automaton with Urgency} is a
	tuple $\left(\states,\actions,\clocks,\trans,s_0,C_0\right)$ where:
	\begin{enumerate*}[label=(\roman*)]
	\item	$\states$ is a denumerable set of \emph{states};
	\item	$\actions$ is a denumerable set of labels partitioned into
			\emph{input labels} $\inactions$ and \emph{output labels}
			$\outactions$, where $\uractions\subseteq\actions$
			are \emph{urgent labels};
	\item	$\clocks$ is a finite set of \emph{clocks} s.t.\
			each $x\in\clocks$ has an associated continuous probability measure
			$\mu_x$ with support on $\R_{>0}$;
	\item	${\trans} \subseteq
				\states\times\clocks\times\actions\times\clocks\times\states$
			is a \emph{transition function};
	\item	$s_0\in\states$ is the \emph{initial state}; and
	\item	$C_0\subseteq\clocks$ are clocks initialized in $s_0$.
	\end{enumerate*}
\end{definition}
Six constraints on $\trans$ ensure that, in closed \iosa obtained from the parallel composition of all automata, nondeterminism is restricted to urgent actions.
These semantic constraints are translated into the syntactic conditions previously mentioned for \ft{s}.
For insights see \cite{DM18}; details on how to represent gates and basic elements with \iosa automata are in \cite{Mon18}.

Modelling fault trees as \iosa allows us to perform Monte Carlo simulation: we generate \emph{traces}, i.e.\ sequences of states $\xbf^0,\xbf^1,\ldots,\xbf^m$ where each $\xbf^j,\xbf^{j+1}$ is the projection on $\states^2$ of an element of $\trans$ from \Cref{def:iosau}.
Then, we estimate dependability metrics via statistical analyses on a set of sampled traces.
}{}

\paragraph{Standard Monte Carlo simulation (SMC).}
Monte Carlo simulation takes random samples from stochastic models to estimate a (dependability) metric of interest. 
For instance, to estimate the unreliability of a tree \Tree we sample $N$ independent traces from its \iosa semantics.
An unbiased statistical estimator for $p=\UNREL$ is the proportion of traces observing a top level event, that is, $\hat{p}_N=\frac{1}{N}\sum_{j=1}^N X^j$ where $X^j=1$ if the $j$-th trace exhibits a top level failure before time $T$ and $X^j=0$ otherwise.
The statistical error of $\hat{p}$ is typically quantified with two numbers $\delta$ and $\varepsilon$ s.t.\ $\hat{p}\in[p-\varepsilon,p+\varepsilon]$ with probability $\delta$.
The interval $\hat{p}\pm\varepsilon$ is called a \emph{confidence interval} (\ci) with coefficient $\delta$ and precision~$2\varepsilon$.

Such procedures scale linearly with the number of tree nodes and cater for a wide range of \pdf{s}, even non-Markovian distributions.
However, they encounter a bottleneck to estimate \emph{rare events}: if $p\approx0$, very few traces observe $X^j=1$.
Therefore, the variance of estimators like $\hat{p}$ becomes huge, and \ci{s} become very broad, easily degenerating to the trivial interval $[0,1]$.
Increasing the number of traces alleviates this problem, but even standard \ci settings---where $\varepsilon$ is relative to $p$---require sampling an unacceptable number of traces \cite{RT09b}.
\ifthenelse{\boolean{longversion}}{%
For instance, choosing $\delta=0.95$ and $\varepsilon=\frac{p}{10}$ (``95\% confidence and 10\% relative error'') requires $N\geqslant\sfrac{384}{p}$ samples.
Thus if $\UNREL\approx10^{-8}$, one needs \mbox{$N\geqslant38400000000$} traces, making the simulation times unacceptably long.}%
{}
Rare event simulation techniques solve this specific problem.

\paragraph{Rare Event Simulation (RES).}
%\label{sec:background:res}
%
\res techniques \cite{RT09b} increase the amount of traces that observe the rare event, e.g.\ a top level event in an \rft.
Two prominent classes of \res techniques are \emph{importance sampling}, which adjusts the \pdf of failures and repairs, and \emph{importance splitting} (\ISPLIT \cite{LLLT09}), which samples more (partial) traces from states that are closer to the rare event.
We focus on \ISPLIT due to its flexibility with respect to the probability distributions.

\ISPLIT can be efficiently deployed as long as the rare event $\gamma$ can be described as a nested sequence of less-rare events $\gamma=\gamma_M\subsetneq\gamma_{M-1}\subsetneq\cdots\subsetneq\gamma_0$. 
This decomposition allows \ISPLIT to study the conditional probabilities $p_k=\mathit{Prob}(\gamma_{k+1}\,|\,\gamma_k)$ separately, to then compute $p=\mathit{Prob}(\gamma) = \prod_{k=0}^{M\text{-}1}\mathit{Prob}(\gamma_{k+1}\,|\,\gamma_k)$.
Moreover, \ISPLIT requires all conditional probabilities $p_k$ to be much greater than $p$, so that estimating each $p_k$ can be done efficiently with \smc.

The key idea behind \ISPLIT is to define the events $\gamma_k$ via a so called \emph{importance function} \mbox{$\IFUN\colon\states\to\N$} that assigns an \emph{importance} to each state~$s\in\states$.
The higher the importance of a state, the closer it is to the rare event $\gamma_M$.
Event $\gamma_k$ collects all states with importance at least $\ell_k$, for certain sequence of \emph{threshold levels} $0=\ell_0 <\ell_1<\cdots<\ell_M$.
Formally: $\gamma_k=\{s\in\states\mid\IFUN[s]\geqslant\ell_k\}$.
\ifthenelse{\boolean{longversion}}{%
In other words, a higher importance is assigned to states from which it is more likely to observe the rare event.
That is, for $s,s'\in\states$ s.t.\ $s\in\gamma_k$ and $s'\in\gamma_{k'}$, one wants $\IFUN[s]<\IFUN[s']$ iff $k<k'$.
Because then, in the nested sequence of events $\gamma_0\supsetneq\cdots\supsetneq\gamma_M$, each step $\gamma_{k-1}\boldsymbol{\rightarrow}\gamma_k$ makes it more likely to observe the rare event.
Therefore, choosing many thresholds ($M\gg0$) all very close to each other, one ensures that $\mathit{Prob}(\gamma_k\,|\,\gamma_{k-1})\gg0$ for all $0<k\leqslant M$.
Simply put, one makes a lot of baby steps in the right direction.}%
{}

To exploit the importance function $\IFUN$ in the simulation procedure, \ISPLIT samples more (partial) traces from states with higher importance.
Two well-known methods are deployed and compared in this paper: Fixed Effort and \restart.
\emph{Fixed Effort} (\fe \cite{Gar00}) samples a predefined amount of traces in each region $\states_k=\gamma_k\setminus\gamma_{k+1}=\{s\in\states\mid\ell_{k+1}>\IFUN(s)\geqslant\ell_k\}$.
Thus, starting at $\gamma_0$ it first estimates the proportion of traces that reach $\gamma_1$, i.e.\ $p_0=\mathit{Prob}(\gamma_1\,|\,\gamma_0)=\mathit{Prob}(\states_0)$.
Next, from the states that reached $\gamma_1$ new traces are generated to estimate $p_1=\mathit{Prob}(\states_1)$, and so on until $p_M$.
Fixed Effort thus requires that
($i$)~each trace has a clearly defined ``end,'' so that estimations of each $p_k$ finish with probability 1, and
($ii$)~all rare events reside in the uppermost region.
\ifthenelse{\boolean{longversion}}{%
In particular, using \fe for steady-state analysis (e.g.\ to estimate \UNAVA) requires regeneration theory \cite{Gar00}, which is hard to apply to non-Markovian models.
}{}

\begin{figure}[ht]
	\vspace{-2ex}
	\centering
	\begin{subfigure}{.4\linewidth}
		\centering
		\includegraphics[width=\linewidth]{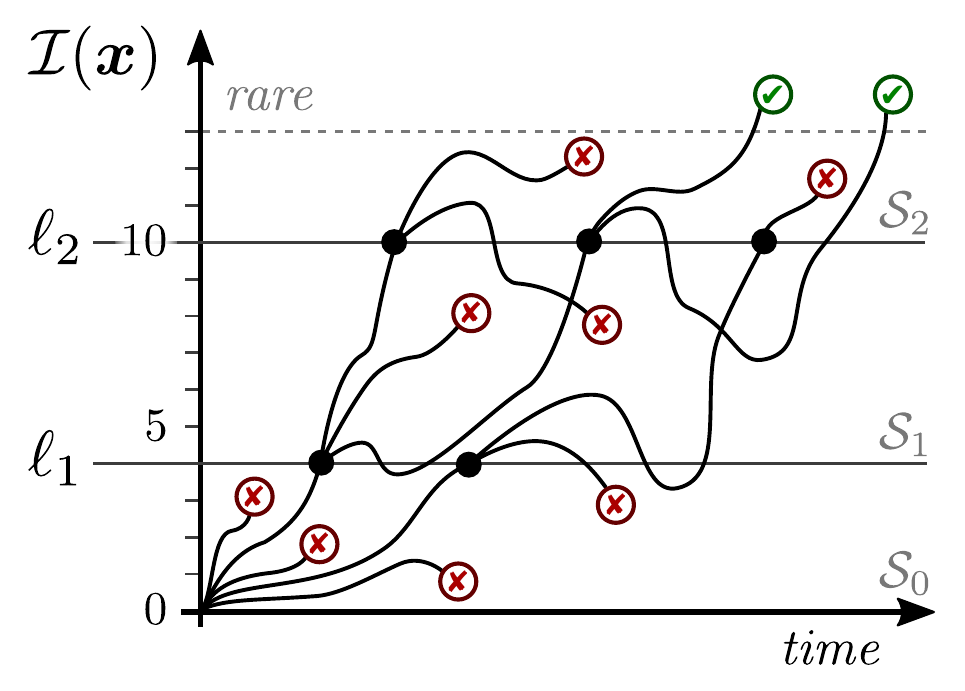}
		\vspace{-1ex}
		\caption{\fen{5} for $\mathit{Prob}(\lnot\avoid\,\mathsf{U}\goal)$}
		\label{fig:isplit:fixed_effort}
	\end{subfigure}
	\qquad
	\begin{subfigure}{.4\linewidth}
		\centering
		\includegraphics[width=\linewidth]{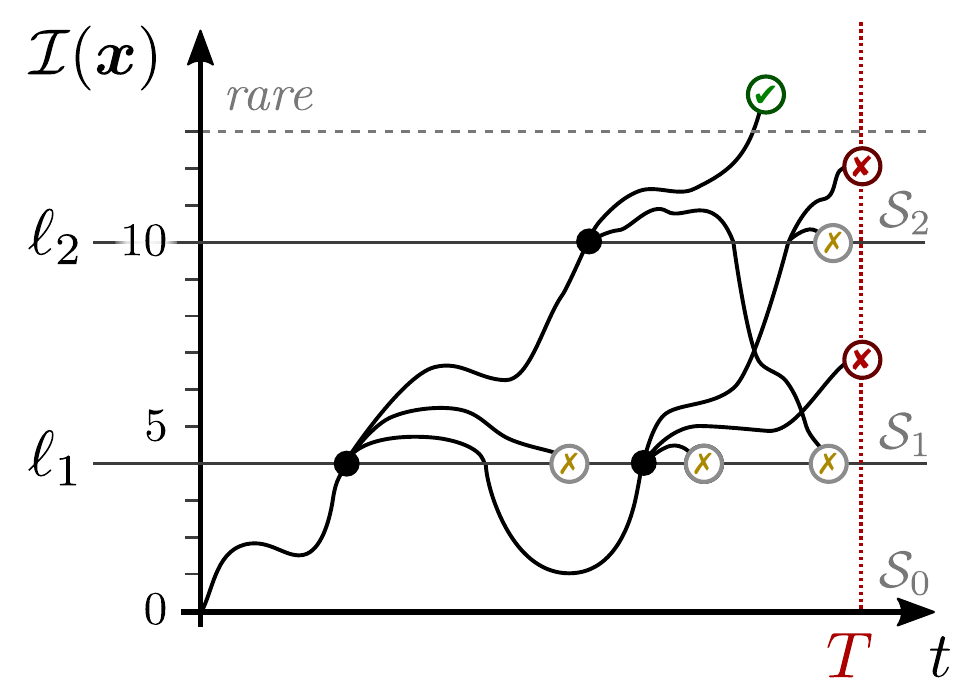}
		\vspace{-1ex}
		\caption{\rstes for \UNREL}
		\label{fig:isplit:restart}
	\end{subfigure}
	\caption{Importance Splitting algorithms Fixed Effort \& \restart}
	\label{fig:isplit}
	\vspace{-2ex}
\end{figure}

\paragraph{Example.}
%%	\begin{wrapfigure}[10]{r}{.4\linewidth}
%%		\vspace{-6ex}
%%		\includegraphics[width=\linewidth]{fixed_effort}
%%		\vspace{-4ex}
%%		\caption{\fen{5} for $\mathit{Prob}(\lnot\avoid\,\mathsf{U}\goal)$}
%%		\label{fig:fixed_effort}
%%	\end{wrapfigure}%
\Cref{fig:isplit:fixed_effort} shows Fixed Effort estimating the probability to visit states labelled \goal\ before others labelled \avoid.
States \goal\ have importance >13, and thresholds $\ell_1,\ell_2=4,10$ partition the state space in regions $\{\states_i\}_{i=0}^2$ s.t.\ all $\goal\in\states_2$.
The effort is 5 simulations per region, for all regions: we call this algorithm \fen{5}.
In region $\states_0$, 2 simulations made it from the initial state to threshold $\ell_1$, i.e.\ they reached some state with importance 4 before visiting a state \avoid.
In $\states_1$, starting from these two states, 3 simulations reached $\ell_2$.
Finally, 2 out of 5 simulations visited states \goal\ in $\states_2$.
Thus, the estimated rare event probability of this run of \fe{\:5} is
$\hat{p}=\prod_{i=1}^2\hat{p_i}=\frac{2}{5}\frac{3}{5}\frac{2}{5}=9.6\times10^{-2}$.

\emph{RESTART} (\rstn{} \cite{VAVA91,VAM+94}) is another \res algorithm, which starts one trace
%%	\begin{wrapfigure}[10]{l}[0pt]{.38\linewidth}
%%		\vspace{-5.3ex}
%%		\centering
%%		\includegraphics[width=\linewidth]{restart}
%%		\vspace{-4ex}
%%		%\caption{\restart used to estimate the system unreliability}
%%		\caption{\smaller\restart for \UNREL}
%%		\label{fig:restart}
%%	\end{wrapfigure}
in $\gamma_0$ and monitors the importance of the states visited.
If the trace up-crosses threshold $\ell_1$, the first state visited in $\states_1$ is saved and the trace is cloned, aka \emph{split}---see \Cref{fig:isplit:restart}.
This mechanism rewards traces that get closer to the rare event.
Each clone then evolves independently, and if one up-crosses threshold $\ell_2$ the splitting mechanism is repeated.
Instead, if a state with importance below $\ell_1$ is visited, the trace is \emph{truncated}
(\raisebox{-2.5pt}{\tikz{\node[shape=circle,draw=Gray,very thick,inner sep=1pt]{\!\smaller\trunc};}} in \Cref{fig:isplit:restart}).
\ifthenelse{\boolean{longversion}}{%
In general, each clone is truncated as soon as it visits a state with importance lower than its level of creation.
}{}
This penalises traces that move away from the rare event.
To avoid truncating all traces, the one that spawned the clones in region $\states_k$ can go below importance $\ell_k$.
To deploy an unbiased estimator for $p$, \restart measures how much split was required to visit a rare state \cite{VAM+94}.
In particular, \restart does not need the rare event to be defined as $\gamma_M$ \cite{VA98}, and it was devised for steady-state analysis \cite{VAVA91} (e.g.\ to estimate \UNAVA) although it can also been used for transient studies as depicted~in~\Cref{fig:isplit:restart}~\cite{VA07}.

\section{Importance Splitting for FTA}
\label{sec:theory}

%This work is based on the insight that the fundamentally structured and stepwise failure of a fault tree, matches the layered state space requirements of \ISPLIT.
The effectiveness of \ISPLIT crucially relies on the choice of the importance function \IFUN as well as the threshold levels $\ell_k$ \cite{LLLT09}.
Traditionally, these are given by domain and/or \res experts, requiring a lot of domain knowledge.
This section presents a technique to obtain \IFUN and the $\ell_k$ automatically for an \rft.

%% %% %% %% %% %% %% %% %% %% %% %% %% %% %% %% %% %% %% %% %% %% %% %% %% %% 
%
\subsection{Compositional importance functions for Fault Trees}
\label{sec:theory:ifun}

By the core idea behind importance splitting, states that are \textup{more likely} to lead to the rare event should have a higher importance.
To achieve this, the key lies in defining an importance function \IFUN and thresholds $\ell_k$ that are sensitive to both the state space $\states$ and the transition \textup{probabilities} of the system.
For us, $\states\subseteq\N^n$ are all possible states of a repairable fault tree (\rft).
Its top event fails when certain nodes fail in certain order, and remain failed before certain repairs occur.
To exploit this for \ISPLIT, the structure of the tree must be embedded into \IFUN.

%%	Because in a trace $y = e_1 e_2 \cdots e_m$, each step $e_{k-1}\Rightarrow e_k$ changes the state $\xbf(t_{k-1})\Rightarrow\xbf(t_k)$ of \Tree.
%%	And the way in which this step affects the probability of observing a future top event depends on the structure of \Tree.

The strong dependence of the importance function \IFUN on the structure of the tree is easy to see in the following example.
Take the \rft \Tree from \Cref{fig:tiny_RWC} and let its current state \xbf be s.t.\ \leaf{P} is failed and \leaf{HVcab} and \leaf{S} are operational.
If the next event is a repair of \leaf{P}, then the new state $\xbf'$ (where all basic elements are operational) is farther from a failure of the top event.
Hence, a good importance function should satisfy $\IFUN[\xbf]>\IFUN[\xbf']$.
Oppositely, if the next event had been a failure of \leaf{S} leading to state $\xbf''$, then one would want that $\IFUN[\xbf]<\IFUN[\xbf'']$.
The key observation is that these inequalities depend on the structure of \Tree as well as on the failures/repairs of basic elements.
\ifthenelse{\boolean{longversion}}{%
Because if instead of an \ANDgate, the top event were a \PANDgate gate (call this tree $\Tree^{\!\!\ast}$), the importance function should behave in the exact opposite way.
That is, in tree $\Tree^{\!\!\ast}$ one wants that $\IFUN[\xbf]>\IFUN[\xbf'']$, since in $\xbf''$ the right child of the top \PANDgate has failed before the left child.
When this happens, \PANDgate gates go into an out-of-order internal state, and cannot output a failure.
So the same step from $\xbf$ to $\xbf''$ has completely different meanings for \Tree and for $\Tree^{\!\!\ast}$, as a result of their structure being different.
}{}

In view of the above, any attempt to define an importance function for an arbitrary fault tree \Tree must put its gate structure in the forefront.
In \Cref{tab:compifun} we introduce a compositional heuristic for this, which defines \emph{local importance functions} distinguished per node type.
%Then, via structural induction, these local functions are aggregated into an importance function for the whole tree.
The importance function associated to node \nodev is $\IFUN_\nodev\colon\N^n\to\N$.
We define the \emph{global importance function} of the tree ($\IFUN_\Tree$ or simply \IFUN) as the local importance function of the top event node of \Tree.

\begin{table}
	\vspace{-2ex}
	\centering%\smaller
	\caption{Compositional importance function for \rft{s}.}
	\label{tab:compifun}
	\vspace{2ex}
	%% !TEX root =  ../main.tex
%
\begingroup
\renewcommand{\arraystretch}{1.8}
\def\treei{\ensuremath{\Tree_i}}
\def\maxI[#1]{\ensuremath{\max^{\IFUN}_{#1}}}
\def\lcm{\ensuremath{\mathrm{lcm}}}
\def\ord{\ensuremath{\mathit{ord}}}
\def\bitsmaller[#1]{\scalebox{1.15}{$#1$}}
%
% https://tex.stackexchange.com/a/12704
\begin{tabular}{>{\centering}m{2cm}@{\qquad}>{\centering\arraybackslash}m{5.8cm}m{1.2cm}}
	\toprule
	\addlinespace[-0.7ex]  % this row a bit shorter
	\type[\nodev] & \multicolumn{1}{c}{$\IFUN_\nodev(\xbf)$} & \\
	\midrule
	\addlinespace[-.7ex]  % this row a bit shorter
	\BE, \SBE &
		\raisebox{.7ex}{$\displaystyle \zbf_\nodev$}\\
	\addlinespace[-1ex]  % this row a bit shorter
	\ANDgate &
		$\lcm_\nodev\cdot \bitsmaller[\sum_{w\in\child[\nodev]}
		\frac{\IFUN_w(\xbf)}{\maxI[w]}]$\\
	\ORgate &
		$\displaystyle \lcm_\nodev\cdot \max_{w\in\child[\nodev]}
		\left\{ \bitsmaller[\frac{\IFUN_w(\xbf)}{\maxI[w]}] \right\}$\\
	$\VOTgate_k$ &
		$\displaystyle \lcm_\nodev\cdot \max_{W\subseteq\child[\nodev], |W|=k}
		\left\{ \bitsmaller[\sum_{\nodew\in W}
		\frac{\IFUN_\nodew(\xbf)}{\maxI[\nodew]}] \right\}$\\
	\addlinespace[.4ex]  % this row a bit taller
	%% $\VOTgate_k$ &
	%% 	$\displaystyle \lcm_\nodev\: \max_{i=1}^{\binom{m}{k}}
	%% 	\left\{ \bitsmaller[\sum_{\nodew\in\mathrm{comb}_i(\child[\nodev])}
	%% 	\frac{\IFUN_\nodew(\xbf)}{\maxI[\nodew]}] \right\}$\\
	%% \addlinespace[.5ex]  % this row a bit taller
	\SPAREgate &
		$\displaystyle \lcm_\nodev\cdot \max\Big(
			\bitsmaller[\sum_{w\in\child[\nodev]}
			\frac{\IFUN_w(\xbf)}{\maxI[w]}] ~,~ \zbf_\nodev\cdot m \Big)$\\
	%\addlinespace[-5ex]  % this row a bit shorter
	\PANDgate &
		$\displaystyle \lcm_\nodev\cdot \max\Big(
			 \bitsmaller[\frac{\IFUN_l(\xbf)}{\maxI[l]}]
			+ \ord\:\bitsmaller[\frac{\IFUN_r(\xbf)}{\maxI[r]}]
                        ~,~ \zbf_\nodev\cdot 2 \Big)$\\[-2ex]
	%% \PANDgate &
	%% 	$\displaystyle \lcm_\nodev\cdot \max\Big( 0 ~,~ \zbf_\nodev
	%% 		+ \bitsmaller[\frac{\IFUN_l(\xbf)}{\maxI[l]}]
	%% 		+ \ord\:\bitsmaller[\frac{\IFUN_r(\xbf)}{\maxI[r]}]\Big)$\\[-2ex]
		& \multicolumn{2}{l}{\smaller
			where~$\ord=1$ if $\xbf_\nodev\in\{1,4\}$ and $\ord=-1$ otherwise}\\
         \multicolumn{3}{c}{with \ $\max^{\IFUN}_{\nodev}=\max_{\xbf\in\states}\IFUN_\nodev(\xbf)$ \ and \ $\mathrm{lcm}_\nodev=\mathrm{lcm}\left\{
	\max^{\IFUN}_\nodew \,\middle|\, \nodew\in\child[\nodev]
\right\}$}\\
	\bottomrule
\end{tabular}
\endgroup

\end{table}

Thus, $\IFUN_\nodev$ is defined in \Cref{tab:compifun} via structural induction in the fault tree.
It is defined so that it assigns to a \emph{failed} node \nodev its \emph{highest importance value}.
%% Thus, \IFUN is defined in \Cref{tab:compifun} via structural induction in the fault tree.
%% This assigns to a \emph{failed} node \nodev its \emph{highest importance value} $\max^{\IFUN}_{\nodev}=\max_{\xbf\in\states}\IFUN_\nodev(\xbf)$.
%That is, state $\xbf\in\argmax_{\xbf}\IFUN_\nodev$ iff $1=\zbf_\nodev=\zbf_\nodev^{\type[\nodev]}(\xbf)$.
%\footnote{This can be proved by structural induction on the tree \cite{THISarXiv}.}.
Functions with this property deploy the most efficient \ISPLIT implementations \cite{LLLT09}, and some \res algorithms (e.g.\ Fixed Effort) require this property \cite{Gar00}.
%
%In our inductive definition of \IFUN, the tree leaves are the base cases:

In the following we explain our definition of $\IFUN_\nodev$.
%\paragraph{Basic elements.}
%
If \nodev is a failed \BE or \SBE, then its importance is $1$; else it is $0$.
This matches the output of the node, thus $\IFUN_\nodev(\xbf)=\zbf_\nodev$.
Intuitively, this reflects how failures of basic elements are positively correlated to top event failures.
%
%\paragraph{Static gates.}
%
The importance of \ANDgate, \ORgate, and $\VOTgate_k$ gates depends exclusively on their input.
The importance of an \ANDgate is the sum of the importance of their children scaled by a normalisation factor.
This reflects that \ANDgate gates fail when all their children fail, and each failure of a child brings an \ANDgate closer to its own failure, hence increasing its importance.
Instead, since \ORgate gates fail as soon as a single child fails, their importance is the maximum importance among its children.
The importance of a $\VOTgate_k$ gate is the sum of the $k$ (out of $m$) children with highest importance value.
%, denoted with $\sum^{\max^k}_{w\in\child[\nodev]}$ in \Cref{tab:compifun}.

Omiting normalisation may yield an undesirable importance function.
To understand why, suppose a binary \ANDgate gate \nodev with children $l$ and $r$, and define $\IFUN^{\text{naive}}_\nodev(\xbf)=\IFUN_l(\xbf)+\IFUN_r(\xbf)$.
Suppose that $\IFUN_l$ takes it highest value in $\max^{\IFUN}_l=2$ while $\IFUN_r$ in $\max^{\IFUN}_r=6$ and assume that states $\xbf$ and $\xbf'$ are s.t.\ $\IFUN_l(\xbf)=1$, $\IFUN_r(\xbf)=0$, $\IFUN_l(\xbf')=0$, $\IFUN_r(\xbf')=3$.
This means that in both states one child of \nodev is ``good-as-new'' and the other is ``half-failed'' and hence the system is equally close to fail in both cases.  Hence we expect $\IFUN^{\text{naive}}_v(\xbf)=\IFUN^{\text{naive}}_v(\xbf')$ when actually $\IFUN^{\text{naive}}_v(\xbf)=1\neq 3=\IFUN^{\text{naive}}_v(\xbf')$.
Instead, $\IFUN_\nodev$ operates with $\frac{\IFUN_l(\xbf)}{\max^{\IFUN}_l}$ and $\frac{\IFUN_r(\xbf)}{\max^{\IFUN}_r}$, which can be interpreted as the ``percentage of failure'' of the  children of \nodev.
To make these numbers integers we scale them by $\mathrm{lcm}_\nodev$, the \emph{least common multiple} of their max importance values.  In our case $\mathrm{lcm}_\nodev=6$ and hence $\IFUN_v(\xbf)=\IFUN_v(\xbf')=3$.
Similar problems arise whit all gates, hence normalization is applied in general.

\SPAREgate gates with $m$ children (including its primary) behave similarly to \ANDgate gates: every failed child brings the gate closer to failure, as reflected in the left operand of the $\max$ in \Cref{tab:compifun}.
However, \SPAREgate{s} fail when their primaries fail and no \SBE{s} are \emph{available}, e.g.\ possibly being used by another \SPAREgate.
This means that the gate could fail in spite of some children being operational.
To account for this we exploit the gate output: multiplying $\zbf_\nodev$ by $m$ we give the gate its maximum value when it fails, even when this happens due to unavailable but operational \SBE{s}.
For a \PANDgate gate \nodev we have to carefully look at the states.
If the left child $l$ has failed, then the right child $r$ contributes positively to the failure of the \PANDgate and hence the importance function of the node \nodev.
If instead the right child has failed first, then the \PANDgate gate will not fail and hence we let it contribute negatively to the importance function of \nodev.
Thus, we multiply $\frac{\IFUN_r(\xbf)}{\max^{\IFUN}_r}$ (the normalized importance function of the right child) by $-1$ in the later case (i.e. when state $\xbf_\nodev\notin\{1,4\}$).
Instead, the left child always contribute positively.
Finally, the max operation is two-fold: on the one hand, $\zbf_\nodev\cdot 2$ ensures that the importance value remains at its maximun while failing (\PANDgate{s} remain failed even after the left child is repaired);
%(the factor ``2'' is a consequence of the 2 children.)
on the other, it ensures that the smallest value posible is 0 while operational (since importance values can not be negative.)

%% %% %% %% %% %% %% %% %% %% %% %% %% %% %% %% %% %% %% %% %% %% %% %% %% %% 
%
\subsection{Automatic importance splitting for FTA}
\label{sec:theory:isplit}

Our compositional importance function is based on the distribution of operational/failed basic elements in the fault tree, and their failure order.
This follows the core idea of importance splitting: the more failed \BE{s}/\SBE{s} (in the right order), the closer a tree is to its top event failure.

However, \ISPLIT is about running more simulations from state with higher \emph{probability} to lead to rare states.
This is only partially reflected by whether basic element $b$ is failed.
Probabilities lie also in the distributions $F_b,R_b,D_b$.
These distributions govern the transitions among states $\xbf\in\states$, and can be exploited for importance splitting.
We do so using the two-phased approach of \cite{BDH15,BDM16}, which in a first (static) phase computes an importance function, and in a second (dynamic) phase selects the thresholds from the resulting importance values.

In our current work, the first phase runs breadth-first search in the \iosa module of each tree node.
This computes node-local importance functions, that are aggregated into a tree-global \IFUN using our compositional function in \Cref{tab:compifun}.

The second phase involves running ``pilot simulations'' on the importance-labelled states of the tree.
Running simulations exercises the fail/repair distributions of \BE{s}/\SBE{s}, imprinting this information in the thresholds $\ell_k$.
Several algorithms can do such \emph{selection of thresholds}.
They operate sequentially, starting from the initial state---a fully operational tree---which has importance $i_0=0$.
For instance, Expected Success~\cite{BDH17} runs $N$ finite-life simulations.
If \mbox{$K<\frac{N}{2}$} simulations reach the next smallest importance $i_1>i_0$, then the first threshold will be $\ell_1=i_1$.
Next, $N$ simulations start from states with importance $i_1$, to determine whether the next importance $i_2$ should be chosen as threshold $\ell_2$, and~so~on.

Expected Success also computes the \emph{effort} per splitting region $\states_k=\{\mbox{$\xbf\in\states$}\mid\ell_{k+1}>\IFUN(\xbf)\geqslant\ell_k\}$.
For Fixed Effort, ``effort'' is the base number of simulations to run in region $\states_k$.
For \restart, it is the number of clones spawned when threshold $\ell_{k+1}$ is up-crossed.
In general, if $K$ out of $N$ pilot simulations make it from $\ell_{k-1}$ to $\ell_k$, then the $k$-th effort is $\left\lceil\frac{N}{K}\right\rceil$.
This is chosen so that, during \res estimations, one simulation makes it from threshold $\ell_{k-1}$~to~$\ell_k$ on average.

Thus, using the method from \cite{BDH15,BDM16} based on our importance function $\IFUN_\Tree$, we compute (automatically) the thresholds and their effort for tree \Tree.
This is all the meta-information required to apply importance splitting \res \cite{Gar00,GOK02,BDH15}.

% Long version has "Implementation" in a whole section.
% Short version shows it here, in half a page.
\ifthenelse{\boolean{longversion}}{}{%else:
\begin{figure}[ht]
	% \vspace{-3ex}
	\centering
	\includegraphics[width=.9\linewidth]{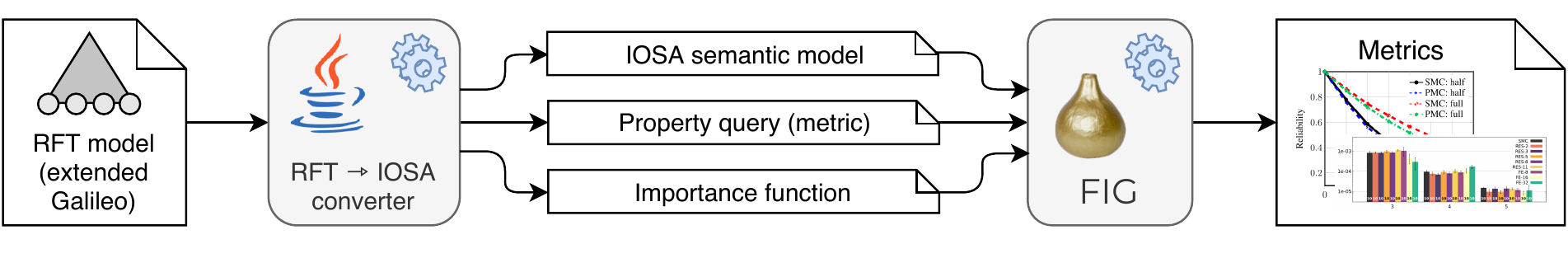}
	\vspace{-2ex}
	\caption{Tool chain}
	\label{fig:toolchain}
	\vspace{-5ex}
\end{figure}
\paragraph{Implementation.}
\Cref{fig:toolchain} outlines a tool chain implemented to deploy the theory described above.
The input model is an \rft, described in the Galileo textual format \cite{SDC99,GALILEO} extended with repairs and arbitrary \pdf{s}.
This \rft file is given as input to a Java converter that produces three outputs: the \iosa semantics of the tree, the property queries for its reliability or availability, and our compositional importance function in terms of variables of the \iosa semantic model.
This information is dumped into a single text file and fed to \fig: a statistical model checker specialised in importance splitting \res.
% \fig builds local importance functions for the (\iosa modules corresponding to the) nodes of the \rft.
% The compositional function of \Cref{tab:compifun} is used to aggregate these local functions into a global function for the whole tree.
\fig interprets this importance function, deploying it into its internal model 
representation, which results in a global function for the whole tree.
\fig can then use \ISPLIT algorithms such as \restart and Fixed Effort, via the automatic methods described above.
The result are confidence intervals that estimate the reliability or availability of the \rft.
In this way, we implemented automatic importance splitting for \fta.
In \cite{THISarXiv} we provide more details about our tool chain and its capabilities.
}

\ifthenelse{\boolean{longversion}}%
	{%%%%%%%%%%%%%%%%%%%%%%%%%%%%%%%%%%%%%%%%%%%%%%%%%%%%%%%%%%%%%%%%%%%%%%%%%%%%%%
%% !TEX root =  ../main.tex
\section{Tool chain implementation}
\label{sec:impl}

We implemented the theory introduced in \Cref{sec:theory} in a full-stack tool chain.
Its input are plain text files in the Galileo textual format \cite{SDC99,CS00,GALILEO}: a widespread syntax to describe fault trees \cite{BD05,DP07,LWK17}.
Galileo was not designed for repairs, and has limited support for non-Markovian distributions: we thus extend it to fit our needs.
\Cref{fig:toolchaindiagram} shows the tool chain:
a converter parses the \rft defined in extended Galileo;
it generates an \iosa model, property queries, and compositional importance function (using \Cref{tab:compifun});
from this input, the \fig tool can implement \res (importance splitting) algorithms, and use them to estimate system unreliability and unavailability.

\begin{figure*}[h]
	\centering
	\includegraphics[width=\linewidth]{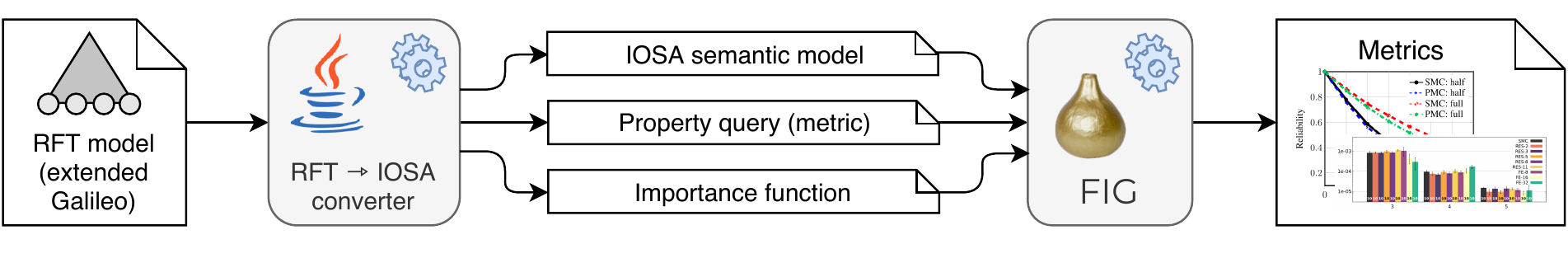}
	\caption{Tool chain}
	\label{fig:toolchaindiagram}
\end{figure*}

%% %% %% %% %% %% %% %% %% %% %% %% %% %% %% %% %% %% %% %% %% %% %% %% %% %% 
%
\subsection{Extensions to Galileo}
\label{sec:impl:galileoext}

Standard Galileo supports exponential, log-normal, and Weibull \pdf{s}.
We use the keyword \code{EXT\_failPDF} to define arbitrary failure distributions.
In \Cref{code:dormPDF}, the \SPAREgate gate (\code{Gate2}) has its primary (\code{BE\_C}) and one spare (\code{BE\_D}), whose resp.\ fail \pdf{s} are Rayleigh ($\sigma=0.06$) and exponential ($\lambda=0.0011$).
%
%%	In \Cref{code:otherFailPDF}, basic elements \code{BE\_A} and \code{BE\_B} (children of the \ORgate gate \code{Gate1}) have resp.\ an exponential \pdf with rate \rarep{6.0}{5}, and an Erlang \pdf with shape $k=3$ and rate $\lambda=350$.
%%	
%%	\begin{lstlisting}[%
%%		caption={Arbitrary failure \pdf in extended Galileo},%
%%		captionpos=b,%
%%		label={code:otherFailPDF},%
%%		frame=single,%
%%		basicstyle=\footnotesize\ttfamily%
%%		%,float,floatplacement=h]
%%	]
%%	toplevel "Gate1";
%%	"Gate1" or "BE_A" "BE_B";
%%	"BE_A" lambda=6.0E-5;
%%	"BE_B" EXT_failPDF=erlang(3,350);
%%	\end{lstlisting}
%%	%
%
We also allow the dormancy \pdf of an \SBE to be independent of its fail \pdf.
For this we add the \code{EXT\_dormPDF} keyword to define an arbitrary dormancy \pdf.
Thus we define the dormancy of \code{BE\_D} as an $\mathrm{Erlang}(k=3,\lambda=9)$ in \Cref{code:dormPDF}.
In the current implementation, a new time to failure is sampled (from the corresponding \pdf) as the \SBE is activated when the primary \BE fails.
This is a simplification since we work with potentially non-Markovian distributions; more realistic implementations are proposed as future work in \Cref{sec:conclu}.

\begin{lstlisting}[%
	caption={\SBE{s} and arbitrary \pdf{s} in extended Galileo},%
	label={code:dormPDF},%
	captionpos=b,%
	frame=single,%
	basicstyle=\footnotesize\ttfamily%
]
toplevel "Gate2";
"Gate2" wsp "BE_C" "BE_D";
"BE_C" EXT_failPDF=rayleigh(6.0E-2);
"BE_D" lambda=1.11E-3 EXT_dormPDF=erlang(3,9);
\end{lstlisting}

Finally, we also extend Galileo with the keywords \mbox{\code{repairbox\_priority}} and \mbox{\code{EXT\_repairPDF}}.
These respectively define arbitrary repair policies for the \RBOX elements, and the repair \pdf{s} of \BE{s} and \SBE{s}.
All \BE{s} in \Cref{code:repairPDF} are repairable, with repair time uniformly distributed on the real intervals $[8,24]$ and $[8,12]$.
The last line of the \namecref{code:repairPDF} defines the \RBOX of the system, which handles one repair at a time.
Its repair policy determines which \BE to choose when more than one is failed at the same time.
For instance, if all \BE{s} fail and the \RBOX ``finishes repairing'' \code{BE\_G}, it will next repair \code{BE\_E} (before \code{BE\_F}).

\begin{lstlisting}[%
	caption={Repairs in extended Galileo},%
	label={code:repairPDF},%
	captionpos=b,%
	frame=single,%
	basicstyle=\footnotesize\ttfamily%
]
toplevel "Gate3";
"Gate3" and "BE_E" "BE_F" "BE_G";
"BE_E" lambda=6.0E-5 EXT_repairPDF=uniform(8,24);
"BE_F" lambda=7.0E-5 EXT_repairPDF=uniform(8,24);
"BE_G" lambda=6.0E-5 EXT_repairPDF=uniform(8,12);
"RB1" repairbox_priority "BE_E" "BE_F" "BE_G";
\end{lstlisting}

%% %% %% %% %% %% %% %% %% %% %% %% %% %% %% %% %% %% %% %% %% %% %% %% %% %% 
%
\subsection{Converter: RFT \raisebox{.3pt}{$\to$} IOSA}
\label{sec:impl:dft2iosa}

We also implemented the \iosa semantics for \rft \cite{Mon18}.
For this we developed a Java textual converter whose input is an \rft defined in extended Galileo.
The converter outputs the \iosa semantics of the tree, and the composition function for the corresponding local importance functions using \Cref{tab:compifun}.

\begingroup
\def\algin{\ensuremath{\mathit{in}}\xspace}
\def\algout{\ensuremath{\mathit{out}}\xspace}
\def\algrft{\ensuremath{\mathit{rft}}\xspace}
\def\algcif{\ensuremath{\mathit{cif}}\xspace}
\begin{algorithm}
	\caption{Conversion from \rft to \iosa}
	\label{alg:conversion}
	\begin{algorithmic}[1] % The number tells where the line numbering starts
		\Procedure{RFTtoIOSA}{\algin,\,\algout{[3]}}
		\State \algrft      $\gets$ \Call{parseRFT}{\algin}\\
		\hspace{12px} \Call{convertDynamicGates}{\algrft}
		\State \algcif      $\gets$ \Call{templateImpFun}{\algrft}
		\State \algout{[0]} $\gets$ \Call{convertTree}{\algrft}
		\State \algout{[0]} $\gets$ \Call{convertRBOX}{\algrft,\,\algout{[0]}}
		\State \algout{[1]} $\gets$ \Call{generateProperties}{\algout{[0]}}
		\State \algout{[2]} $\gets$ \Call{generateImpFun}{\algcif,\,\algout{[0]}}
		\EndProcedure
	\end{algorithmic}
\end{algorithm}
\endgroup

The conversion procedure is show as \Cref{alg:conversion}.
The \acronym{ast} parsed from the \rft is first processed to convert the dynamic \PANDgate and \FDEPgate gates as described in \Cref{sec:theory:ifun}.
From the resulting tree we implement (a template of) the importance function following \Cref{tab:compifun}.
Next, the \iosa for each tree node is computed following \cite{Mon18}.
Once all automata names were thus defined, the \RBOX is built and added to the (now final) \iosa semantics.
Also the property queries (system unreliability and unavailability) are then defined.
Finally, the importance function template is filled with the \iosa automata names, and returned as a complete importance function.

Regarding the queries, unreliability and unavailability are encoded as variants of \acronym{pctl} \cite{HJ94} and \acronym{csl} \cite{BKH99} that \fig can take as input.
For instance, say we want to estimate the unreliability of the \rft defined in \Cref{code:repairPDF} at $T=15.5$.
If the converter defined the variable \code{count}, internal to the \iosa module corresponding to the \code{Gate3} \ANDgate, then \textsc{generateProperties()} produces a \acronym{pctl} query as in \Cref{code:reliabilityProp}.

\begin{lstlisting}[%
	caption={Unreliability property query for \Cref{code:repairPDF}},%
	label={code:reliabilityProp},%
	captionpos=b,%
	frame=single,%
	basicstyle=\footnotesize\ttfamily%
]
properties
  P( U<=15.5 Gate3.count==3)
endproperties
\end{lstlisting}

%% %% %% %% %% %% %% %% %% %% %% %% %% %% %% %% %% %% %% %% %% %% %% %% %% %% 
%
\subsection{FIG: RES to estimate rare dependability metrics}
\label{sec:impl:fig}

The \fig tool was devised to study temporal logic queries of \iosa models \cite{Bud17}, described either in their native syntax or in the \acronym{jani} model exchange format \cite{BDH+17}.
%We extended \fig for this work to further estimate time-bounded reachability, i.e.\ system reliability.
Using \res embedded in statistical model checking, \fig computes (arbitrary) \ci{s} that estimate the degree to which a model complies to a property specification.

\fig was designed for automatic \res, implementing the algorithms from \cite{BDH15,BDM16} to derive an importance function from the system model.
It can select an aggregation operator, to compose the local functions computed for the modules of the system.
This, however, depends on the property query, and does not lead to high quality importance functions for \fta, where the structure is in the tree and not in the query---see \cite{Bud17} and our discussion in \Cref{sec:theory:ifun}.

\fig can also be input a composition function, 
%described in terms of the names of the modules composing the system.
to aggregate the local importance functions of the system modules.
This is the feature used by our $\rft\rightarrow\iosa$ converter, as detailed in \Cref{sec:impl:dft2iosa}
Thus, from the \iosa model and the importance function produced by our converter, \fig performs \res to compute \ci{s} around the dependability metrics queried.

}%
	{}%
\ifthenelse{\boolean{longversion}}%
	{%%%%%%%%%%%%%%%%%%%%%%%%%%%%%%%%%%%%%%%%%%%%%%%%%%%%%%%%%%%%%%%%%%%%%%%%%%%%%%
%% !TEX root =  ../main.tex
\section{Experimental evaluation}
\label{sec:expe}

%% %% %% %% %% %% %% %% %% %% %% %% %% %% %% %% %% %% %% %% %% %% %% %% %% %% 
%
\subsection{General setup}
\label{sec:expe:setting}

Using our tool chain, we have verified the efficiency of the theory introduced in \Cref{sec:theory}.
We experimented on 26 repairable non-Markovian \dft{s} using different simulation algorithms:
\begin{enumerate*}[label={\bfseries\arabic*.}]
%% NOTE: keep enumeration in-line, we gain 3 lines (we do need those lines)
\item	Standard Monte Carlo (\smc);
\item	\restart with thresholds selected via the Sequential Monte Carlo
		algorithm \cite{Bud17,BDH17} for different splitting values
		(\rstn{$n$} for $n=2,3,5,8,11$);
\item	\restart with thresholds selected via the Expected Success
		algorithm \cite{BDH17} (\rstes); and
\item	Fixed Effort \cite{Gar00,BDH17} for different number of runs
		performed in each importance region (\fen{$n$} for $n=8,12,16,24,32$).
\end{enumerate*}
\res algorithms were implemented using the importance function defined in \Cref{tab:compifun}, by following the theory from \cite{BDH15,BDM16,BDH17} to choose thresholds and splitting values automatically.

\begingroup
\def\reg{\textsuperscript{\textregistered}\space}
We ran our experiments in two types of nodes of a SLURM cluster running 64-bit Linux (Ubuntu, kernel~3.13.0-168): \emph{korenvliet} nodes have CPUs Intel\reg Xeon\reg E5-2630 v3 @ 2.40\,GHz, and 64\,GB of DDR4 @ 1600\,MHz RAM memory; \emph{caserta} has CPUs Intel\reg Xeon\reg E7-8890 v4 @ 2.20\,GHz, and 2\,TB of DDR4 @ 1866\,MHz RAM memory.
\endgroup

%% %% %% %% %% %% %% %% %% %% %% %% %% %% %% %% %% %% %% %% %% %% %% %% %% %% 
%
\subsection{Division of experimental instances}
\label{sec:expe:instances}

We experimented on seven case studies.
These were originally Markovian and without repairs \cite{DBB90,VSDFMR02,GSS15,RRBS17}.
To turn them into non-Markovian \rft{s} we added \RBOX elements and modified its fail and repair \pdf{s} as detailed in \Cref{sec:expe:casestudies}.
Moreover, to delineate the performance boost of our theory in the analysis of rare dependability metrics, we tested each case study in increasingly resilient configurations.
For this, we parameterised them: a higher value of the parameter in a case study implies a more resilient system, i.e.\ smaller unavailability or unreliability values.
The values of the parameters are given in \Cref{tab:expe:overview} and described below. 
%We estimated unavailability and unreliability ($T=10^3$) for these trees: results range from \rarep{1.20}{2} to \rarep{2.02}{8}---see \Cref{tab:expe:overview}.
\Cref{fig:expe:results:ava,fig:expe:results:rel} show that, the rarer the metric, the more efficient our \res implementation becomes w.r.t.\ \smc.
% \msnote{and how about the other techniques??}
% ^^^ which other techniques? I don't understand the question

%%	Our experiments span seven parametrized case studies. These case studies were originally designed to validate methods with Markovian failure rates, and no repairs \cite{AddReferences}. 
%%	To make them suitable for our setting, we made them repairable and non-Markovian and repairable: we added \RBOX elements and modified its fail and repair \pdf{s}. Further, to investigate the performance for rare dependability metrics, we tested each case study in increasingly resilient configurations. We have parameterised these cases, where higher parameter values correspond a more resilient system, i.e., smaller unavailability and unreliability values.
%%	The values of the parameters are given in  Table~\ref{tab:expe:overview}, and described below. 
%%	Figure~\ref{WhereAreTheseResults} shows that, the rarer the metric, the more efficient our \res implementation becomes w.r.t.\ \smc. \msnote{and how about the other techniques??}

\begin{table}
	\vspace{-3ex}
	\centering
	\caption{General overview of experimental setting}
	\label{tab:expe:overview}
	\vspace{2ex}.  
	\smaller
	%\sffamily
	%% !TEX root =  ../main.tex
%
\begingroup
\def\head#1{\textbf{#1}}
\def\lenvcol{2em}
\def\lenfcol{5em}
\def\lenlcol{9em}
\begin{tabular}{lr@{\quad}c@{~~~}c@{~~~}r@{\;}l@{\quad~}l}
	%%%%%%%%%%%%%%%%%%%%%%%%%%%%%%%%%%%%%%%%%%%%%%%%%%%%%%%%%%%%
	\toprule
	\multirow{2}{\lenvcol}{\!\head{Metric}}
		& \multirow{2}{*}{\parbox{\lenfcol}{\raggedleft%
		                  \head{Case}\phantom{j}\\[-.2ex]\head{study}}}
		& \multicolumn{4}{l}{~~~~~\head{Difficulty}}
		& \multirow{2}{*}{\parbox{\lenlcol}{\centering%
		                  \head{Simulation}\\[-.2ex]\head{algorithms}}}\\
	 & & P. & Est.\phantom{$\big($} & \multicolumn{2}{l}{TO.} & \\
	\toprule
	%%%%%%%%%%%%%%%%%%%%%%%%%%%%%%%%%%%%%%%%%%%%%%%%%%%%%%%%%%%%
	\multirow{16}{\lenvcol}{\rotatebox{90}{\uline{UN\,AVAILABILITY}}}
		%%%%%%%%%%%%%%%%%%%%%%
		& \multirow{3}{*}{VOT}
			& 2 & \rarep{8.47}{4} &  5&m &
		  \multirow{3}{*}{\parbox{\lenlcol}{\smc\\
		                  \rstn{$\{2,3,5,8,11\}$}}}\\
	    &   & 3 & \rarep{1.94}{5} & 30&m & \\
	    &   & 4 & \rarep{4.70}{7} &  3&h & \\
	\cmidrule(l){2-7}
		%%%%%%%%%%%%%%%%%%%%%%%
		& \multirow{5}{*}{HECS}
			& 1 & \rarep{6.26}{3} &  5&s &
		  \multirow{5}{*}{\parbox{\lenlcol}{\smc\\\rstes\\
		                  \rstn{$\{2,5,8,11\}$}}}\\
		&   & 2 & \rarep{6.11}{5} & 20&s & \\
	    &   & 3 & \rarep{1.56}{6} &  2&m & \\
	    &   & 4 & \rarep{1.16}{7} & 10&m & \\
	    &   & 5 & \rarep{2.02}{8} &  1&h & \\
	\cmidrule(l){2-7}
		%%%%%%%%%%%%%%%%%%%%%%%
		& \multirow{4}{*}{RC}
			& 3 & \rarep{3.73}{5} & 30&s &
		  \multirow{4}{*}{\parbox{\lenlcol}{\smc\\\rstes\\
		                  \rstn{$\{2,5,8,11\}$}}}\\
	    &   & 4 & \rarep{3.39}{6} &  5&m & \\
	    &   & 5 & \rarep{5.07}{7} & 30&m & \\
	    &   & 6 & \rarep{1.02}{7} &  2&h & \\
	\cmidrule(l){2-7}
		%%%%%%%%%%%%%%%%%%%%%%%
		& \multirow{4}{*}{RWC}
			& 1 & \rarep{4.88}{4} & 30&s &
		  \multirow{4}{*}{\parbox{\lenlcol}{\smc\\
		                  \rstn{$\{2,5,8,11\}$}}}\\
	    &   & 2 & \rarep{3.15}{5} &  5&m & \\
	    &   & 3 & \rarep{3.03}{6} & 30&m & \\
	    &   & 4 & \rarep{4.55}{7} &  2&h & \\
	\midrule[.7pt]
	%%%%%%%%%%%%%%%%%%%%%%%%%%%%%%%%%%%%%%%%%%%%%%%%%%%%%%%%%%%%
	\multirow{17}{\lenvcol}{\rotatebox{90}{\uline{UN\,RELIABILITY}}}
		%%%%%%%%%%%%%%%%%%%%%%
		& \multirow{3}{*}{DSPARE}
			& 3 & \rarep{7.03}{4} &  5&m &
		  \multirow{3}{*}{\parbox{\lenlcol}{\smc\\
		                  \rstn{$\{2,3,5,8,11\}$}\\
		                  \fen{$\{8,16,32\}$}}}\\
	    &   & 4 & \rarep{6.08}{5} & 30&m & \\
	    &   & 5 & \rarep{7.31}{6} &  3&h & \\
	\cmidrule(l){2-7}
		%%%%%%%%%%%%%%%%%%%%%%%
		& \multirow{4}{*}{HECS}
			& 2 & \rarep{1.98}{3} & 20&s &
		  \multirow{4}{*}{\parbox{\lenlcol}{\smc\\
		                  \rstn{$\{2,5,8,11\}$}\\
		                  \fen{$\{8,16,32\}$}}}\\
	    &   & 3 & \rarep{3.60}{5} &  5&m & \\
	    &   & 4 & \rarep{2.35}{6} & 30&m & \\
	    &   & 5 & \rarep{2.61}{7} &  3&h & \\
	\cmidrule(l){2-7}
		%%%%%%%%%%%%%%%%%%%%%%
		& \multirow{3}{*}{\parbox{\lenfcol}{\raggedleft FTPP\\(triad)}}
			& 4 & \rarep{1.20}{2} & 30&s &
		  \multirow{3}{*}{\parbox{\lenlcol}{\smc\\
		                  \rstn{$\{2,5\}$}\\
		                  \fen{$\{8,12,16,24\}$}}}\\
	    &   & 5 & \rarep{2.49}{4} &  4&m & \\
	    &   & 6 & \rarep{6.34}{7} & 40&m & \\
	\cmidrule(l){2-7}
		%%%%%%%%%%%%%%%%%%%%%%%
		& \multirow{4}{*}{HVC}
			& 4 & \rarep{1.11}{2} & 90&s &
		  \multirow{4}{*}{\parbox{\lenlcol}{\smc\\
		                  \rstn{$\{2,5,8,11\}$}\\
		                  \fen{$8$}}}\\
	    &   & 5 & \rarep{4.61}{4} &  5&m & \\
	    &   & 6 & \rarep{3.44}{5} & 30&m & \\
	    &   & 7 & \rarep{4.17}{6} &  2&h & \\
	\cmidrule(l){2-7}
		%%%%%%%%%%%%%%%%%%%%%%%
		& \multirow{3}{*}{RWC}
			& 2 & \rarep{7.03}{4} &  5&m &
		  \multirow{3}{*}{\parbox{\lenlcol}{\smc\\
		                  \rstn{$\{2,5,8,11\}$}}}\\
	    &   & 3 & \rarep{6.08}{5} & 30&m & \\
	    &   & 4 & \rarep{7.31}{6} &  2&h & \\
	\bottomrule
\end{tabular}
\endgroup

	\vspace{-3ex}
\end{table}

For each parametric case study we compare the simulation algorithms mentioned above.
We estimate system unavailability and unreliability at time $T=10^3$.
%; their values range from \rarep{1.20}{2} to \rarep{2.02}{8}. 
For each combination of metric, fault tree, and algorithm---an \emph{instance}---we computed \ci{s} of 95\% confidence level around the point estimate for the metric.
To do so, we ran simulations with \fig for predefined wall-clock runtimes (that depend on the case and parameter as detailed in \Cref{tab:expe:overview}), and built 10 \ci{s} for each instance.
\
We then compared the average width of the \ci{s} per instance.
The algorithm with the most precise (narrowest) intervals was the \emph{most efficient} to compute that metric on that tree.
In \Cref{sec:expe:results} we show that for the most resilient configurations of all case studies, \res algorithms implemented from our importance functions are more efficient (and as automatic) than \smc.

An overview of the full experimental setting is given in \Cref{tab:expe:overview}.
The parameterised configurations of all case studies are detailed in \textbf{Difficulty}.
Its sub-columns are:
\begin{enumerate*}[label=]
	\item[[\,P.]]
	that gives the parameter value of each case
	study---see~\Cref{sec:expe:casestudies};
	\item[[\,TO.]]
	for ``Time-Out,'' i.e.\ the simulation runtime,
	higher for the more resilient configurations of a case study
	to let the algorithms sample some rare event; and
	\item[[\,Est.]]
	that gives the point estimate averaged over all values%
	\footnote{We removed outliers using a modified Z-score with $m=2$ \cite{IH93}.},
	ranging over all simulation algorithms and the 10 (repeated and independent)
	computations performed for each tree and metric.
\end{enumerate*}

Note that the Time-Out chosen for a (parameterised) case study may be insufficient for certain algorithms to observe any rare event, e.g.\ for \smc.
If that happens, the stochastic model checker \fig reports a ``null estimate'' $[0,0]$.
Moreover, the simulation of random events depend on the \acronym{rng}---and the seed---used by \fig, so different runs may yield different \ci{s}.
To account for these factors when assessing the outcome of each instance, we computed each \ci 10 times.
This gives us three dimensions to assess the performance of an algorithm in an instance:
\begin{enumerate*}[label=\mbox{(\emph{\roman*}\hspace{.5pt})}]
\item	how many times did it yield a not-null estimate,
\item	what was the average width of the resulting \ci{s} for that case study
		and parameter (considering not-null estimates only), and
\item	what was the variance of those widths.
\end{enumerate*}

For example, running simulations for 2 minutes, we estimated the unavailability of the parameterised case study ``HECS-3.\!\!''
Using \smc we computed 10 independent \ci{s}.
The same was done for each of \rstn{2,5,8,11,\textsc{es}}.
Results are shown as whisker-bar plots in \Cref{sec:expe:results}.
Each bar corresponds to the \ci{s} computed for an instance, i.e.\ a specific algorithm on one case study with certain parameter.
The height of the bar is the mean \ci width for the 10 iterations of the algorithm (discarding null estimates).
The whiskers on top of it are the standard deviation of these widths, and a bold number at its base (e.g.~%
\scalebox{.6}{\colorbox[HTML]{440154}{\color{white}\dejavu{\bfseries{~3~}}}}%
\scalebox{.6}{\colorbox[HTML]{21908D}{\color{white}\dejavu{\textbf{10}}}}%
\,) indicates how many iterations of \algo yielded not-null estimates.

\begin{figure}[ht]
	\centering
	\includegraphics[width=\textwidth]{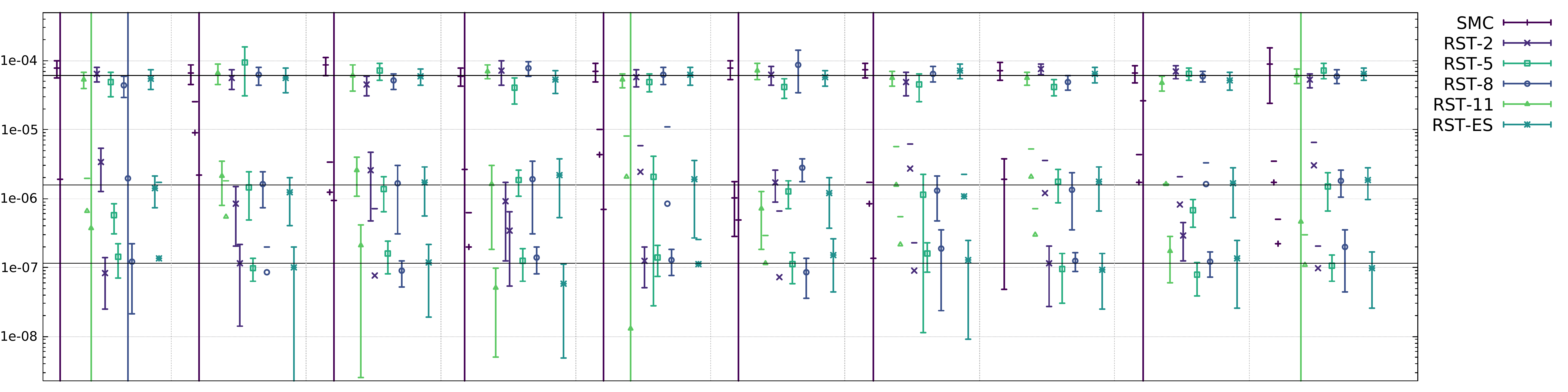}
	\vspace{-2ex}
	\caption{\ci{s} for unavailability of HECS}
	\label{fig:HECS_ava_ci}
	~\\[2ex]
	\includegraphics[width=\textwidth]{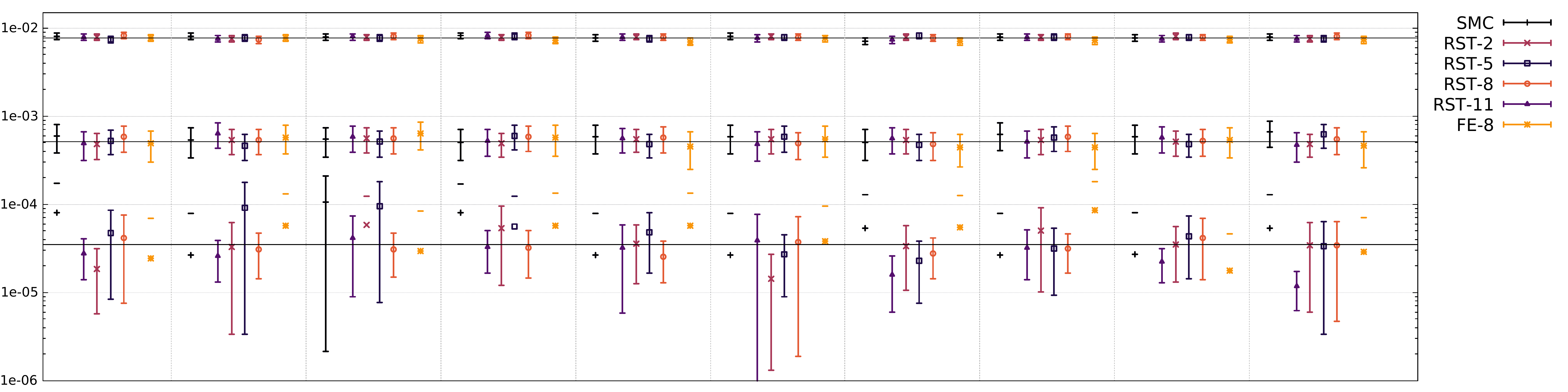}
	\vspace{-2ex}
	\caption{\ci{s} for unreliability of HVC}
	\label{fig:HVC_rel_ci}
\end{figure}

The \ci{s} themselves---whose width we compare in \Cref{sec:expe:results}---for unavailability of HECS-2, HECS-3, and HECS-4, are shown in \Cref{fig:HECS_ava_ci}.
The three horizontal lines are their corresp.\ unavailability values: \rarep{6.11}{5}, \rarep{1.56}{6}, and \rarep{1.16}{7}.
Here we tested algorithms \smc, \rstes, and \rstn{$\{2,5,8,11\}$}.
To explore \res diversity, yet keep the amount of experimentation manageable, in other cases we tested different algorithms---see %
%the last column in
\Cref{tab:expe:overview}.

The 10 \ci{s} computed per instance are separated in \Cref{fig:HECS_ava_ci} by vertical gray dashed lines.
The \ci{s} are the coloured vertical error-bars.
Some of them are the trivial real interval $[0,1]$ and appear as vertical coloured lines; e.g.\ all iterations of \smc for HECS-4 except for the 4\textsuperscript{th}, 8\textsuperscript{th}, and 10\textsuperscript{th}.
Sometimes only one extreme of the interval is not trivial, e.g.\ the 4\textsuperscript{th} and 10\textsuperscript{th} iterations of \smc for HECS-4 which respectively yielded $[0,\rarep{6.31}{7}]$ and $[0,\rarep{5.07}{7}]$.
When not even a point estimate was computed for an iteration, the \ci is missing completely from the plot, e.g.\ the 8\textsuperscript{th} iteration of \smc for HECS-4.

\Cref{fig:HVC_rel_ci} shows the \ci{s} computed in the same way for unreliability of the HVC case study.
In this case we experimented with the algorithms \smc, \fen{8}, and \rstn{$\{2,5,8,11\}$}.
It can be seen that for HVC-6 (the downmost horizontal line at \rarep{3.44}{5}) only the third iteration of \smc yielded a complete \ci, and it is very wide.
In contrast, algorithms like \rstn{2} and \rstn{8} always converged to reasonable \ci estimates in 30~m of simulation runtime for this system configuration.
This is the trend with all experiments: as expected, the more rare the metric, the wider the \ci{s} computed in the time limit via \smc, and at some point it becomes infeasible to converge to non-trivial \ci{s}.
In contrast, \res algorithms---implemented from our importance function---can still compute \ci{s} in the most extreme situations experimented with our case studies.
This is conveyed in \Cref{sec:expe:results} via whisker-bar plots, that show the average width of \ci{s} achieved per instance.

%% %% %% %% %% %% %% %% %% %% %% %% %% %% %% %% %% %% %% %% %% %% %% %% %% %% 
%
\subsection{The case studies}
\label{sec:expe:casestudies}

We briefly describe the seven parametric case studies:
VOT and DSPARE were devised for this work, to check whether \res is efficient on such tree structure and probe different simulation runtime limits;
FTPP and HECS were taken from the literature on \fta \cite{DBB90,VSDFMR02}; and
RC, HVC, and RWC concern industrial railroad systems \cite{GKS+14,GSS15,RRBS17}.
The structure of all these systems is presented in \Cref{fig:expe:casestudies}; the fail, repair, and dormancy \pdf{s} of their \BE{s} and \SBE{s} are given in \Cref{tab:expe:BEs}.

\begin{figure}
	\centering
	\caption{Case studies used for experimentation}
	\label{fig:expe:casestudies}

	\vspace{4ex}
	\begin{subfigure}[t]{.315\linewidth}
		\includegraphics[width=\linewidth]{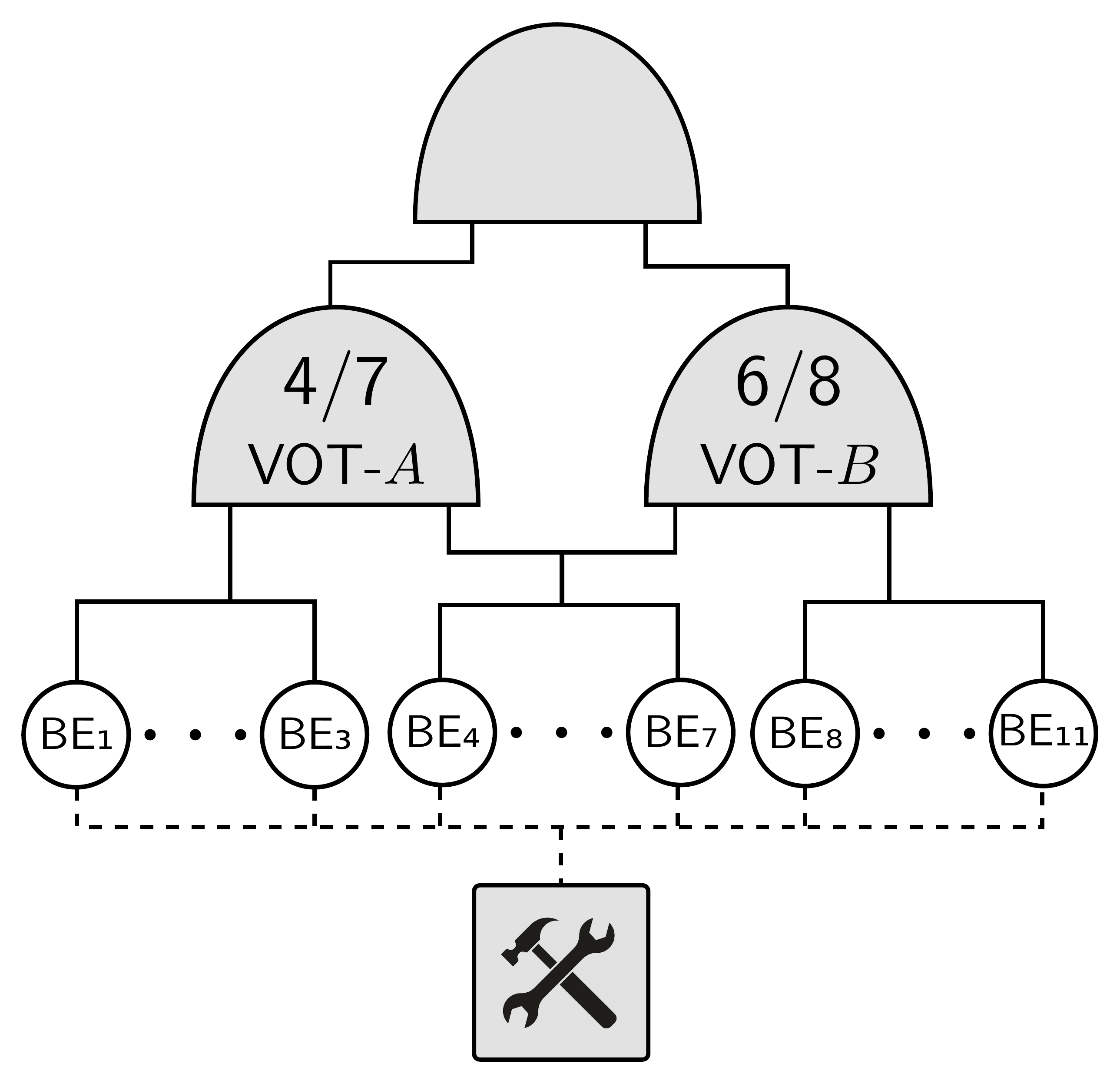}
		\caption{VOT}
		\label{fig:expe:casestudies:vot}
	\end{subfigure}
	\quad
	\raisebox{2ex}{%
	\begin{subfigure}[b]{.294\linewidth}
		\includegraphics[width=\linewidth,trim=0 0 0 0]{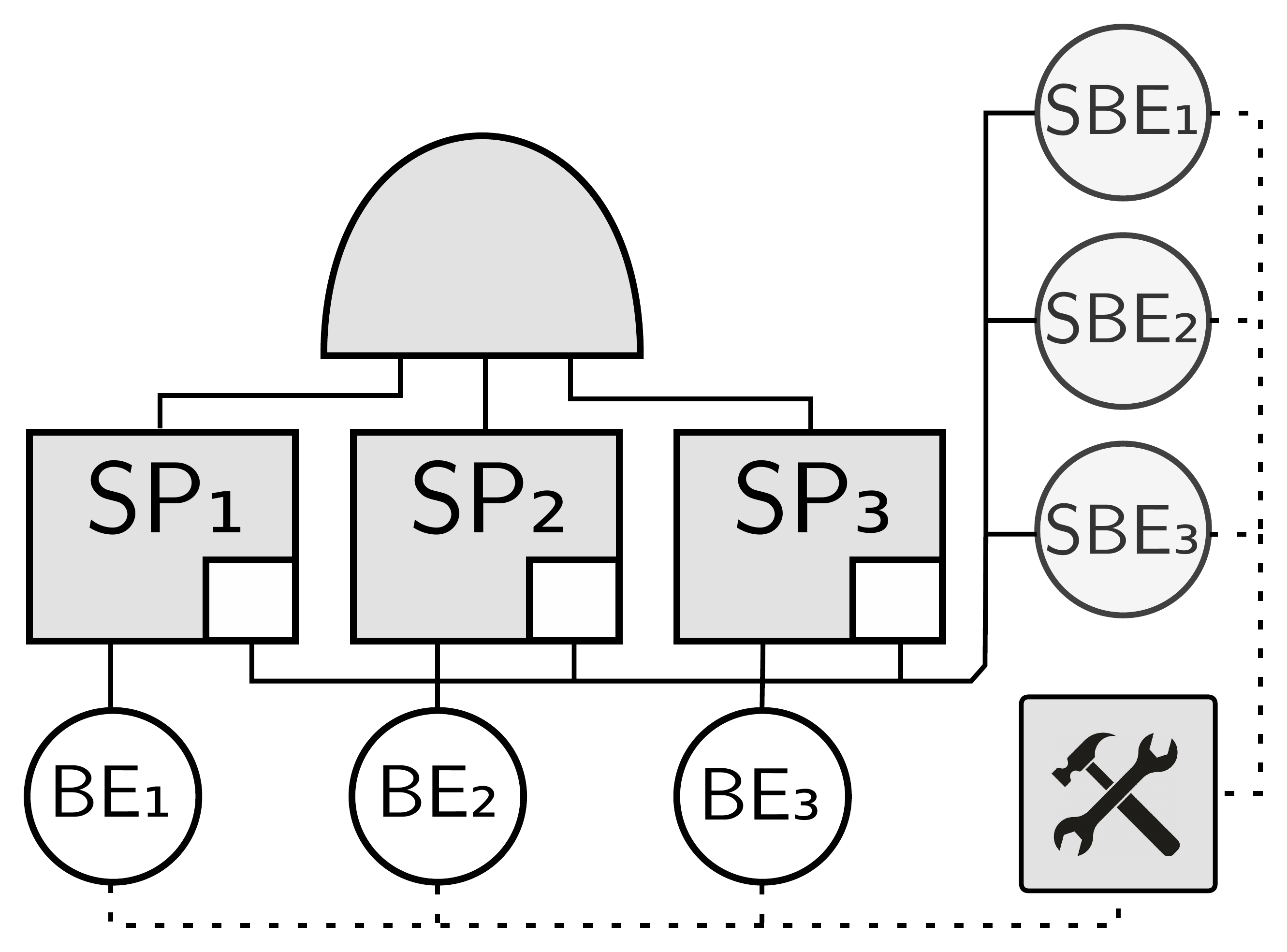}
		\caption{DSPARE}
		\label{fig:expe:casestudies:dspare}
	\end{subfigure}}
	\quad
	\raisebox{-3ex}{%
	\begin{subfigure}[t]{.28\linewidth}
		\includegraphics[width=\linewidth]{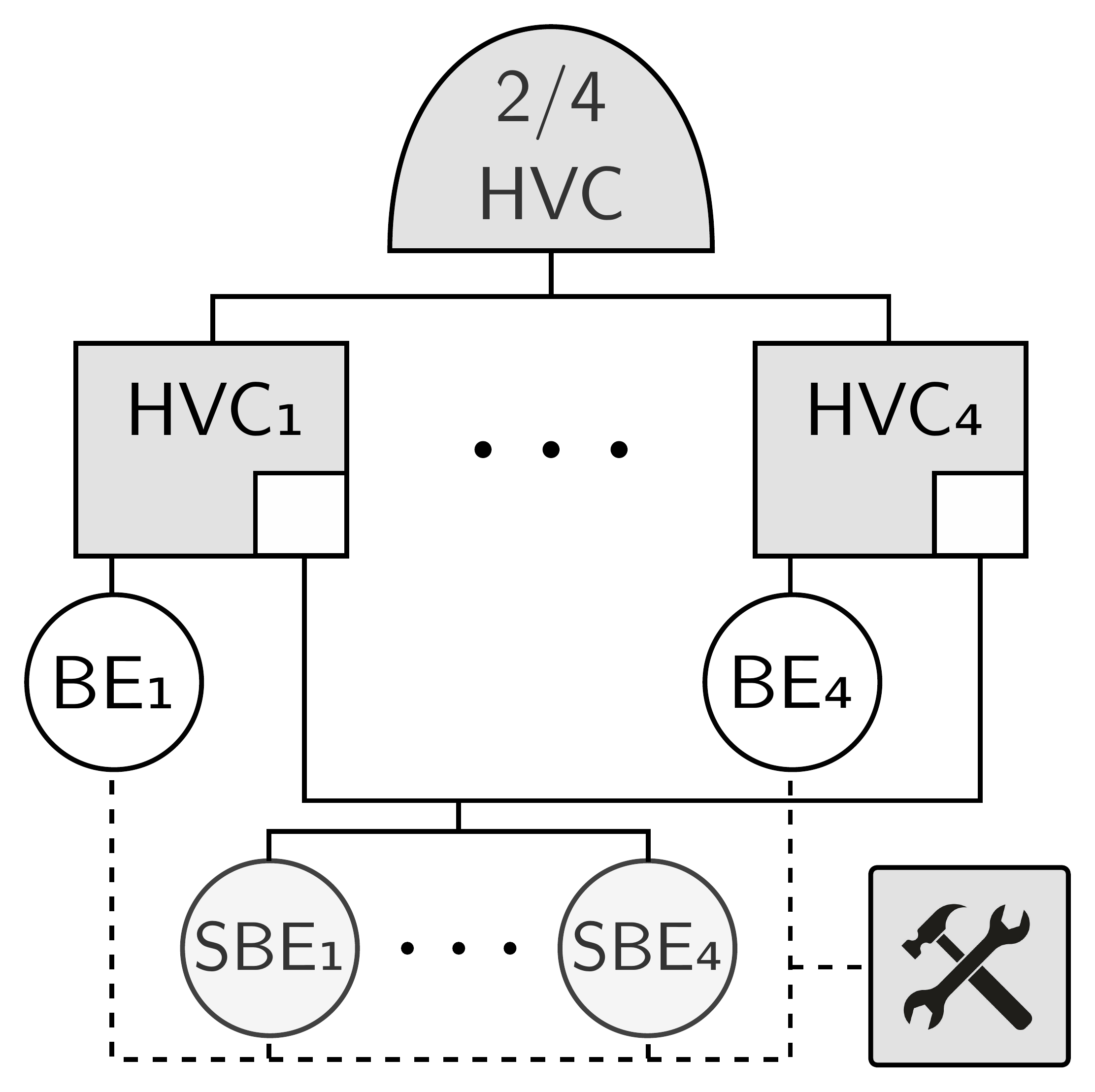}
		\caption{HVC}
		\label{fig:expe:casestudies:hvc}
	\end{subfigure}}

	\vspace{-3ex}
	\begin{subfigure}[b]{.63\linewidth}
		\includegraphics[width=\linewidth]{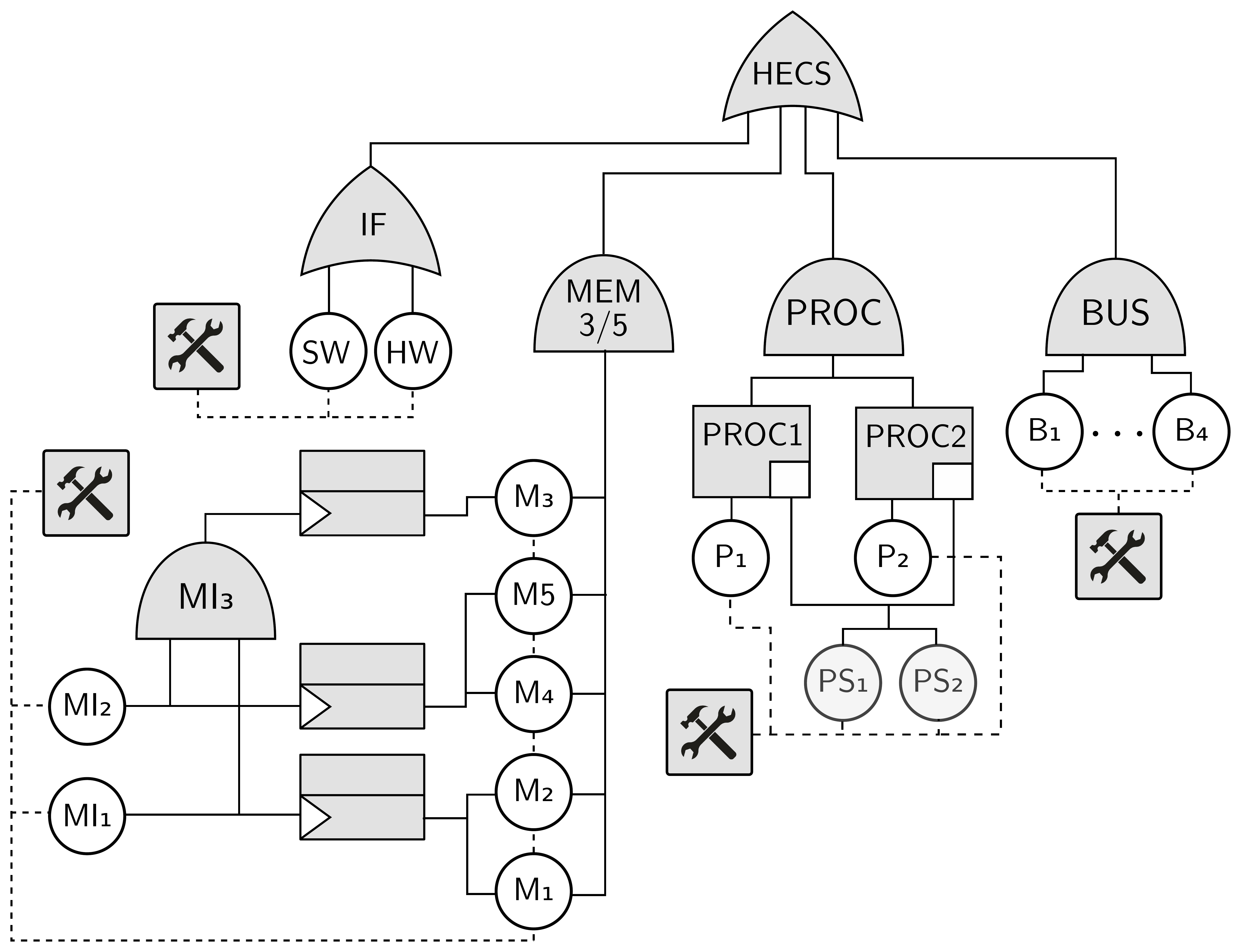}
		\caption{HECS}
		\label{fig:expe:casestudies:hecs}
	\end{subfigure}
	\quad
	\raisebox{7ex}{
	\begin{subfigure}[b]{.28\linewidth}
		\includegraphics[width=\linewidth]{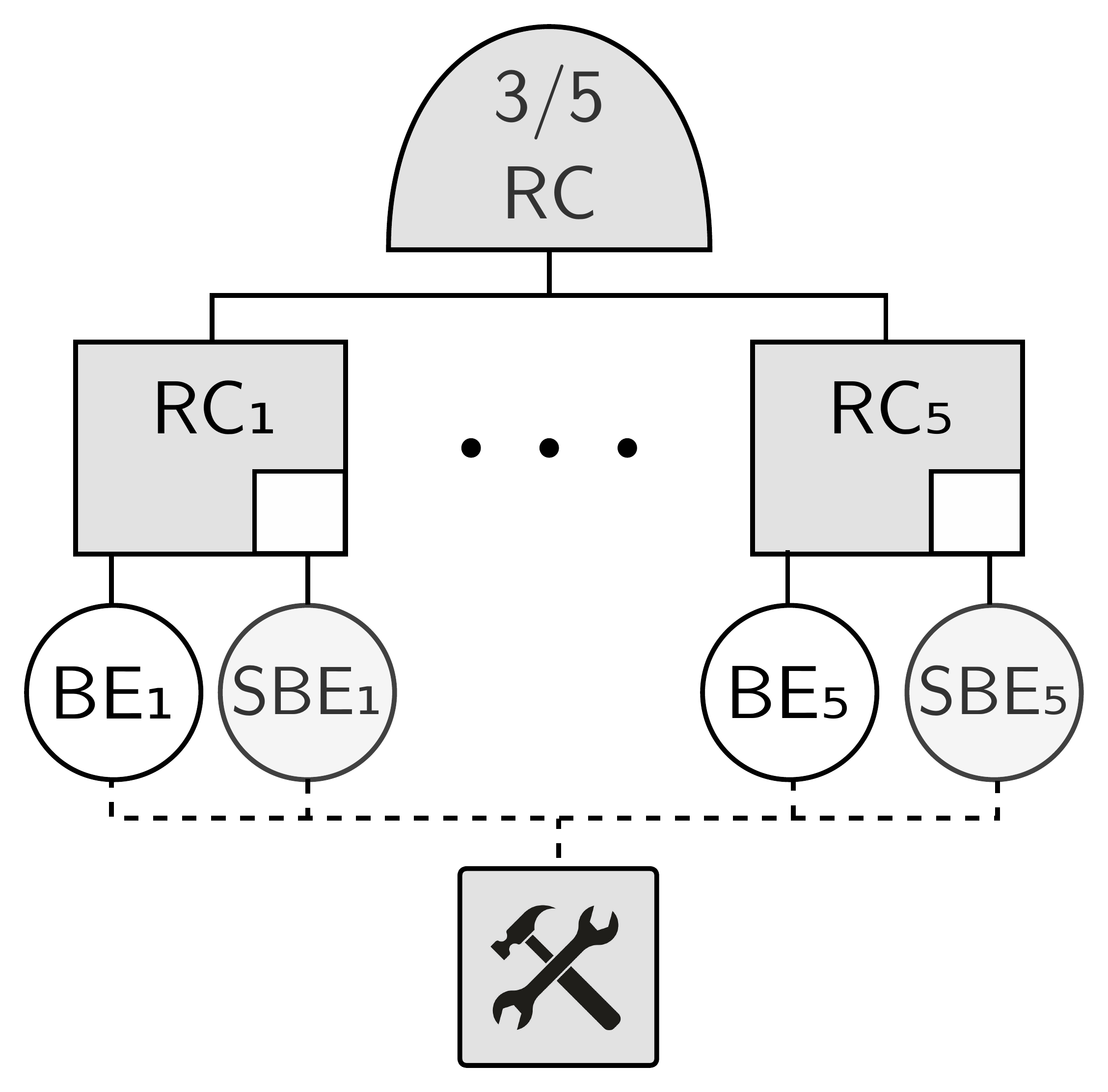}
		\caption{RC}
		\label{fig:expe:casestudies:rc}
	\end{subfigure}}
	
	\begin{subfigure}[b]{.45\linewidth}
		\includegraphics[width=\linewidth]{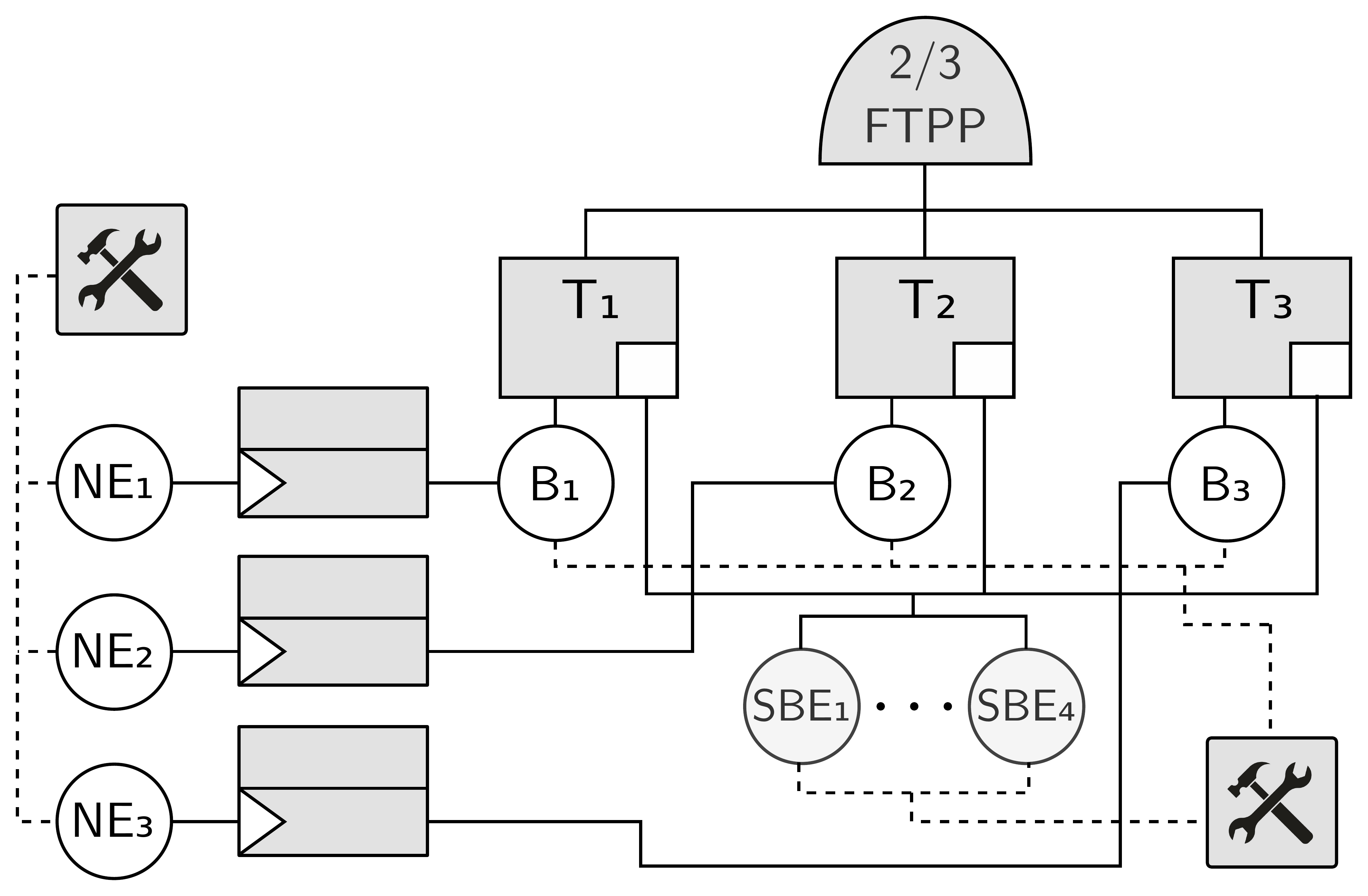}
		\caption{FTPP}
		\label{fig:expe:casestudies:ftpp}
	\end{subfigure}
	\quad~~
	\raisebox{2ex}{%
	\begin{subfigure}[b]{.5\linewidth}
		\includegraphics[width=\linewidth]{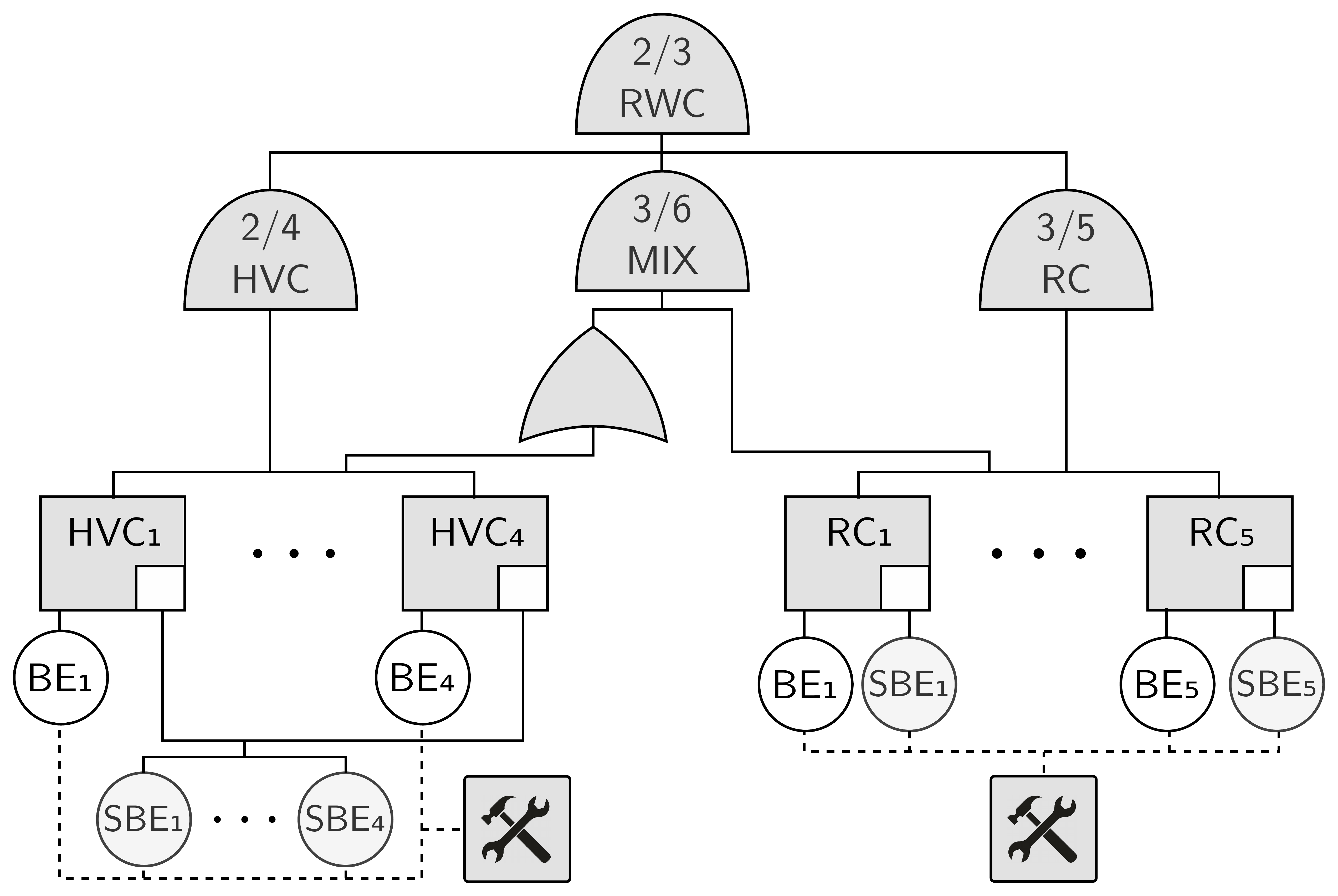}
		\caption{RWC}
		\label{fig:expe:casestudies:rwc}
	\end{subfigure}\hspace{-2em}}
\end{figure}

\paragraph{VOT}(\Cref{fig:expe:casestudies:vot})
~
The first case study is a binary \ANDgate gate whose children are \VOTgate gates, whose children are basic elements.
All \BE{s} are connected to a single \RBOX, which first repairs children of \VOTgate-$A$ and, if all are operational, then repairs any failed child of \VOTgate-$B$.
\VOTgate-$A$ is a $\VOTgate_{k_A}$ gate with $n_A$ children, and analogously for \VOTgate-$B$, where \mbox{$n_A=n_B-1$}, \mbox{$k_A=n_A-3$}, and \mbox{$k_B=n_B-2$}.
From the $n_A$ children of \VOTgate-$A$, $n_B-4$ \BE{s} are also children of \VOTgate-$B$.
VOT is parameterised on $n_B = 8,9,10$ resp. for VOT-$\{2,3,4\}$.
Thus in VOT-2 in \Cref{fig:expe:casestudies:vot}, \VOTgate-$A$ is a $\VOTgate_4$ with $7$ children, and $\VOTgate_B$ is a $\VOTgate_6$ with $8$ children.

\paragraph{DSPARE}(\Cref{fig:expe:casestudies:dspare})
~
This is a ternary \ANDgate gate whose children are \SPAREgate gates, whose children are basic elements.
\SBE{s} are shared among all \SPAREgate{s}: each gate has a unique primary \BE and $n$ spare \BE{s}.
The parameterisation is on $n\in\{3,4,5\}$: \Cref{fig:expe:casestudies:dspare} shows DSPARE-3.
%All \BE{s} (and \SBE{s}) fail and repair (and dormancy) times are the same, as described in \Cref{tab:expe:BEs}.
All basic elements are connected to a single \RBOX, whose priority is to first repair failed \SBE{s} and, if all are operational, then repair failed \BE{s}.

\paragraph{HECS}(\Cref{fig:expe:casestudies:hecs})
~
The Hypothetical Example Computer System is a classic case study from \fta literature \cite{VSDFMR02}.
We study the variant with two memory-unit interfaces, that affect the memory banks via functional dependencies.
Furthermore, we define one \RBOX for each subsystem: Interface, Memory, Processors, and Bus.
HECS is parameterised on the number of parallel buses (\BE{s} \leaf{B_k}) and shared spare processors (\SBE{s} \leaf{PS_b}).
For $n\in\{1,2,3,4,5\}$ HECS-$n$ has $n$ shared spare processors and $2\,n$ parallel buses; \Cref{fig:expe:casestudies:hecs} depicts HECS-2.

\paragraph{FTPP}(\Cref{fig:expe:casestudies:ftpp})
~
The Fault Tolerant Parallel Processor is another classic \ft.
We implemented the grouped cold-spare variant from \cite{DBB90}, where all triads depend on all network elements, and there is an independent \SBE per triad.
We study an individual triad: the tree root is thus a $\VOTgate_2$ with 3 \SPAREgate{s} as children---see \Cref{fig:expe:casestudies:ftpp}.
We defined independent repair boxes for the network and processing elements; the \RBOX in charge of processors prioritises the repair of primary \BE{s}.
FTPP is parameterised on the number of (shared) \SBE{s} of the triad.
For $n\in\{4,5,6\}$ FTPP-$n$ has $n$ shared \SBE{s}; \Cref{fig:expe:casestudies:ftpp} depicts FTPP-4.

\paragraph{RC}(\Cref{fig:expe:casestudies:rc})
~
The Relay Cabinets subsystem is one of the components of the railways cabinets example in \Cref{fig:tiny_RWC}.
It is a $\VOTgate_k$ gate with $k+2$ children: \SPAREgate{s} with one (independent) \SBE besides their primary \BE.
There is a single \RBOX for all basic elements, which prioritises repairs of primary \BE{s}.
RC is parameterised in the number of \SPAREgate{s} that need to fail to cause a top event: $k\in\{3,4,5,6\}$; \Cref{fig:expe:casestudies:rc} depicts RC-3.

\paragraph{HVC}(\Cref{fig:expe:casestudies:hvc})
~
High Voltage Cabinets is the other main component of the railways cabinet example.
This is a $\VOTgate_2$ gate with 4 \SPAREgate children.
Here however the \SBE{s} are shared among all \SPAREgate{s}: HVC is parameterised in the amount of these \SBE{s}, $n\in\{4,5,6,7\}$, with \Cref{fig:expe:casestudies:hvc} depicting HVC-4.
The single \RBOX is analogous to the one in RC.

\paragraph{RWC}(\Cref{fig:expe:casestudies:rwc})
~
The full Railways Cabinet case study combines RC and HVC with a \VOTgate.
The \SPAREgate{s} of RC are direct children of this gate, whereas the high voltage cabinets are interfaced via an \ORgate.
%All basic elements of the subsystems are kept unmodified; the two repair boxes are also kept unmodified and separate for RC and HVC.
The parameterisation of RWC, $m\in\{1,2,3,4\}$, combines those of its subsystems.
RWC-$m$ uses RC-$(m+1)$ and HVC-$(m+2)$; \Cref{fig:expe:casestudies:rwc} depicts RWC-2.

\begin{table}
	\vspace{-3ex}
	\centering
	\caption{Basic elements of the case studies}
	\label{tab:expe:BEs}
	\vspace{2ex}
	\smaller
	%% !TEX root =  ../main.tex
%
\begingroup
\def\casestudy#1{\multicolumn{4}{l}{\!\!#1:\phantom{$\big(\!$}}\\}
\def\dir#1{\ensuremath{\mathrm{dir}(#1)}}
\def\exp#1{\ensuremath{\mathrm{exp}(#1)}}
\def\erl#1#2{\ensuremath{\mathrm{erl}(#1,#2)}}
\def\uni#1#2{\ensuremath{\mathrm{uni}(#1,#2)}}
\def\ray#1{\ensuremath{\mathrm{ray}(#1)}}
\def\nor#1#2{\ensuremath{\mathrm{nor}(#1,#2)}}
\def\lnor#1#2{\ensuremath{\mathrm{lnor}(#1,#2)}}
\def\weib#1#2{\ensuremath{\mathrm{wei}(#1,#2)}}
\begin{tabular}{c@{~~~}l@{~~}l@{~~}l}
	\toprule
	\parbox{4em}{\centering Basic\\[-.5ex]element\\[.7ex]}
		& Fail time \pdf
		& Rep.\;\pdf
		& Dorm.\;\pdf\\
	\toprule
	\casestudy{VOT}
	\BE-$A$     & \lnor{4.37}{0.33}     & \uni{0.4}{0.95}   \\
	\BE-$B$     & \weib{4.5}{0.0125}    & \uni{0.4}{0.95}   \\
	\casestudy{DSPARE}
	\BE         & \exp{0.07}            & \uni{1.0}{2.0}    \\
	\SBE        & \exp{0.07}            & \uni{1.0}{2.0}    & \exp{0.035} \\
	\casestudy{HECS}
	\leaf{SW}   & \exp{\rarep{4.5}{12}} & \uni{28.0}{56.0}  \\
	\leaf{HW}   & \exp{\rarep{1.0}{10}} & \uni{28.0}{56.0}  \\
	\leaf{MI_i} & \exp{\rarep{5.0}{9}}  & \uni{21.0}{28.0}  \\
	\leaf{M_j}  & \exp{\rarep{6.0}{8}}  & \uni{21.0}{28.0}  \\
	\leaf{B_k}  & \exp{\rarep{8.7}{4}}  & \lnor{4.45}{0.24} \\
	\leaf{P_a}  & \exp{\rarep{1.0}{3}}  & \lnor{4.45}{0.24} \\
	\leaf{PS_b} & \exp{\rarep{1.5}{3}}  & \lnor{4.45}{0.24} & \dir{\aleph} \\
	\casestudy{FTPP}
	\leaf{NE_i} & \lnor{6.5}{0.5}       & \nor{150.0}{50.0} \\
	\leaf{B_j}  & \exp{\rarep{2.8}{2}}  & \nor{15.0}{3.0}   \\
	$\SBE_k$    & \exp{\rarep{2.8}{2}}  & \nor{15.0}{3.0}   & \dir{\aleph} \\
	\casestudy{RC}
	$\BE_i$     & \exp{0.04}            & \nor{2.0}{0.7}    \\
	$\SBE_j$    & \exp{0.04}            & \nor{2.0}{0.7}    & \exp{0.5}\\ 
	\casestudy{HVC}
	$\BE_i$     & \ray{1.999}           & \uni{0.15}{0.45}  \\
	$\SBE_j$    & \ray{1.999}           & \uni{0.15}{0.45}  & \erl{3.0}{0.25}\\ 
	\bottomrule
\end{tabular}

\vspace{1ex}
\raggedleft
\begin{tabular}{l@{\quad}l}
	\uline{Abbrev}:     & \uline{Distribution}:\\[.5ex]
	\dir{x}             & $\mathrm{Dirac}(x)$\\
	\exp{\lambda}       & $\mathrm{exponential}(\lambda)$\\
	\erl{k}{\lambda}    & $\mathrm{Erlang}(k,\lambda)$\\
	\uni{a}{b}          & $\mathrm{uniform}([a,b]_\R)$\\
	\ray{\sigma}        & $\mathrm{Rayleigh}(\sigma)$\\
	\weib{k}{\lambda}   & $\mathrm{Weibull}(k,\lambda)$\\
	\nor{\mu}{\sigma}   & $\mathrm{normal}(\mu,\sigma)$\\
	\lnor{\mu}{\sigma}  & $\mathrm{log\text{-}normal}(\mu,\sigma)$\\
\end{tabular}
\endgroup

	\vspace{-3ex}
\end{table}

%% %% %% %% %% %% %% %% %% %% %% %% %% %% %% %% %% %% %% %% %% %% %% %% %% %% 
%
\subsection{Results of experimentation: comparing CI widths}
\label{sec:expe:results}

\begin{figure}
	\centering
	\caption{\ci precision for system unavailability}
	\label{fig:expe:results:ava}
	
	\begin{subfigure}[b]{\linewidth}
		\centering
		\includegraphics[width=.9\linewidth]{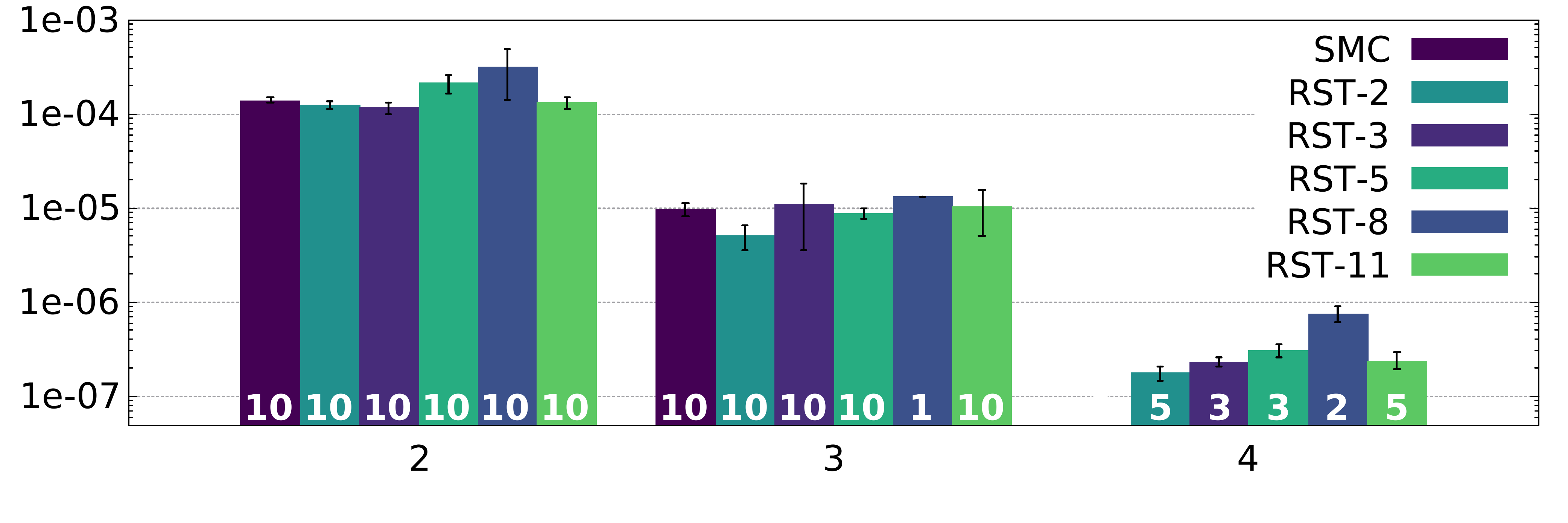}
		\vspace{-2ex}
		\caption{VOT}
		\label{fig:expe:results:ava:vot}
	\end{subfigure}
	
	\begin{subfigure}[b]{\linewidth}
		\centering
		\includegraphics[width=.9\linewidth]{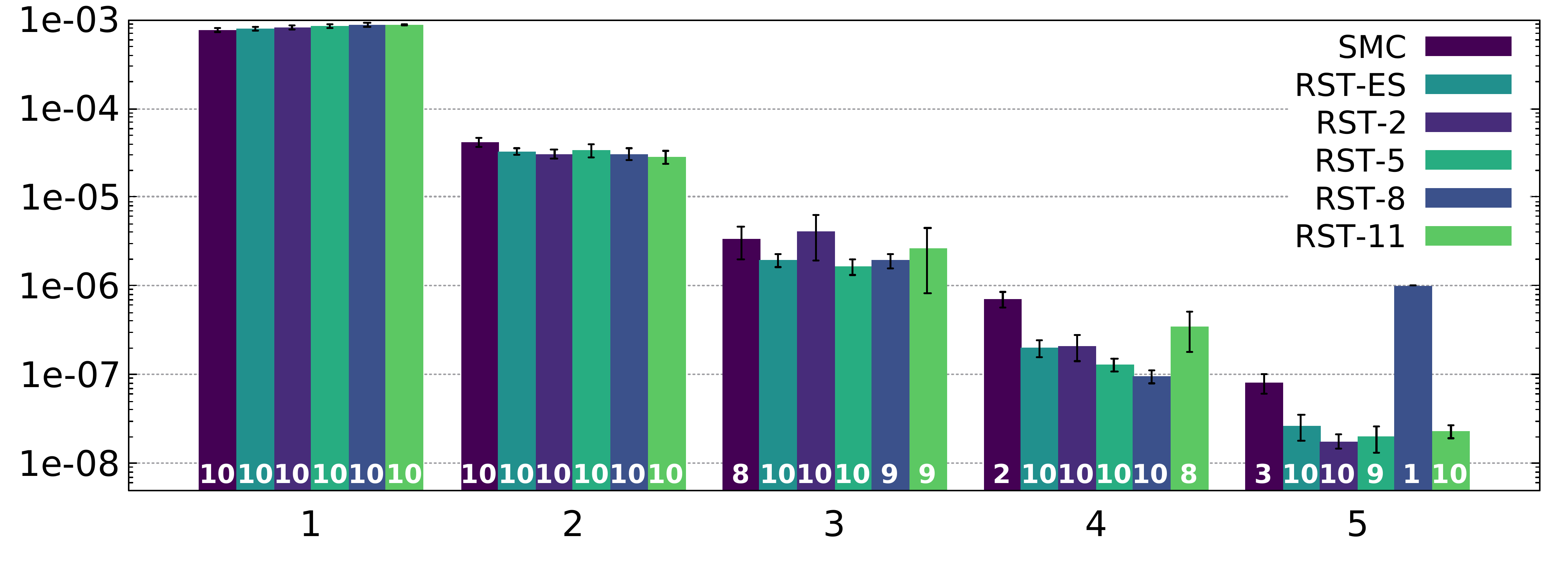}
		\vspace{-2ex}
		\caption{HECS}
		\label{fig:expe:results:ava:hecs}
	\end{subfigure}
	
	\begin{subfigure}[b]{\linewidth}
		\centering
		\includegraphics[width=.9\linewidth]{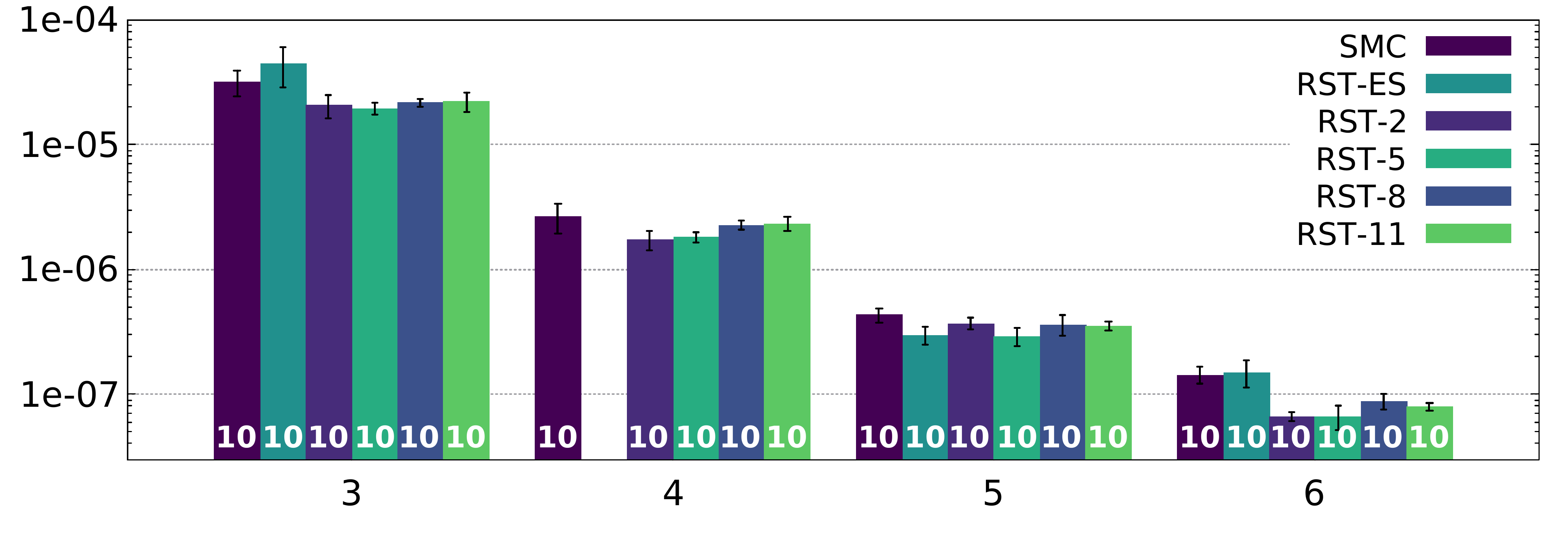}
		\vspace{-2ex}
		\caption{RC}
		\label{fig:expe:results:ava:rc}
	\end{subfigure}
	
	\begin{subfigure}[b]{\linewidth}
		\centering
		\includegraphics[width=.9\linewidth]{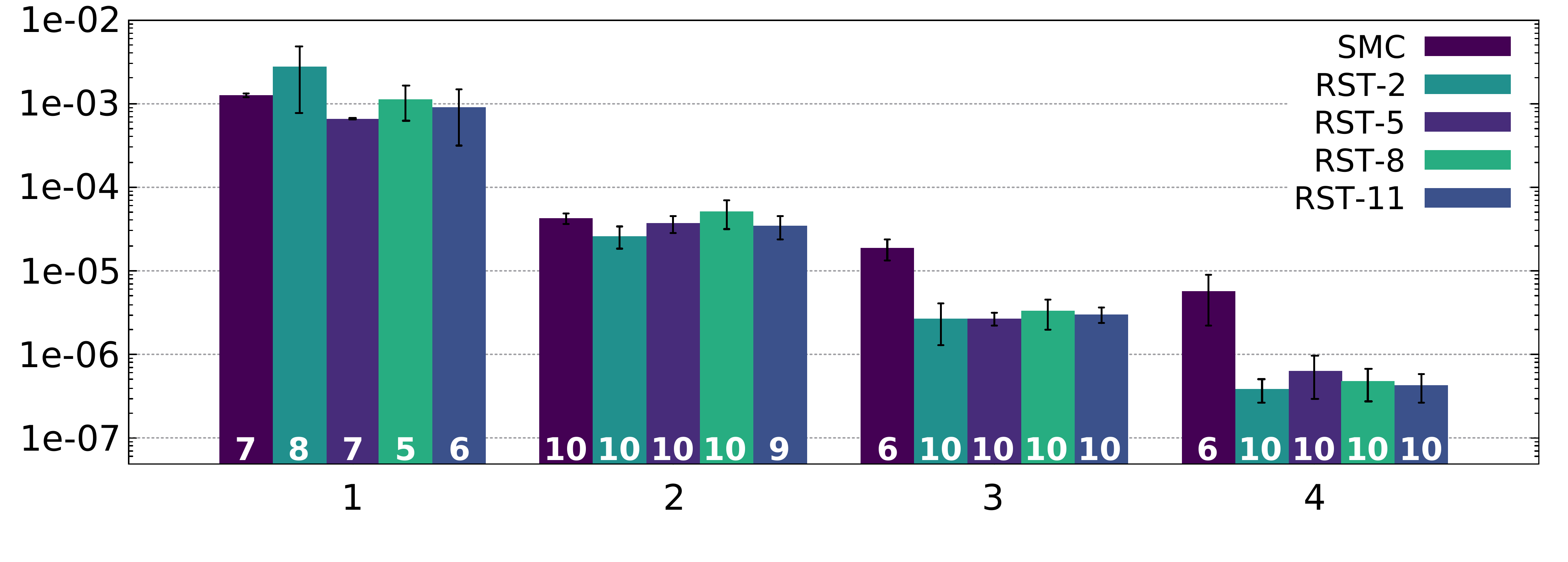}
		\vspace{-2ex}
		\caption{RWC}
		\label{fig:expe:results:ava:rwc}
	\end{subfigure}
	\vspace{-2ex}
\end{figure}

Using \smc and \restart we computed \UNAVA for VOT-$\{2,3,4\}$, HECS-$\{1,2,3,4,5\}$, RC-$\{3,4,5,6\}$, and RWC-$\{1,2,3,4\}$.
\fe was not used since it requires regeneration theory for steady-state analysis \cite{Gar00}, which is not always feasible with non-Markovian models.
The mean widths of the \ci{s} achieved per instance (ulting 95\% confidence level) are shown in \Cref{fig:expe:results:ava}.

For example for VOT-2 (\Cref{fig:expe:results:ava:vot}), 10 independent computations with \smc ran in caserta for 5~min, and all converged to not-null \ci{s} (\,\scalebox{.7}{\colorbox[HTML]{440154}{\color{white}\dejavu{\bfseries10}}}\,).
The mean width of these \ci{s} was \rarep{1.40}{4} and their standard deviation \rarep{7.96}{6}.
For \mbox{VOT-3}, all \smc computations yielded not-null \ci{s} (after 30~min) with an average precision of \rarep{9.62}{6} and standard deviation \rarep{1.52}{6}.
For VOT-4 all \smc simulations yielded null \ci{s} after 3~hours of simulation.
Instead, \rstn{2} converged to 10, 10, and 5 not-null \ci{s} resp.\ for VOT-\{2,3,4\}, with mean widths (and standard deviation):
\rarep{1.24}{4} (\rarep{1.19}{5}),
\rarep{5.09}{6} (\rarep{1.48}{6}), and
\rarep{1.79}{7} (\rarep{3.19}{8}).
Thus for the VOT case study, \rstn{2} was consistently more efficient than \smc, and the efficiency gap increased as \UNAVA became rarer.

This trend repeats in all experiments: as expected, the rarer the metric, the wider the \ci{s} computed in the time limit, until at some point it becomes very hard to converge to not-null \ci{s} at all (specially for \smc).
For the least resilient configuration of each case study, \smc can be competitive or even more efficient than some \ISPLIT variants.
For instance for VOT-1 and HECS-1 in \Cref{fig:expe:results:ava:vot,fig:expe:results:ava:hecs}, all computations converged to not-null \ci{s} for all algorithms, but \smc exhibits less variable \ci widths, viz.\ smaller whiskers.
This is reasonable: truncating and splitting traces in \restart adds
\begin{enumerate*}[label=(\emph{\roman*}\hspace{.5pt})]
\item	simulation overhead that may not pay off to estimate not-so-rare events,
		and on top of it
\item	correlations of cloned traces that share a common history,
		increasing the variability among independent runs.
\end{enumerate*}
On the other hand and as expected, \smc looses this competitiveness for all case studies as failures become rarer, here when $\UNAVA\leqslant\rarep{1.0}{5}$.
This holds nicely for the biggest case studies:
HECS-5\!\!\!%
\footnote{\rstn{8} for HECS-5 escapes this trend: analysing the execution logs it was found that \fig crashed during the second computation.\phantom{..\ so yeah, fuck it.}\hspace{-8em}}%
(a 42-nodes \rft whose \iosa has 126-not-clock variables $\approx\expnum{2.89}{}{38}$ states, with 57 clocks of exponential, uniform, and log-normal \pdf{s})
and RWC-4
(42 nodes, 181 variables $\approx\expnum{6.93}{}{73}$ states, 62 clocks of exponential, Erlang, Rayleigh, uniform, and normal \pdf{s}).

We also estimated the \UNREL[1000] of DSPARE-\{3,4,5\}, RWC-\{2,3,4\}, FTPP-\{4,5,6\}, HVC-\{4,\ldots,7\}, and HECS-\{2,\ldots,5\} using \smc, \restart, and \fe.
For HVC (only) we ran 20 experiments per tree, 10 in each cluster node.
\Cref{fig:expe:results:rel} shows the results.
%Results are shown in \Crefrange{fig:expe:results:rel:dspare}{fig:expe:results:rel:hecs}.

\begin{figure}
	\centering
	\caption{\ci precision for system unreliability}
	\label{fig:expe:results:rel}
	
	\begin{subfigure}[b]{\linewidth}
		\centering
		\includegraphics[width=.8\linewidth]{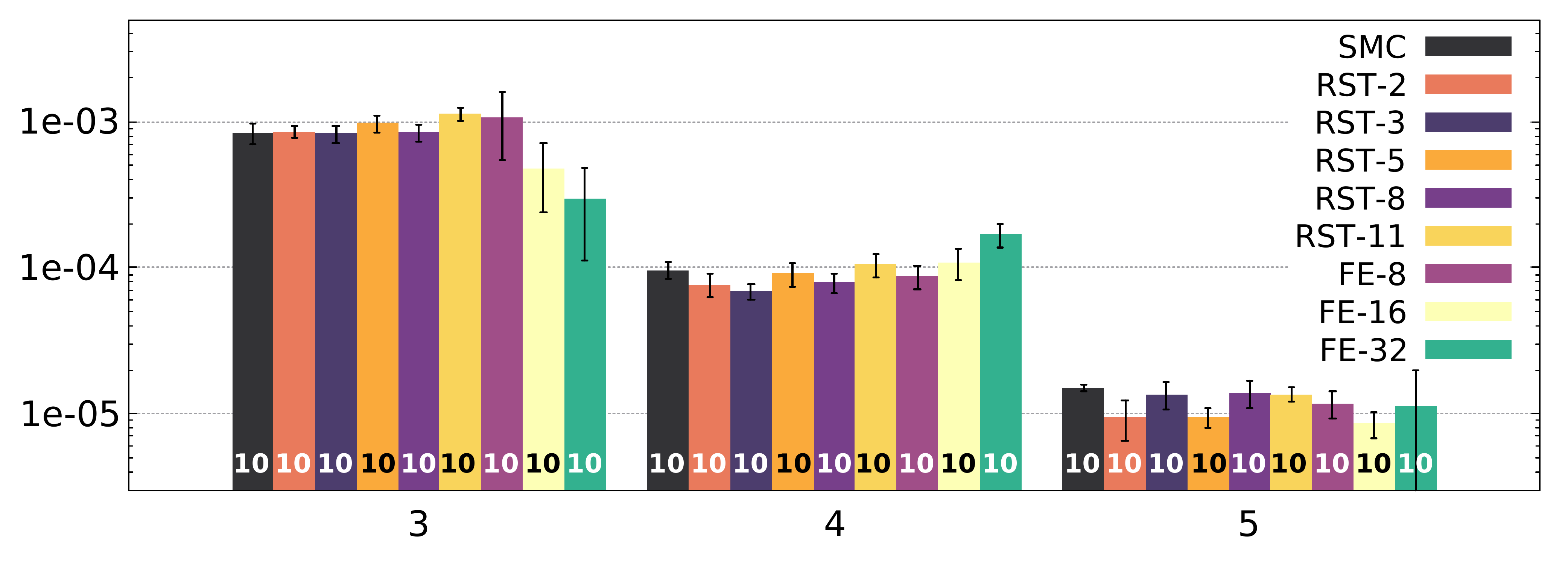}
		\vspace{-2.5ex}
		\caption{DSPARE}
		\label{fig:expe:results:rel:dspare}
	\end{subfigure}
	
	\begin{subfigure}[b]{\linewidth}
		\centering
		\includegraphics[width=.8\linewidth]{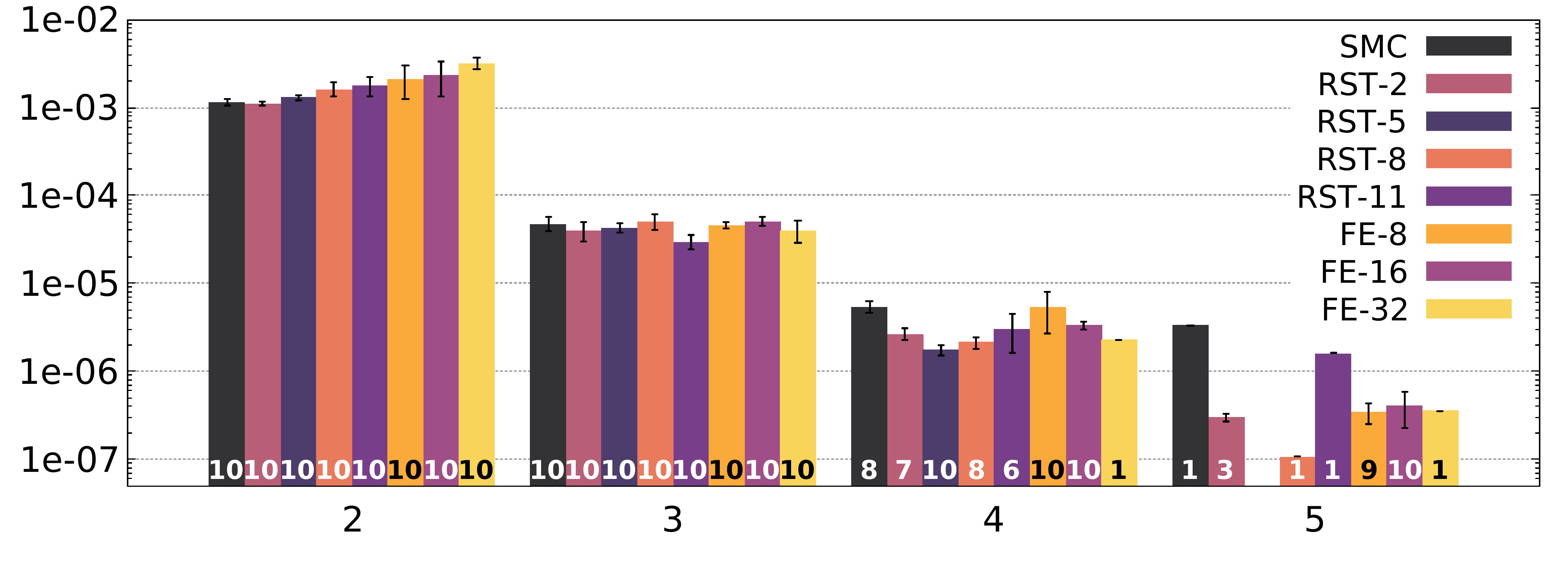}
		\vspace{-2.5ex}
		\caption{HECS}
		\label{fig:expe:results:rel:hecs}
	\end{subfigure}
	
	\begin{subfigure}[b]{\linewidth}
		\centering
		\includegraphics[width=.8\linewidth]{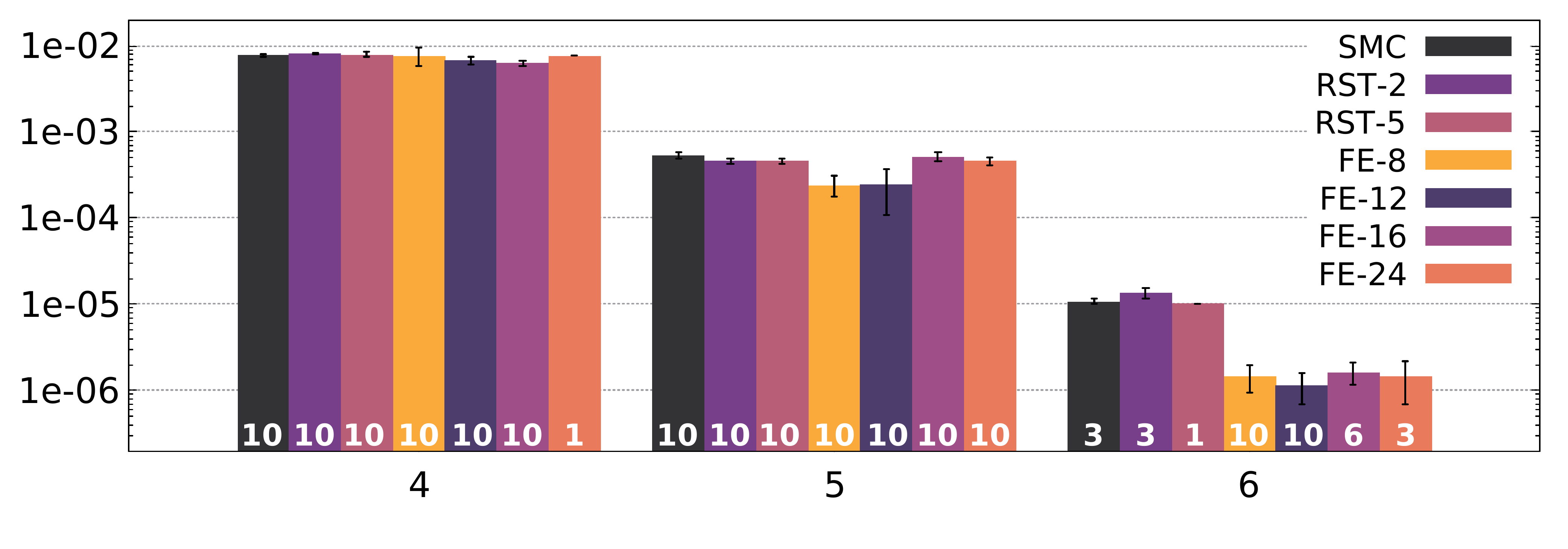}
		\vspace{-2.5ex}
		\caption{FTPP}
		\label{fig:expe:results:rel:ftpp}
	\end{subfigure}
	
	\begin{subfigure}[b]{\linewidth}
		\centering
		\includegraphics[width=.8\linewidth]{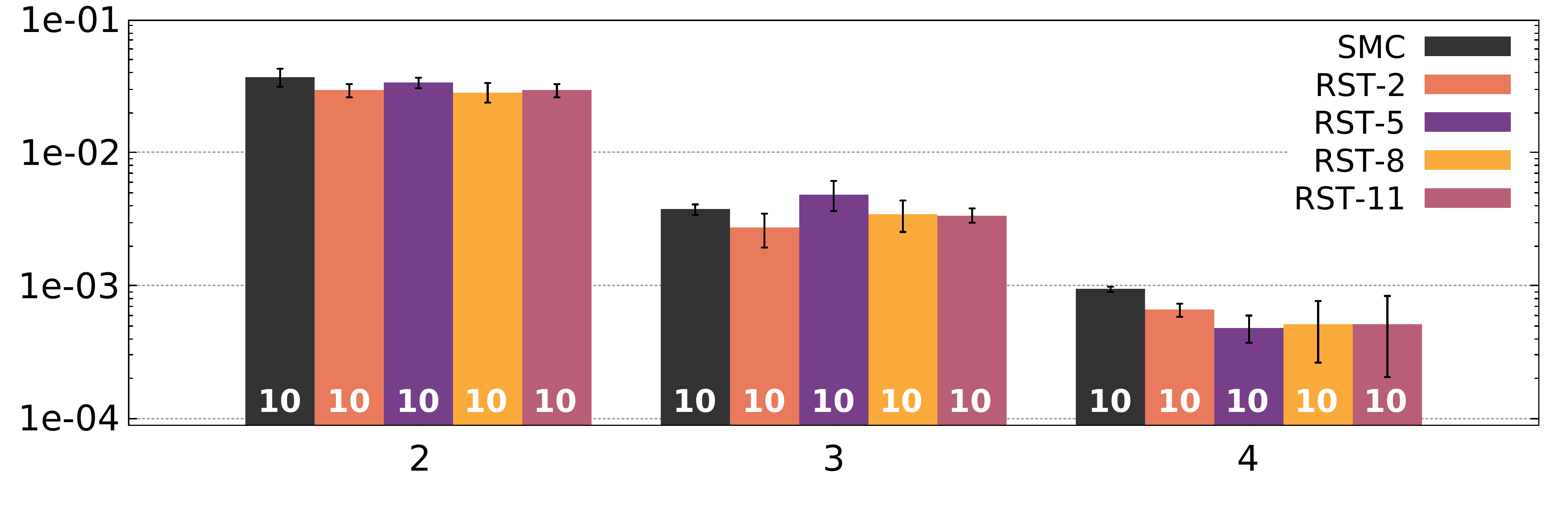}
		\vspace{-2.5ex}
		\caption{RWC}
		\label{fig:expe:results:rel:rwc}
	\end{subfigure}
	
	\begin{subfigure}[b]{\linewidth}
		\centering
		\includegraphics[width=.8\linewidth]{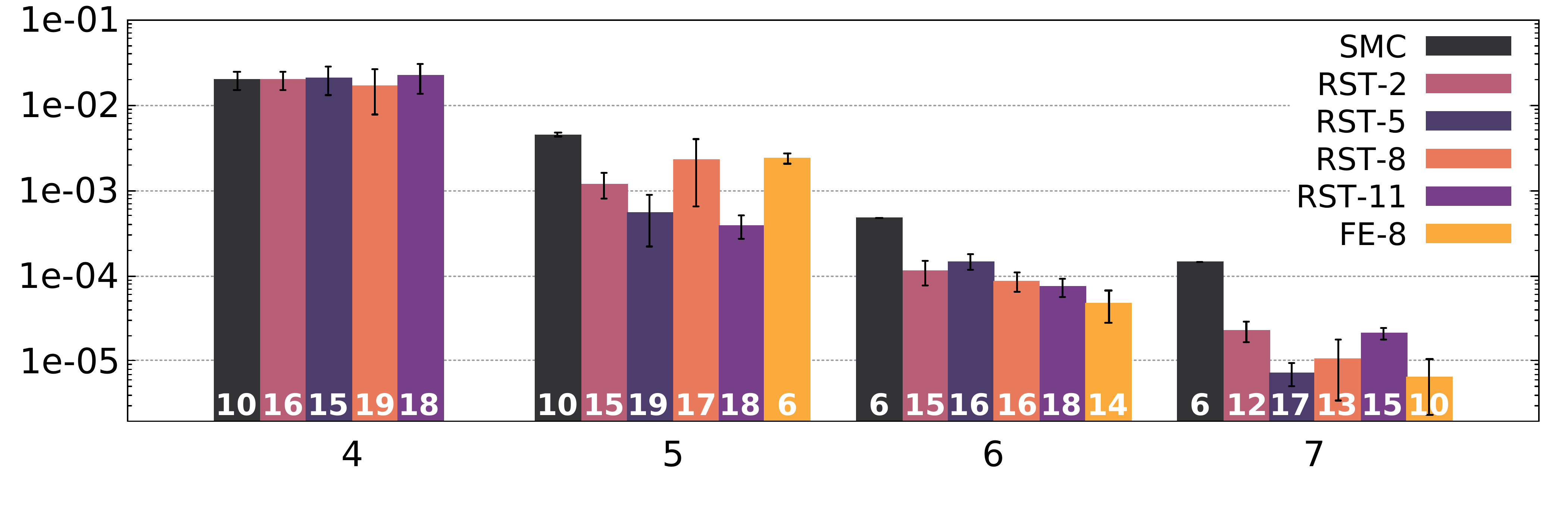}
		\vspace{-2.5ex}
		\caption{HVC}
		\label{fig:expe:results:rel:hvc}
	\end{subfigure}
	\vspace{-2ex}
\end{figure}

The overall trend shown for unreliability estimations is similar to the previous unavailability cases.
Here however it was possible to use Fixed Effort, since every simulation has a clearly defined end at time $T=10^3$.
It is interesting thus to compare the efficiency of \restart vs. \fe: we note for example that some variants of \fe performed considerably better than any other approach in the most resilient configurations of FTPP and HECS.
It is nevertheless difficult to draw general conclusions from \Crefrange{fig:expe:results:rel:dspare}{fig:expe:results:rel:hecs}, since some variants that performed best in a case study---e.g.\ \fen{16} in HECS---did worse in others---e.g.\ FTPP, where the best algorithms were \fen{8,12}.
Furthermore, \fen{8}, which is always better than \smc when $\UNREL[1000]<10^{-3}$, did not perform very well in HVC, where the algorithms that achieved the narrowest and most not-null \ci{s} were \rstn{5,11}.
Such cases notwithstanding, \fe is a solid competitor of \restart in our benchmark.

Another relevant point of study is the optimal effort $e$ for \rstn{e} or \fen{e}, which shows no clear trend in our experiments.
Here, $e$ is a ``global effort'' used by these algorithms, equal for all $\states_k$ regions.
$e$ also alters the way in which the thresholds selection algorithm Sequential Monte Carlo (\acronym{seq} \cite{BDM16}) selects the $\ell_k$.
The lack of guidelines to select a value for $e$ that works well across different systems was raised in \cite{Bud17}.
This motivated the development of Expected Success (\acronym{es} \cite{BDH17}), which selects efforts individually per $\states_k$ (or $\ell_k$).
Thus, in \rstes, a trace up-crossing threshold $\ell_k$ is split according to the individual effort $e_k$ selected by \acronym{es}.
In the benchmark of \cite{BDH17}, which consists mostly of queueing systems, \acronym{es} was shown superior to \acronym{seq}.
However, experimental outcomes on \dft{s} in this work are different: for \UNAVA, \rstes yielded mildly good results for HECS and RC; for the other case studies and for all \UNREL[1000] experiments, \rstes always yielded null \ci{s}.
It was found that the effort selected for most thresholds $\ell_k$ was either too small---so splitting in $e_k$ was not enough for the \rstes trace to reach $\ell_{k+1}$---or too large---so there was a splitting/truncation overhead.
This point is further addressed in the conclusions.

Beyond comparisons among the specific algorithms, be these for \res or for selecting thresholds, it seems clear that our approach to \fta via \ISPLIT deploys the expected results.
For each parameterised case study \cs[p], we could find a value of the parameter $\mathfrak{p}$ where the level of resilience is such, that \smc is less efficient than our automatically-constructed \ISPLIT framework.
This is particularly significant for big \dft{s} like HECS and RWC, whose complex structure could be exploited by our importance function.

}%
	{\input{sections/experiments_short}}%
%
%%%%%%%%%%%%%%%%%%%%%%%%%%%%%%%%%%%%%%%%%%%%%%%%%%%%%%%%%%%%%%%%%%%%%%%%%%%%%%
%% !TEX root =  ../main.tex
\section{Related work}
\label{sec:relwork}

%\commentMS{One idea per sentence. Also: easy info can be in longer sentences. Complex info in short ones.}

% DFTs and rFTs: dynamic gates + repairs + arbitrary PDFs
Most work on \dft analysis assumes discrete \cite{VSDFMR02,BRFH10} or exponentially distributed \cite{CBS07,KS17} components failure.
Furthermore, components repair is seldom studied in conjunction with dynamic gates \cite{bobbio2004parametric,BRFH10,ruijters2015fault,KS17,LWK17}.
In this work we address repairable \dft{s}, whose failure and repair times can follow arbitrary \pdf{s}.
More in detail, \rft{s} were first formally introduced as stochastic Petri nets in \cite{bobbio2004parametric,DBLP:conf/dsn/RaiteriIFV04}.
Our work stands on \cite{Mon18}, which reviews \cite{DBLP:conf/dsn/RaiteriIFV04} in the context of stochastic automata with arbitrary \pdf{s}.
In particular we also address non-Markovian continuous distributions: in \Cref{sec:expe} we experimented with exponential, Erlang, uniform, Rayleigh, Weibull, normal, and log-normal \pdf{s}.
Furthermore and for the first time, we consider the application of \cite{DBLP:conf/dsn/RaiteriIFV04,Mon18} to study rare events.

% RES: importance sampling and why not
Much effort in \res has been dedicated to study highly reliable systems, deploying either importance splitting or sampling.
%\cite{GSHNG92,Hei95,VA98,NSN01,Rid05,XLL07,LLLT09,BDM16,VA18,RRBS19}
Typically, importance sampling can be used when the system takes a particular shape.
For instance, a common assumption is that all failure (and repair) times are exponentially distributed with parameters $\lambda^i$, for some $\lambda\in\R$ and $i\in\N_{>0}$.
In these cases, a favourable change of measure can be computed analytically \cite{GSHNG92,Hei95,NSN01,Rid05,XLL07,RRBS19}.

% RES: importance splitting and how we're better
In contrast, when the fail/repair times follow less-structured distributions, importance splitting is more easily applicable.
As long as a full system failure can be broken down into several smaller components failures, an importance splitting method can be devised.
Of course, its efficiency relies heavily on the choice of importance function.
This choice is typically done ad hoc for the model under study \cite{VA98,LLLT09,VA18}.
In that sense \cite{JLS13,JLST15,BDH15,BDM16} are among the first to attempt a heuristic derivation of all parameters required to implement splitting.
This is based on formal specifications of the model and property query (the dependability metric).
Here we extended \cite{BDH15,BDM16,Bud17}, using the structure of the fault tree to define composition operands.
With these operands we aggregate the automatically-computed local importance functions of the tree nodes.
This aggregation results in an importance function for the whole model.
%, with which we deploy automatic \res for the computation of dependability metrics in \rft{s}.

%
%%%%%%%%%%%%%%%%%%%%%%%%%%%%%%%%%%%%%%%%%%%%%%%%%%%%%%%%%%%%%%%%%%%%%%%%%%%%%%
%% !TEX root =  ../main.tex
\section{Conclusions}
\label{sec:conclu}

We have presented a theory to deploy \emph{automatic importance splitting} (\ISPLIT) for fault tree analysis of repairable dynamic fault trees (\rft{s}).
This Rare Event Simulation approach supports arbitrary probability distributions of components failure and repair.
The core of our theory is an importance function $\IFUN_\Tree$ defined structurally on the tree. % in \Cref{tab:compifun}.
From such function we implemented \ISPLIT algorithms, and used them to estimate the \emph{unreliability} and \emph{unavailability} of highly-resilient \rft{s}.
%Our algorithms are as automatic as standard approaches to \fta by Monte Carlo simulation.
Departing from classical approaches, that define importance functions ad hoc using expert knowledge, our theory computes all metadata required for \res from the model and metric specifications.
Nonetheless, we have shown that for a fixed simulation time budget and in the most resilient \rft{s}, diverse \ISPLIT algorithms can be automatically implemented from $\IFUN_\Tree$, and always converge to narrower confidence intervals than standard Monte Carlo simulation.

There are several paths open for future development.
First and foremost, we are looking into new ways to define the importance function, e.g.\ to cover more general categories of \ft{s} such as fault maintenance trees \cite{RGD+16}.
It would also be interesting to look into possible correlations among specific \res algorithms and tree structures, that yield the most efficient estimations for a particular metric.
Moreover, we have defined $\IFUN_\Tree$ based on the tree structure alone.
It would be interesting to further include stochastic information in this phase, and not only afterwards during the thresholds-selection phase.

Regarding thresholds, the relatively bad performance of the Expected Success algorithm shows a spot for improvement.
In general, we believe that enhancing its statistical properties should alleviate the behaviour mentioned in \Cref{sec:expe:results}.
Moreover, techniques to increase trace independence during splitting (e.g.\ resampling) could further improve the performance of the \ISPLIT algorithms.
Finally, we are investigating enhancements in \iosa and our tool chain, to exploit the ratio between fail and dormancy \pdf{s} of \SBE{s} in warm~\SPAREgate~gates.

\section*{Acknowledgments}

The authors thank Jos\'e and Manuel Vill\'en-Altamirano for fruitful discussions, who helped to better understand the application scope of our approach.

%%	% Following needed for IEEE documentclass only;
%%	% LNCS has acknowledgements in title (not needed here)
%%	This work was partially funded by NWO, NS, and ProRail project 15474 (\emph{SEQUOIA}), ERC grant 695614 (\emph{POWVER}), EU project 102112 (\emph{SUCCESS}), ANPCyT PICT-2017-3894 (\emph{RAFTSys}), and SeCyT project 33620180100354CB (\emph{ARES}).

%

%%	\bibliographystyle{../../splncs04}
%%	\bibliography{\jobname}  % assume bib file homonymous to source file

\begin{thebibliography}{10}
\providecommand{\url}[1]{\texttt{#1}}
\providecommand{\urlprefix}{URL }
\providecommand{\doi}[1]{https://doi.org/#1}

\bibitem{ABCHS18}
Abate, A., Budde, C.E., Cauchi, N., Hoque, K.A., Stoelinga, M.: Assessment of
  maintenance policies for smart buildings: Application of formal methods to
  fault maintenance trees. {PHM} Society European Conference  \textbf{4}(1)
  (2018)

\bibitem{BKH99}
Baier, C., Katoen, J., Hermanns, H.: Approximate symbolic model checking of
  continuous-time markov chains. In: {CONCUR}~1999. pp. 146--161 (1999).
  \doi{10.1007/3-540-48320-9\_12}

\bibitem{Bay70}
Bayes, A.J.: Statistical techniques for simulation models. Australian computer
  journal  \textbf{2}(4),  180--184 (1970)

\bibitem{BRFH10}
Beccuti, M., Raiteri, D., Franceschinis, G., Haddad, S.: Non deterministic
  repairable fault trees for computing optimal repair strategy. In:
  {VALUETOOLS}~2008 (2010). \doi{10.4108/ICST.VALUETOOLS2008.4411}

\bibitem{BM09}
Blanchet, J., Mandjes, M.: Rare event simulation for queues. In: Rubino and
  Tuffin  \cite{RT09}, pp. 87--124. \doi{10.1002/9780470745403.ch5}

\bibitem{BBK09}
Blom, H.A.P., Bakker, G.J.B., Krystul, J.: Rare event estimation for a
  large-scale stochastic hybrid system with air traffic application. In: Rubino
  and Tuffin  \cite{RT09}, pp. 193--214. \doi{10.1002/9780470745403.ch9}

\bibitem{bobbio2004parametric}
Bobbio, A., Raiteri, D.: Parametric fault trees with dynamic gates and repair
  boxes. In: {RAMS}'04. pp. 459--465 (2004). \doi{10.1109/RAMS.2004.1285491}

\bibitem{BCH+08}
{Boudali}, H., {Crouzen}, P., {Haverkort}, B.R., {Kuntz}, M., {Stoelinga}, M.:
  Architectural dependability evaluation with arcade. In: {DSN}'08. pp.
  512--521 (2008). \doi{10.1109/DSN.2008.4630122}

\bibitem{BD05}
Boudali, H., Dugan, J.B.: A new bayesian network approach to solve dynamic
  fault trees. In: Annual Reliability and Maintainability Symposium, 2005.
  Proceedings. pp. 451--456 (2005). \doi{10.1109/RAMS.2005.1408404}

\bibitem{Bud17}
Budde, C.E.:
  \href{https://git.snt.utwente.nl/buddece/CEB_thesis/raw/master/thesis.pdf}{Automation
  of Importance Splitting Techniques for Rare Event Simulation}. Ph.D. thesis,
  Universidad Nacional de C\'ordoba, C\'ordoba, Argentina (2017)

\bibitem{BDH17}
Budde, C.E., D'Argenio, P.R., Hartmanns, A.: Better automated importance
  splitting for transient rare events. In: {SETTA}. LNCS, vol. 10606, pp.
  42--58. Springer (2017)

\bibitem{BDH15}
Budde, C.E., D'Argenio, P.R., Hermanns, H.: Rare event simulation with fully
  automated importance splitting. In: {EPEW} 2015. {LNCS}, vol.~9272, pp.
  275--290. Springer (2015). \doi{10.1007/978-3-319-23267-6\_18}

\bibitem{BDM16}
Budde, C.E., D'Argenio, P.R., Monti, R.E.: Compositional construction of
  importance functions in fully automated importance splitting. In:
  {VALUETOOLS}~2016. pp. 30--37 (2017). \doi{10.4108/eai.25-10-2016.2266501}

\bibitem{BDH+17}
Budde, C.E., Dehnert, C., Hahn, E.M., Hartmanns, A., Junges, S., Turrini, A.:
  {JANI}: Quantitative model and tool interaction. In: {TACAS}. {LNCS}, vol.
  10206, pp. 151--168. Springer (2017). \doi{10.1007/978-3-662-54580-5\_9}

\bibitem{CSD00}
Coppit, D., Sullivan, K.J., Dugan, J.B.: Formal semantics of models for
  computational engineering: a case study on dynamic fault trees. In:
  {ISSRE}~2000. pp. 270--282 (2000). \doi{10.1109/ISSRE.2000.885878}

\bibitem{CS00}
Coppit, D., Sullivan, K.J.: Galileo: A tool built from mass-market
  applications. In: Software Engineering, 2000. Proceedings of the 2000
  International Conference on. pp. 750--753. IEEE (2000)

\bibitem{CBS07}
Crouzen, P., Boudali, H., Stoelinga, M.: Dynamic fault tree analysis using
  input/output interactive markov chains. In: {DSN}~2007. pp. 708--717 (2007).
  \doi{10.1109/DSN.2007.37}

\bibitem{DArgenioLM16:formats}
D'Argenio, P.R., Lee, M.D., Monti, R.E.: Input/output stochastic automata -
  compositionality and determinism. In: {FORMATS}~2016. {LNCS}, vol.~9884, pp.
  53--68 (2016). \doi{10.1007/978-3-319-44878-7\_4}

\bibitem{DM18}
D'Argenio, P.R., Monti, R.E.: {I}nput/{O}utput {S}tochastic {A}utomata with
  {U}rgency: Confluence and weak determinism. In: {ICTAC}. {LNCS}, vol. 11187,
  pp. 132--152. Springer (2018). \doi{10.1007/978-3-030-02508-3\_8}

\bibitem{DP07}
Distefano, S., Puliafito, A.: Dependability modeling and analysis in dynamic
  systems. In: 2007 IEEE International Parallel and Distributed Processing
  Symposium. pp.~1--8 (2007). \doi{10.1109/IPDPS.2007.370601}

\bibitem{DBB90}
Dugan, J.B., Bavuso, S.J., Boyd, M.A.: Fault trees and sequence dependencies.
  In: Annual Proceedings on Reliability and Maintainability Symposium. pp.
  286--293 (1990). \doi{10.1109/ARMS.1990.67971}

\bibitem{GOK02}
Garvels, M.J.J., van Ommeren, J.K.C.W., Kroese, D.P.: On the importance
  function in splitting simulation. European Transactions on Telecommunications
   \textbf{13}(4),  363--371 (2002). \doi{10.1002/ett.4460130408}

\bibitem{Gar00}
Garvels, M.J.J.: The splitting method in rare event simulation. Ph.D. thesis,
  Department of Computer Science, University of Twente (2000)

\bibitem{GSHNG92}
Goyal, A., Shahabuddin, P., Heidelberger, P., Nicola, V.F., Glynn, P.W.: A
  unified framework for simulating markovian models of highly dependable
  systems. IEEE Transactions on Computers  \textbf{41}(1),  36--51 (1992).
  \doi{10.1109/12.123381}

\bibitem{GSS15}
Guck, D., Spel, J., Stoelinga, M.: {DFTCalc}: Reliability centered maintenance
  via fault tree analysis (tool paper). In: {ICFEM}~2015. pp. 304--311.
  Springer (2015)

\bibitem{GKS+14}
Guck, D., Katoen, J.P., Stoelinga, M., Luiten, T., Romijn, J.: Smart railroad
  maintenance engineering with stochastic model checking. In: Proceedings of
  the Second International Conference on Railway Technology: Research,
  Development and Maintenance, Railways 2014. Civil-Comp Proceedings,
  Civil-Comp Press (2014). \doi{10.4203/ccp.104.299}

\bibitem{HJ94}
Hansson, H., Jonsson, B.: A logic for reasoning about time and reliability.
  Formal Aspects of Computing  \textbf{6}(5),  512--535 (1994).
  \doi{10.1007/BF01211866}

\bibitem{Hei95}
Heidelberger, P.: Fast simulation of rare events in queueing and reliability
  models. {ACM} Trans. Model. Comput. Simul.  \textbf{5}(1),  43--85 (1995).
  \doi{10.1145/203091.203094}

\bibitem{IH93}
Iglewicz, B., Hoaglin, D.: How to Detect and Handle Outliers. {ASQC} basic
  references in quality control, {ASQC} {Q}uality {P}ress (1993)

\bibitem{JLS13}
Jegourel, C., Legay, A., Sedwards, S.: Importance splitting for statistical
  model checking rare properties. In: {CAV}. pp. 576--591. Springer Berlin
  Heidelberg (2013)

\bibitem{JLST15}
J{\'{e}}gourel, C., Legay, A., Sedwards, S., Traonouez, L.: Distributed
  verification of rare properties with lightweight importance splitting
  observers. CoRR  \textbf{abs/1502.01838} (2015),
  \url{http://arxiv.org/abs/1502.01838}

\bibitem{JGKRS15}
Junges, S., Guck, D., Katoen, J.P., Rensink, A., Stoelinga, M.: Fault trees on
  a diet. In: {SETTA}~2015. {LNCS}, vol.~9409, pp. 3--18. Springer (2015).
  \doi{10.1007/978-3-319-25942-0\_1}

\bibitem{JGKS16}
Junges, S., Guck, D., Katoen, J., Stoelinga, M.: Uncovering dynamic fault
  trees. In: {DSN}~2016. pp. 299--310. {IEEE} (2016). \doi{10.1109/DSN.2016.35}

\bibitem{KH51}
Kahn, H., Harris, T.E.: Estimation of particle transmission by random sampling.
  National Bureau of Standards applied mathematics series  \textbf{12},  27--30
  (1951)

\bibitem{KS17}
Katoen, J.P., Stoelinga, M.: Boosting Fault Tree Analysis by Formal Methods,
  pp. 368--389. Springer (2017). \doi{10.1007/978-3-319-68270-9\_19}

\bibitem{LLLT09}
L'Ecuyer, P., Le~Gland, F., Lezaud, P., Tuffin, B.: Splitting techniques. In:
  Rubino and Tuffin  \cite{RT09}, pp. 39--61. \doi{10.1002/9780470745403.ch3}

\bibitem{LWK17}
Liu, Y., Wu, Y., Kalbarczyk, Z.: Smart maintenance via dynamic fault tree
  analysis: A case study on singapore {MRT} system. In: {DSN}~2017. pp.
  511--518 (2017). \doi{10.1109/DSN.2017.50}

\bibitem{Mon18}
Monti, R.E.: Stochastic Automata for Fault Tolerant Concurrent Systems. Ph.D.
  thesis, Universidad Nacional de C\'ordoba, Argentina (2018)

\bibitem{NSN01}
Nicola, V.F., Shahabuddin, P., Nakayama, M.K.: Techniques for fast simulation
  of models of highly dependable systems. IEEE Transactions on Reliability
  \textbf{50}(3),  246--264 (2001). \doi{10.1109/24.974122}

\bibitem{DBLP:conf/dsn/RaiteriIFV04}
Raiteri, D., Iacono, M., Franceschinis, G., Vittorini, V.: Repairable fault
  tree for the automatic evaluation of repair policies. In: {DSN}~2004. pp.
  659--668 (2004). \doi{10.1109/DSN.2004.1311936}

\bibitem{Rid05}
Ridder, A.: Importance sampling simulations of markovian reliability systems
  using cross-entropy. Annals of Operations Research  \textbf{134}(1),
  119--136 (2005). \doi{10.1007/s10479-005-5727-9}

\bibitem{RT09b}
Rubino, G., Tuffin, B.: Introduction to rare event simulation. In: Rare Event
  Simulation Using {M}onte {C}arlo Methods \cite{RT09}, pp. 1--13.
  \doi{10.1002/9780470745403.ch1}

\bibitem{RT09}
Rubino, G., Tuffin, B. (eds.): Rare Event Simulation Using {M}onte {C}arlo
  Methods. John Wiley \& Sons, Ltd (2009)

\bibitem{RGD+16}
Ruijters, E., Guck, D., Drolenga, P., Peters, M., Stoelinga, M.: Maintenance
  analysis and optimization via statistical model checking - evaluating a train
  pneumatic compressor. In: {QEST}~2016. {LNCS}, vol.~9826, pp. 331--347.
  Springer (2016)

\bibitem{RGNS16}
Ruijters, E., Guck, D., van Noort, M., Stoelinga, M.: Reliability-centered
  maintenance of the electrically insulated railway joint via fault tree
  analysis: {A} practical experience report. In: {DSN}~2016. pp. 662--669.
  {IEEE} Computer Society (2016)

\bibitem{RRBS17}
Ruijters, E., Reijsbergen, D., de~Boer, P.T., Stoelinga, M.: {R}are {E}vent
  {S}imulation for {D}ynamic {F}ault {T}rees. In: Computer Safety, Reliability,
  and Security. pp. 20--35. Springer International Publishing (2017)

\bibitem{RRBS19}
Ruijters, E., Reijsbergen, D., de~Boer, P.T., Stoelinga, M.: Rare event
  simulation for dynamic fault trees. Reliability Engineering \& System Safety
  \textbf{186},  220--231 (2019). \doi{10.1016/j.ress.2019.02.004}

\bibitem{ruijters2015fault}
Ruijters, E., Stoelinga, M.: Fault tree analysis: A survey of the
  state-of-the-art in modeling, analysis and tools. Computer science review
  \textbf{15},  29--62 (2015)

\bibitem{Sch77}
Schassberger, R.: {Insensitivity of Steady-State Distributions of Generalized
  Semi-Markov Processes. Part I}. Ann. Probab.  \textbf{5}(1),  87--99 (1977).
  \doi{10.1214/aop/1176995893}

\bibitem{GALILEO}
Sullivan, K.J., Dugan, J.B.: {G}alileo user's manual \& design overview.
  \url{https://www.cse.msu.edu/~cse870/Materials/FaultTolerant/manual-galileo.htm}
  (1998), v2.1-alpha

\bibitem{SDC99}
Sullivan, K., Dugan, J., Coppit, D.: The {G}alileo fault tree analysis tool.
  In: 29th Annual International Symposium on Fault-Tolerant Computing (Cat.
  No.99CB36352). pp. 232--235 (1999). \doi{10.1109/FTCS.1999.781056}

\bibitem{VSDFMR02}
Vesely, W., Stamatelatos, M., Dugan, J., Fragola, J., Minarick, J., Railsback,
  J.: Fault tree handbook with aerospace applications. {NASA} Office of Safety
  and Mission Assurance  (2002), version 1.1

\bibitem{VA98}
Vill\'en-Altamirano, J.: {RESTART} method for the case where rare events can
  occur in retrials from any threshold. Int.\ J.\ Electron.\ Commun.
  \textbf{52},  183--189 (1998)

\bibitem{VA07}
Vill\'en-Altamirano, J.: Importance functions for {RESTART} simulation of
  highly-dependable systems. Simulation  \textbf{83}(12),  821--828 (2007).
  \doi{10.1177/0037549707081257}

\bibitem{VA18}
Vill{\'e}n-Altamirano, J.: {RESTART} vs splitting: A comparative study.
  Performance Evaluation  \textbf{121-122},  38--47 (2018).
  \doi{10.1016/j.peva.2018.02.002}

\bibitem{VAM+94}
Vill{\'e}n-Altamirano, M., Mart{\'i}nez-Marr{\'o}n, A., Gamo, J.,
  Fern{\'a}ndez-Cuesta, F.: Enhancement of the accelerated simulation method
  {RESTART} by considering multiple thresholds. In: Proc.\
  14\textsuperscript{th} Int.\ Teletraffic Congress, Teletraffic Science and
  Engineering, vol.~1, pp. 797--810. Elsevier (1994).
  \doi{10.1016/B978-0-444-82031-0.50084-6}

\bibitem{VAVA91}
Vill\'en-Altamirano, M., Vill\'en-Altamirano, J.: {RESTART}: a method for
  accelerating rare event simulations. In: Queueing, Performance and Control in
  ATM (ITC-13). pp. 71--76. Elsevier (1991)

\bibitem{XLL07}
Xiao, G., Li, Z., Li, T.: Dependability estimation for non-markov
  consecutive-k-out-of-n: F repairable systems by fast simulation. Reliability
  Engineering \& System Safety  \textbf{92}(3),  293--299 (2007).
  \doi{10.1016/j.ress.2006.04.004}

\end{thebibliography}

\end{document}